\documentclass[a4paper,fleqn,usenatbib]{mnras}

\usepackage{mathptmx}

\usepackage[T1]{fontenc}
\usepackage{ae,aecompl}

\usepackage{graphicx}	
\usepackage{amsmath}	
\usepackage{amssymb}	

\usepackage[skip=0pt]{caption}

\title[Atomic data and nebular abundances]{The impact of atomic data selection on nebular abundance determinations}

\author[Juan de Dios \& Rodr\'iguez]{
Leticia Juan de Dios,\thanks{E-mail: leticiajd@inaoep.mx}
M\'onica Rodr\'iguez,
\\
Instituto Nacional de Astrof\'isica \'Optica y Electr\'onica, Luis Enrique Erro 1, Tonantzintla 72840, Puebla, Mexico\\
}

\date{Accepted 2017 April 11. Received 2017 April 7; in original form 2017 February 24}

\pubyear{2017}

\begin{document}
\label{firstpage}
\pagerange{\pageref{firstpage}--\pageref{lastpage}}
\maketitle

\begin{abstract}
Atomic data are an important source of systematic uncertainty in our determinations of nebular chemical abundances. However, we do not have good estimates of these uncertainties since it is very difficult to assess the accuracy of the atomic data involved in the calculations. We explore here the size of these uncertainties by using 52 different sets of transition probabilities and collision strengths, and all their possible combinations, to calculate the physical conditions and the total abundances of O, N, S, Ne, Cl, and Ar for a sample of planetary nebulae and \ion{H}{ii} regions. We find that atomic data variations introduce differences in the derived abundance ratios as low as 0.1--0.2~dex at low density, but that reach or surpass 0.6--0.8~dex at densities above 10$^{4}$ cm$^{-3}$ in several abundance ratios, like O/H and N/O. Removing from the 52 datasets the four datasets that introduce the largest differences, the total uncertainties are reduced, but high density objects still reach uncertainty factors of four for their values of O/H and N/O. We identify the atomic data that introduce most of the uncertainty, which involves the ions used to determine density, namely, the transition probabilities of the S$^{+}$, O$^{+}$, Cl$^{++}$, and Ar$^{+3}$ density diagnostic lines, and the collision strengths of Ar$^{+3}$. Improved calculations of these data will be needed in order to derive more reliable values of chemical abundances in high density nebulae. In the meantime, our results can be used to estimate the uncertainties introduced by atomic data in nebular abundance determinations.
\end{abstract}

\begin{keywords}
atomic data - \ion{H}{ii} regions - planetary nebulae: general
\end{keywords}



\section{Introduction}

The determination of chemical abundances in ionized nebulae requires the use of atomic data, such as transition probabilities and collision strengths, whose values are derived from theoretical calculations. The accuracy of these data is difficult to assess, since experimental measurements are generally unavailable and cannot be easily obtained. When there are several atomic data calculations for the same ion, it is not unusual to find important differences between them. For example, \citet{GMZ97} compare transition probabilities from various sources and find differences around 10--20 per cent for most transitions, but that reach or exceed 50 per cent in some cases. Similarly, \citet{GRHK14} find differences of up to 50 per cent when comparing their collision strengths for S$^{++}$ with previous calculations. The discrepancies can be due to errors in the atomic codes or to the different approximations that must necessarily be made in order to perform the calculations, making the assessment of atomic data a very complex issue \citep{Aggarwal13}.

Several works have explored the effects of using different atomic data in the determination of physical conditions and chemical abundances in ionized nebulae, finding that the choice of data can have a significant impact on the results \citep[e.g.][]{rod99, Copetti2002, Wang04, Luridiana12, Stasinska13, Mendoza14}. However, a single set of atomic data can affect all the calculations, and different combinations of atomic data will affect the results in complex ways. To the best of our knowledge, no previous work has explored the combined effect of using many different datasets for all the observed ions, propagating the differences through all the calculations.

Here we use {\sc pyneb}, a code for the analysis of nebular emission lines developed by \citet*{Pyneb}, and the 52 atomic datasets of transition probabilities and collision strengths that it includes for nine ions, in order to explore the dispersion introduced by atomic data in the determination of physical conditions and chemical abundances in a sample of 36 PNe and 8 \ion{H}{ii} regions that have the best available spectra in the literature. Assuming that the atomic data calculated by different authors with different approximations provide good estimates of atomic data uncertainties, our results can be used to constrain the systematic uncertainties introduced by atomic data in the physical conditions and chemical abundances of ionized nebulae.

\section{Observational data}

Most determinations of chemical abundances in ionized nebulae are based on the analysis of optical spectra. Hence, we have selected a sample of 36 PNe and eight \ion{H}{ii} regions with available optical spectra which are deep enough to allow the measurement of a large number of emission lines that can be used to derive physical conditions and chemical abundances. The sample objects cover a wide range in characteristics such as excitation degree and electron density, and in that sense can be considered representative of the population of ionized nebulae. We used as initial selection criteria the presence in the published spectra of lines belonging to the four most used density diagnostics, [\ion{S}{ii}] $\lambda6716/\lambda6731$, [\ion{O}{ii}] $\lambda3726/\lambda3729$, [\ion{Cl}{iii}] $\lambda5518/\lambda5538$, and [\ion{Ar}{iv}] $\lambda4711/\lambda4740$, and the two most used temperature diagnostics, [\ion{N}{ii}] $\lambda5755/(\lambda6548+\lambda6583)$ and [\ion{O}{iii}] $\lambda4363/(\lambda4959+\lambda5007)$. Besides these diagnostics, the spectra of the sample allow us to derive the ionic abundances of O$^{+}$, O$^{++}$, N$^{+}$, Cl$^{++}$, Ar$^{++}$, Ar$^{+3}$, Ne$^{++}$, S$^{+}$, and S$^{++}$ from collisionally excited lines, in addition to the total abundances of the corresponding elements. Since the initial selection lacked nebulae having both low density, $n_{\rm{e}}\la10^3$~cm$^{-3}$, and a low degree of ionization, $\log(\mbox{O}^+/\mbox{O}^{++})>0.6$, we added to the sample the Galactic \ion{H}{ii} regions M16 and M20, even though the [\ion{Ar}{iv}] density diagnostic is not available for these objects.

The 36 PNe and the references for their spectra are IC~2165 \citep{Hyung94},  NGC~6153 \citep{Liu00}, IC~4846 \citep{Hyung01}, IC~5217 \citep{Hyung01b}, M~1-42, M~2-36 \citep{Liu01}, IC~418 \citep{Sharpee03}, Hu~1-2, NGC~6210, NGC~6572, NGC~6720, NGC~6741, NGC~6826, NGC~6884, NGC~7662 \citep{Liu04}, NGC~6803 \citep{Wesson05}, Cn~2-1, H~1-50, He~2-118, M~1-20, M~2-4, M~3-21, M~3-32, NGC~6439, NGC~6567, NGC~6620 \citep{Wang07}, NGC~2867 \citep{GarciaR09}, NGC~7009 \citep{Fang11}, Cn~1-5, He~2-86, M~1-61, M~3-15, NGC~5189, NGC~6369, PC14, and Pe~1-1 \citep{GarciaR12}. The eight selected \ion{H}{ii} regions and their references are M42 \citep{Esteban04}, 30~Doradus \citep{PeimbertA03}, NGC~3576 \citep{GarciaR04}, M16, M20, NGC~3603 \citep{GarciaR06}, M8, and M17 \citep{GarciaR07}.

\section{Atomic data}

The emissivities of the collisionally excited lines that we will be using can be obtained by solving the equations of statistical equilibrium for the lower laying 5--8 levels of each ion \citep{Osterbrock}. All downward radiative transitions, electron collisional excitations and de-excitations are included in the calculations. Therefore, the results of these calculations depend on the values used for the Einstein coefficients, $A_{ij}$, and the effective collision strengths, $\Upsilon_{ij}$ (the Maxwellian-averaged collision strengths). Recombination can also contribute to the population of the upper levels of some transitions. This effect can be important for at least one of the lines we are using, [\ion{N}{ii}]~$\lambda5755$ \citep{Rubin86,Liu00}, as we discuss in Section~\ref{sec:physical conditions}.

Table~\ref{tab:atomic_data} shows the atomic datasets of transition probabilities and collision strengths that are included in {\sc pyneb} for the ions that we are considering in our calculations. For ease of comparison, we adopt the same notation used in {\sc pyneb} for the atomic data references. The datasets established by default in {\sc pyneb} are labelled with the letter D; the other datasets are assigned a number $n$ that we will use for identification purposes below. Note that some datasets include data from different works and that not all datasets are independent. For example, the sets of collisions strengths for O$^+$ identified as 8 and 9 in Table~\ref{tab:atomic_data} both include data from \citet{P76} and \citet{McLB93} but these data are used for different transitions in each dataset.

\begin{table}
\renewcommand{\arraystretch}{0.63}
	\caption{Atomic data used in our calculations. The letter D identifies the default atomic data in {\sc pyneb}.}
	\begin{tabular}{l c l c l}
\hline
Ion		&	$n$	&	$A_{ij}$ 	&	$n$	&	$\Upsilon_{ij}$	\\
\hline
N$^{+}$ 	&	D	&	GMZ97-WFD96		&	D	&	T11	\\
		&	1	&	FFT04			&	3	&	HB04	\\
		&	2	&	NR79-WFD96		&	4	&	LB94	\\
		&		&				&		&		\\
O$^{+}$ 	&	D	&	Z82-WFD96		&	D	&	P06-T07	\\
		&	5	&	FFT04			&	7	&	Kal09	\\
		&	6	&	WFD96			&	8	&	P76-McLB93-v1	\\
		&		&				&	9	&	P76-McLB93-v2	\\
		&		&				&	10	&	T07	\\
		&		&				&		&		\\
O$^{++}$ 	&	D	&	SZ00-WFD96		&	D	&	AK99	\\
		&	11	&	FFT04			&	13	&	LB94	\\
		&	12	&	GMZ97-WFD96		&	14	&	Pal12-AK99	\\
		&		&				&	15	&	SSB14	\\
		&		&				&		&		\\
Ne$^{++}$ 	&	D	&	GMZ97			&	D	&	McLB00	\\
		&		&				&	16	&	BZ94	\\
		&		&				&		&		\\	
S$^{+}$ 	&	D	&	PKW09			&	D	&	TZ10	\\
		&	17	&	TZ10-PKW09		&	20	&	RBS96	\\
		&	18	&	VVF96-KHOC93		&		&		\\
		&	19	&	VVF96-MZ82a$^1$	&		&		\\
		&		&				&		&		\\	
S$^{++}$ 	&	D	&	PKW09			&	D	&	TG99	\\
		&	21	&	LL93-HSC95-MZ82b-KS86	&	23	&	GMZ95	\\
		&	22	&	MZ82b-HSC95-LL93	&	24	&	GRHK14	\\
		&		&				&	25	&	HRS12	\\
		&		&				&		&		\\	
Cl$^{++}$ 	&	D	&	M83-KS86		&	D	&	BZ89	\\
		&	26	&	Fal99			&	28	&	M83	\\
		&	27	&	M83			&		&		\\
		&		&				&		&		\\	
Ar$^{++}$ 	&	D	&	M83-KS86		&	D	&	GMZ95	\\
		&	29	&	MB09			&	31	&	MB09	\\
		&	30	&	MZ83			&		&		\\
		&		&				&		&		\\	
Ar$^{3+}$ 	&	D	&	MZ82a$^2$		&	D	&	RB97	\\
		&	32	&	MZ82a-KS86		&	33	&	M83	\\
		&		&				&	34	&	ZBL87	\\
\hline
\end{tabular}
References for the atomic data: AK99: \citet{AK99}, BZ89: \citet{BZ89}, BZ94: \citet{BZ94}, Fal99: \citet{Fal99}, FFT04: \citet{FFT04}, GMZ95: \citet{GMZ95}, GMZ97: \citet{GMZ97}, GRHK14: \citet{GRHK14}, HB04: \citet{HB04}, HRS12: \citet{HRS12}, HSC95: \citet{HSC95}, Kal09: \citet{Kal09}, KHOC93: \citet{KHOC93}, KS86: \citet{KS86}, LB94: \citet{LB94}, LL93: \citet{LL93}, M83: \citet{M83}, MB09: \citet{MB09}, McLB93: \citet{McLB93}, McLB00: \citet{McLB00}, MZ82a: \citet{MZ82a}, MZ82b: \citet{MZ82b}, MZ83: \citet{MZ83}, NR79: \citet{NR79}, P06: \citet{P06}, P76: \citet{P76}, Pal12: \citet{Pal12}, PKW09: \citet{PKW09}, , RB97: \citet{RB97}, RBS96: \citet{RBS96}, SH95: \citet{SH95}, SSB14: \citet{SSB14}, SZ00: \citet{SZ00}, T07 : \citet{T07}, T11: \citet{T11}, TG99: \citet{TG99}, TZ10: \citet{TZ10}, VVF96: \citet{VVF96}, WFD96: \citet{WFD96}, Z82: \citet{Z82}, ZBL87: \citet{ZBL87}. 

$^1$ Identified as VVF96-M82a in {\sc pyneb}.\\
$^2$ Identified as MZ82 in {\sc pyneb}.
\label{tab:atomic_data}
\end{table}

\begin{figure}
	\includegraphics[width=\columnwidth]{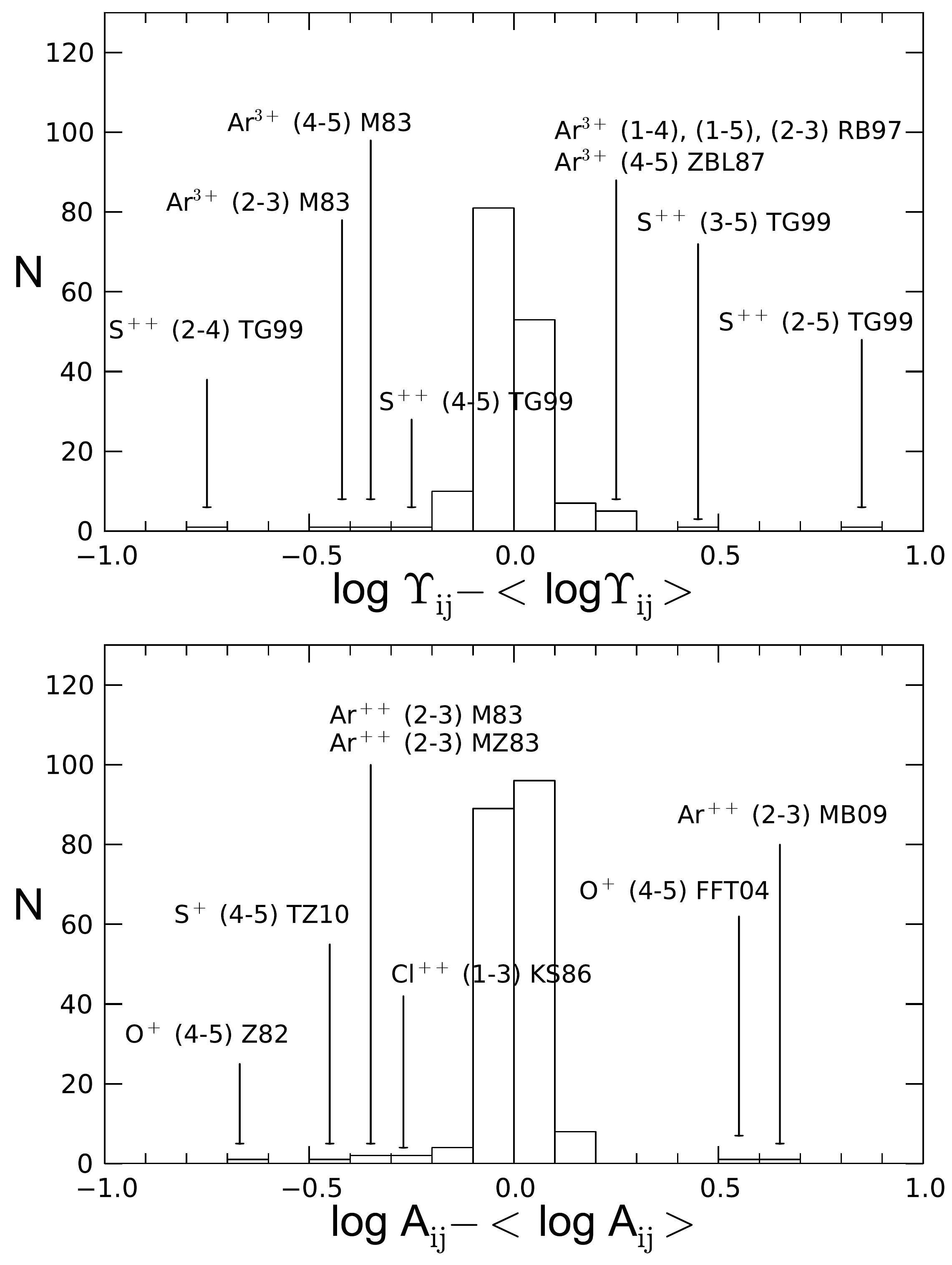}
	\vspace{-0.4cm}
    \caption{Histograms of the differences from the geometric mean of the collision strengths (upper panel) and transition probabilities (lower panel) for each transition where more than two calculations are available in {\sc pyneb}. The largest differences are labelled with the notation used in Table~\ref{tab:atomic_data}. The numbers between parentheses identify the transitions. Three of the collision strengths calculated by \citet{TG99} for S$^{++}$ are outside the plotted range (see text).}
    \label{fig:Histogram_atomic_dif}
\end{figure}

For each transition where three or more calculations of the collision strengths or transition probabilities are available in {\sc pyneb}, we have calculated the geometric mean of these atomic data. Fig.~\ref{fig:Histogram_atomic_dif} shows the distributions of the deviations from these geometric means of the collision strengths (for an electron temperature $T_{\rm{e}}=10000$~K, top panel) and the transition probabilities (bottom panel). Most results can be seen to be within 0.1~dex of the mean value but the data for some transitions have large differences. We identify in Fig.~\ref{fig:Histogram_atomic_dif} the data that show the largest departures from the mean values. Besides, there are three values of the collision strengths calculated by \citet{TG99} for S$^{++}$ that fall outside of the range plotted in Fig.~\ref{fig:Histogram_atomic_dif}. They correspond to the transitions 1--2, 3--4, and 1--5, and have $\log\Upsilon_{ij}-<\log\Upsilon_{ij}>=-1.01$, $-1.19$, and $-1.65$, respectively. Most of the differences have a large impact on the calculations, especially so for the data involved in density determinations, as we show below.

The ionic and total abundances are derived with respect to hydrogen, and the He$^{+}$ and He$^{++}$ abundances are used to determine the ionization correction factor for oxygen. Hence, we need the effective recombination coefficients for H$^{+}$, He$^{+}$, and He$^{++}$. We use for these ions the data available in {\sc pyneb}, namely, the calculations of \citet{SH95} for H$^{+}$, \citet{Porter12,Porter13} for He$^{+}$, and \citet{SH95} for He$^{++}$.

\section{Analysis}

We want to explore the impact of the atomic data differences on the determinations of physical conditions and chemical abundances in our sample of \ion{H}{ii} regions and PNe. In principle, this requires that we perform all calculations for each object using all possible combinations of the atomic data listed in Table~\ref{tab:atomic_data}, obtaining in this way distributions of values for the electron temperatures, electron densities, and ionic and total abundances for each object. The total number of combinations of the 52 sets of atomic data in Table~\ref{tab:atomic_data} is 67,184,640. However, most of these combinations will not introduce significant differences in the results. In particular, the atomic data used for Ne$^{++}$, S$^{++}$, or Ar$^{++}$ will only affect the abundance determination of the corresponding ion, since these ions are not used for the determination of physical conditions. However, since all the calculations depend on the values used for the electron density and temperature, the atomic data used for the other ions, N$^+$, O$^+$, O$^{++}$, S$^+$, Cl$^{++}$, and Ar$^{3+}$, will affect all the determinations of physical conditions and ionic and total abundances. If we restrict the number of combinations of atomic data to be explored to those involving the second set of ions, we get 466,560 combinations. This is still a large number and, again, not all these combinations will introduce significant differences in our results.

We decided to simplify the procedure in the following way. For four objects, the PNe H~1-50 and NGC~5189 and the \ion{H}{ii} regions 30~Doradus and M17, we performed all the necessary calculations. We then used these results to explore the amount of sampling of atomic data combinations that is required in order to reproduce to within ten per cent the shapes of the final distributions of physical conditions and ionic and total abundances.

We start the sampling with the atomic data used by default in {\sc pyneb} and perform all the calculations. We follow by randomly selecting one set of atomic data from the 52 listed in Table~\ref{tab:atomic_data} and use it to repeat all the calculations, keeping fixed the other datasets from the previous calculation. The procedure is then repeated. At each iteration, all atomic data, except for the set that is replaced by another set of atomic data, are left fixed. The changes made in each iteration are kept for the next cycle until the sampling is completed.

We find that with around 44,000 to 48,000 iterations we can reproduce the distributions of physical conditions and ionic and total abundances in the four reference objects to better than ten per cent (using bin sizes of 0.05 dex in density, 0.01 dex in temperature, and 0.02 dex in ionic and total abundances). Therefore, we decided to set the total number of iterations to 50,000 in order to analize the remaining objects in the sample.

\subsection{Physical conditions}
\label{sec:physical conditions}

We estimate the electron densities in each object using the intensity ratios [\ion{S}{ii}]~$\lambda6716/\lambda6731$, [\ion{O}{ii}]~$\lambda3726/\lambda3729$, [\ion{Cl}{iii}]~$\lambda5518/\lambda5538$, and [\ion{Ar}{iv}]~$\lambda4711/\lambda4740$. The exceptions are the \ion{H}{ii} regions M16, M17, and M20, where we could not use the last diagnostic because the [\ion{Ar}{iv}]~$\lambda4740$ line was not measured (in M16 and M20) or was too weak to lead to a viable diagnostic. We use the average of the densities obtained with the four (or three) diagnostics in each calculation, although we also explore in Section~\ref{sec:densities} below the effect of using just one density diagnostic.

We consider two ionizations zones in the nebula, which are characterized by the two temperatures derived from the line intensity ratios [\ion{N}{ii}]~$\lambda5755/(\lambda6548+\lambda6583)$ and [\ion{O}{iii}]~$\lambda4363/(\lambda4959+\lambda5007)$, although they have the same average density. We use $T_{\rm{e}}$[\ion{N}{ii}] for all calculations involving the singly ionized ions, excepting He$^{+}$, and $T_{\rm{e}}$[\ion{O}{iii}] for the remaining ions.

The upper level of [\ion{N}{ii}]~$\lambda5755$, and hence the intensity of this line, can be significantly affected by the recombination of N$^{++}$ in some objects \citep{Rubin86}. When there are measurements of \ion{N}{ii} recombination lines in the optical or [\ion{N}{iii}] collisionally excited lines in the infrared or ultraviolet ranges, one can calculate the N$^{++}$ abundance and correct for this effect using an expression derived by \citet{Liu00}. Of those objects in the sample that have estimates of this correction, the PN NGC~7662 shows the largest difference in $T_{\rm{e}}$[\ion{N}{ii}], with its value decreasing 1000~K after subtracting the recombination contribution from the intensity of [\ion{N}{ii}]~$\lambda5755$ \citep{Liu04}. We repeated our calculations for this PN using [\ion{N}{iii}]~57$\mu$m to estimate the N$^{++}$ abundance with the transition probabilities of \citet{GMZ98} and the collision strengths of \citet{BP92}. Some of the new distributions of physical conditions and ionic and total abundances for NGC~7662 are unaffected, some others are displaced to lower values by up to 0.06~dex. The widths of the affected distributions are also smaller, but by less than 0.01~dex. We conclude that the corrections to $T_{\rm{e}}$[\ion{N}{ii}] from recombination effects would not change our results in any significant way and we ignore them in what follows.

\subsection{Ionic and total abundances}
\label{sec:ionic abundances}

We determine the ionic abundances of O$^{+}$, O$^{++}$, N$^{+}$, Cl$^{++}$, Ar$^{++}$, Ar$^{+3}$, Ne$^{++}$, S$^{+}$, and S$^{++}$ using the brightest collisionally excited lines that can be observed in the optical range: [\ion{O}{ii}]~$\lambda\lambda3726+29$, [\ion{O}{iii}]~$\lambda4959$ (we do not use [\ion{O}{iii}]~$\lambda5007$ because in some objects it was saturated), [\ion{N}{ii}]~$\lambda\lambda6548+83$, [\ion{Cl}{iii}]~$\lambda\lambda5518+38$, [\ion{Ar}{iii}]~$\lambda7136$, [\ion{Ar}{iv}]~$\lambda\lambda4711+40$, [\ion{Ne}{iii}]~$\lambda\lambda3869,3968$, [\ion{S}{ii}]~$\lambda\lambda6716+31$, and [\ion{S}{iii}]~$\lambda6312$. The ionic abundances of He$^{+}$ and He$^{++}$ have been estimated using the recombination lines \ion{He}{i}~$\lambda6678$ and \ion{He}{ii}~$\lambda4686$.

The total abundances of oxygen, chlorine, argon, neon, and sulphur are obtained using the ionization correction factors (ICFs) calculated by \citet{DelIng14}, which are based on a large grid of photoionization models. Although these ICFs were calculated for PNe, they can also be applied to \ion{H}{ii} regions \citep{DelIng15}. The ICF for oxygen depends on the ratio $\mbox{He}^{++}/(\mbox{He}^{+}+\mbox{He}^{++})$. For the other elements, the ICFs are based on the values derived for $\mbox{O}^{++}/(\mbox{O}^{+}+\mbox{O}^{++})$.

For nitrogen we use the classical ICF based on the similarity of the ionization potentials of N$^{+}$ and O$^{+}$, $\mbox{N}/\mbox{O}=\mbox{N}^+/\mbox{O}^+$, since it seems to be working better than the one proposed by \citet{DelIng14} for this element \citep[see][]{DelIng15}.

\section{Results and discussion}

As a result of the procedure described above, we obtain for each object a distribution of 50,000 values for all the parameters that we are calculating: $n_{\rm{e}}$[\ion{S}{ii}], $n_{\rm{e}}$[\ion{O}{ii}], $n_{\rm{e}}$[\ion{Cl}{iii}], $n_{\rm{e}}$[\ion{Ar}{iv}], the average of the previous values $\langle{n_{\rm{e}}}\rangle$, $T_{\rm{e}}$[\ion{N}{ii}], $T_{\rm{e}}$[\ion{O}{iii}], and the ionic and total abundances of O, N, Cl, Ar, Ne, and S. The distributions we find for the physical conditions and element abundances have irregular shapes and can reach extreme values, as shown below. The determination of the average value in these kinds of distributions will be affected in a disproportionate way by a few extreme values. However, in these cases one can use the median of the distribution as a more robust estimate of the central value. On the other hand, irregular distribution shapes are not well described by parameters like the standard deviation. Therefore, we use the medians and total widths to characterize the distributions. These quantities are listed in Tables~\ref{tab:physical_conditions}, \ref{tab:ionic abundances}, and \ref{tab:total_abundances} for all the objects in the sample. The results for the abundance ratios relative to oxygen are shown in Table~\ref{tab:total_abundances2}. Note that the distributions often do not sample in a continuous way the full range identified in the tables.

\begin{table*}
\begin{minipage}{15cm}
\renewcommand{\arraystretch}{1.2}
\caption{Median and spread of the distributions of densities and temperatures implied by different combinations of atomic data.}
\begin{tabular}{ l c c c c c c c c}
\hline					
Object & $\log n_{\rm{e}}$[\ion{S}{ii}] & $\log n_{\rm{e}}$[\ion{O}{ii}]& $\log n_{\rm{e}}$[\ion{Cl}{iii}]& $\log n_{\rm{e}}$[\ion{Ar}{iv}]& $\log\left\langle n_{\rm{e}}\right\rangle$ & $\log T_{\rm{e}}$[\ion{N}{ii}]& $\log T_{\rm{e}}$[\ion{O}{iii}]& Ref. \\
\hline													\multicolumn{9}{l}{Planetary nebulae}\\
Cn~1-5   & $3.69^{+0.28}_{-0.18}$ & $3.59^{+0.15}_{-0.40}$ & $3.67^{+0.12}_{-0.16}$ & $4.13^{+0.62}_{-0.13}$ & $3.84^{+0.44}_{-0.18}$ & $3.93^{+0.02}_{-0.06}$ & $3.94^{+0.01}_{-0.02}$ & 1\\
Cn~2-1   & $3.80^{+0.36}_{-0.20}$ & $3.92^{+0.18}_{-0.40}$ & $4.01^{+0.22}_{-0.14}$ & $4.48^{+0.57}_{-0.20}$ & $4.13^{+0.45}_{-0.19}$ & $4.04^{+0.03}_{-0.11}$ & $4.00^{+0.01}_{-0.03}$ & 2\\
H~1-50   & $3.85^{+0.42}_{-0.19}$ & $3.81^{+0.20}_{-0.44}$ & $4.07^{+0.23}_{-0.13}$ & $4.26^{+0.56}_{-0.18}$ & $4.09^{+0.36}_{-0.23}$ & $4.06^{+0.03}_{-0.08}$ & $4.04^{+0.01}_{-0.03}$ & 2\\
He~2-86  & $4.28^{+0.31}_{-0.17}$ & $3.89^{+0.30}_{-0.54}$ & $4.38^{+0.48}_{-0.16}$ & $4.79^{+0.61}_{-0.24}$ & $4.53^{+0.52}_{-0.28}$ & $3.96^{+0.07}_{-0.15}$ & $3.92^{+0.01}_{-0.04}$ & 1\\
He~2-118 & $3.70^{+0.22}_{-0.20}$ & $3.91^{+0.25}_{-0.48}$ & $4.22^{+0.22}_{-0.19}$ & $4.62^{+0.54}_{-0.21}$ & $4.26^{+0.43}_{-0.21}$ & $4.13^{+0.05}_{-0.13}$ & $4.09^{+0.01}_{-0.03}$ & 2\\
Hu~1-2   & $3.66^{+0.18}_{-0.17}$ & $3.72^{+0.16}_{-0.37}$ & $3.77^{+0.10}_{-0.15}$ & $3.66^{+0.55}_{-0.12}$ & $3.73^{+0.25}_{-0.21}$ & $4.11^{+0.03}_{-0.03}$ & $4.29^{+0.01}_{-0.02}$ & 3\\
IC~418   & $4.27^{+0.28}_{-0.16}$ & $3.95^{+0.28}_{-0.53}$ & $4.20^{+0.22}_{-0.19}$ & $3.79^{+0.62}_{-0.11}$ & $4.14^{+0.27}_{-0.35}$ & $3.96^{+0.04}_{-0.06}$ & $3.94^{+0.01}_{-0.02}$ & 4\\
IC~2165  & $3.53^{+0.12}_{-0.17}$ & $3.64^{+0.14}_{-0.35}$ & $3.60^{+0.09}_{-0.14}$ & $3.82^{+0.51}_{-0.18}$ & $3.66^{+0.30}_{-0.19}$ & $4.11^{+0.02}_{-0.03}$ & $4.16^{+0.01}_{-0.02}$ & 5\\
IC~4846  & $3.83^{+0.33}_{-0.20}$ & $4.06^{+0.31}_{-0.54}$ & $4.06^{+0.15}_{-0.18}$ & $3.98^{+0.61}_{-0.12}$ & $4.06^{+0.29}_{-0.30}$ & $4.07^{+0.04}_{-0.06}$ & $4.02^{+0.01}_{-0.02}$ & 6\\
IC~5217  & $3.67^{+0.19}_{-0.17}$ & $3.62^{+0.14}_{-0.34}$ & $3.67^{+0.10}_{-0.14}$ & $3.80^{+0.56}_{-0.16}$ & $3.70^{+0.31}_{-0.19}$ & $4.13^{+0.03}_{-0.04}$ & $4.03^{+0.01}_{-0.02}$ & 7\\
M~1-20   & $3.99^{+2.45}_{-0.22}$ & $3.95^{+0.32}_{-0.65}$ & $4.07^{+0.24}_{-0.23}$ & $4.13^{+0.61}_{-0.19}$ & $4.21^{+1.64}_{-0.36}$ & $4.02^{+0.05}_{-0.38}$ & $3.99^{+0.01}_{-0.16}$ & 2\\
M~1-42   & $3.08^{+0.06}_{-0.15}$ & $3.13^{+0.10}_{-0.27}$ & $3.29^{+0.12}_{-0.10}$ & $2.79^{+0.59}_{-0.20}$ & $3.12^{+0.18}_{-0.18}$ & $3.95^{+0.01}_{-0.01}$ & $3.96^{+0.01}_{-0.02}$ & 8\\
M~1-61   & $4.35^{+0.52}_{-0.10}$ & $4.22^{+0.58}_{-0.69}$ & $4.35^{+0.41}_{-0.17}$ & $4.74^{+0.60}_{-0.23}$ & $4.59^{+0.47}_{-0.33}$ & $4.03^{+0.09}_{-0.16}$ & $3.95^{+0.02}_{-0.04}$ & 1\\
M~2-4    & $3.79^{+0.37}_{-0.18}$ & $3.71^{+0.16}_{-0.42}$ & $3.94^{+0.20}_{-0.13}$ & $4.09^{+0.59}_{-0.16}$ & $3.96^{+0.36}_{-0.22}$ & $3.99^{+0.03}_{-0.06}$ & $3.93^{+0.01}_{-0.02}$ & 2\\
M~2-36   & $3.59^{+0.18}_{-0.17}$ & $3.59^{+0.14}_{-0.37}$ & $3.86^{+0.12}_{-0.18}$ & $3.69^{+0.63}_{-0.11}$ & $3.72^{+0.30}_{-0.20}$ & $3.97^{+0.02}_{-0.03}$ & $3.92^{+0.01}_{-0.02}$ & 8\\
M~3-15   & $3.91^{+0.51}_{-0.21}$ & $3.95^{+0.25}_{-0.51}$ & $4.09^{+0.16}_{-0.19}$ & $4.03^{+0.60}_{-0.15}$ & $4.07^{+0.34}_{-0.28}$ & $4.03^{+0.04}_{-0.07}$ & $3.92^{+0.01}_{-0.02}$ & 1\\
M~3-21   & $4.01^{+0.15}_{-0.17}$ & $3.79^{+0.20}_{-0.46}$ & $4.16^{+0.27}_{-0.15}$ & $4.53^{+0.60}_{-0.18}$ & $4.25^{+0.50}_{-0.21}$ & $4.07^{+0.04}_{-0.14}$ & $3.99^{+0.01}_{-0.03}$ & 2\\
M~3-32   & $3.52^{+0.10}_{-0.16}$ & $3.63^{+0.14}_{-0.32}$ & $3.02^{+0.03}_{-0.07}$ & $3.13^{+0.64}_{-0.10}$ & $3.43^{+0.20}_{-0.22}$ & $4.22^{+0.02}_{-0.02}$ & $3.94^{+0.01}_{-0.02}$ & 2\\
NGC~2867 & $3.59^{+0.16}_{-0.18}$ & $3.53^{+0.13}_{-0.34}$ & $3.67^{+0.15}_{-0.12}$ & $3.95^{+0.60}_{-0.12}$ & $3.74^{+0.37}_{-0.17}$ & $4.05^{+0.02}_{-0.05}$ & $4.08^{+0.01}_{-0.02}$ & 9\\
NGC~5189 & $3.08^{+0.06}_{-0.15}$ & $3.01^{+0.10}_{-0.24}$ & $3.12^{+0.10}_{-0.08}$ & $3.33^{+0.51}_{-0.21}$ & $3.14^{+0.31}_{-0.16}$ & $3.98^{+0.01}_{-0.01}$ & $4.06^{+0.01}_{-0.02}$ & 1\\
NGC~6153 & $3.56^{+0.16}_{-0.17}$ & $3.53^{+0.13}_{-0.34}$ & $3.71^{+0.11}_{-0.15}$ & $3.52^{+0.63}_{-0.10}$ & $3.60^{+0.27}_{-0.19}$ & $4.01^{+0.02}_{-0.02}$ & $3.96^{+0.01}_{-0.02}$ & 10\\
NGC~6210 & $3.61^{+0.18}_{-0.18}$ & $3.64^{+0.15}_{-0.37}$ & $3.59^{+0.14}_{-0.11}$ & $3.90^{+0.62}_{-0.12}$ & $3.73^{+0.37}_{-0.18}$ & $4.04^{+0.02}_{-0.04}$ & $3.98^{+0.01}_{-0.02}$ & 3\\
NGC~6369 & $3.66^{+0.18}_{-0.17}$ & $3.52^{+0.13}_{-0.32}$ & $3.67^{+0.14}_{-0.11}$ & $3.81^{+0.61}_{-0.11}$ & $3.70^{+0.33}_{-0.17}$ & $4.12^{+0.02}_{-0.04}$ & $4.03^{+0.01}_{-0.02}$ & 1\\
NGC~6439 & $3.68^{+0.25}_{-0.18}$ & $3.56^{+0.13}_{-0.37}$ & $3.80^{+0.17}_{-0.12}$ & $3.94^{+0.57}_{-0.16}$ & $3.79^{+0.35}_{-0.19}$ & $3.98^{+0.02}_{-0.04}$ & $4.01^{+0.01}_{-0.02}$ & 2\\
NGC~6567 & $3.84^{+0.37}_{-0.20}$ & $3.91^{+0.23}_{-0.48}$ & $4.01^{+0.22}_{-0.13}$ & $4.02^{+0.61}_{-0.12}$ & $4.02^{+0.33}_{-0.25}$ & $4.04^{+0.03}_{-0.06}$ & $4.02^{+0.01}_{-0.02}$ & 2\\
NGC~6572 & $4.25^{+0.23}_{-0.16}$ & $4.10^{+0.38}_{-0.59}$ & $4.31^{+0.37}_{-0.14}$ & $4.32^{+0.60}_{-0.15}$ & $4.36^{+0.37}_{-0.29}$ & $4.06^{+0.06}_{-0.11}$ & $4.01^{+0.01}_{-0.03}$ & 3\\
NGC~6620 & $3.40^{+0.10}_{-0.17}$ & $3.34^{+0.10}_{-0.31}$ & $3.57^{+0.10}_{-0.15}$ & $3.52^{+0.56}_{-0.17}$ & $3.46^{+0.29}_{-0.17}$ & $3.95^{+0.01}_{-0.02}$ & $3.98^{+0.01}_{-0.02}$ & 2\\
NGC~6720 & $2.70^{+0.08}_{-0.12}$ & $2.60^{+0.11}_{-0.20}$ & $2.76^{+0.03}_{-0.07}$ & $3.13^{+0.50}_{-0.25}$ & $2.83^{+0.34}_{-0.16}$ & $4.01^{+0.01}_{-0.01}$ & $4.02^{+0.01}_{-0.02}$ & 3\\
NGC~6741 & $3.71^{+0.24}_{-0.18}$ & $3.67^{+0.16}_{-0.39}$ & $3.85^{+0.11}_{-0.18}$ & $4.01^{+0.54}_{-0.17}$ & $3.84^{+0.34}_{-0.20}$ & $4.03^{+0.02}_{-0.05}$ & $4.09^{+0.01}_{-0.03}$ & 3\\
NGC~6803 & $3.88^{+0.53}_{-0.18}$ & $3.65^{+0.15}_{-0.38}$ & $4.07^{+0.24}_{-0.13}$ & $3.69^{+0.62}_{-0.11}$ & $3.96^{+0.30}_{-0.26}$ & $4.03^{+0.03}_{-0.05}$ & $3.98^{+0.01}_{-0.02}$ & 11\\
NGC~6826 & $3.29^{+0.06}_{-0.16}$ & $3.21^{+0.09}_{-0.28}$ & $3.24^{+0.08}_{-0.12}$ & $3.48^{+0.57}_{-0.17}$ & $3.31^{+0.34}_{-0.15}$ & $4.02^{+0.02}_{-0.01}$ & $3.97^{+0.01}_{-0.02}$ & 3\\
NGC~6884 & $3.90^{+0.49}_{-0.20}$ & $3.78^{+0.19}_{-0.43}$ & $3.89^{+0.12}_{-0.18}$ & $4.11^{+0.60}_{-0.13}$ & $3.99^{+0.38}_{-0.23}$ & $4.05^{+0.03}_{-0.07}$ & $4.04^{+0.01}_{-0.03}$ & 3\\
NGC~7009 & $3.63^{+0.17}_{-0.17}$ & $3.65^{+0.14}_{-0.36}$ & $3.64^{+0.14}_{-0.11}$ & $3.78^{+0.57}_{-0.16}$ & $3.69^{+0.31}_{-0.19}$ & $4.08^{+0.02}_{-0.04}$ & $3.99^{+0.01}_{-0.02}$ & 12\\
NGC~7662 & $3.49^{+0.11}_{-0.16}$ & $3.44^{+0.11}_{-0.30}$ & $3.40^{+0.08}_{-0.12}$ & $3.56^{+0.51}_{-0.20}$ & $3.47^{+0.27}_{-0.18}$ & $4.10^{+0.02}_{-0.02}$ & $4.12^{+0.01}_{-0.02}$ & 3\\
PC14     & $3.61^{+0.18}_{-0.17}$ & $3.55^{+0.13}_{-0.36}$ & $3.63^{+0.11}_{-0.15}$ & $3.78^{+0.62}_{-0.11}$ & $3.67^{+0.34}_{-0.18}$ & $4.01^{+0.02}_{-0.04}$ & $3.97^{+0.01}_{-0.02}$ & 1\\
Pe~1-1   & $4.20^{+0.28}_{-0.15}$ & $4.14^{+0.49}_{-0.69}$ & $4.56^{+0.78}_{-0.22}$ & $4.85^{+0.59}_{-0.26}$ & $4.75^{+0.49}_{-0.44}$ & $3.96^{+0.12}_{-0.17}$ & $3.98^{+0.03}_{-0.05}$ & 1\\
\multicolumn{9}{l}{\ion{H}{ii} regions}\\
30~Doradus & $2.63^{+0.06}_{-0.14}$ & $2.56^{+0.11}_{-0.20}$ & $2.42^{+0.09}_{-0.14}$ & $2.84^{+0.46}_{-0.39}$ & $2.62^{+0.29}_{-0.21}$ & $4.02^{+0.01}_{-0.01}$ & $4.00^{+0.01}_{-0.02}$ & 13\\
M8  & $3.20^{+0.05}_{-0.16}$ & $3.13^{+0.10}_{-0.27}$ & $3.33^{+0.10}_{-0.13}$ & $3.40^{+0.64}_{-0.10}$ & $3.28^{+0.34}_{-0.15}$ & $3.92^{+0.01}_{-0.01}$ & $3.91^{+0.01}_{-0.02}$ & 14\\
M16 & $3.14^{+0.05}_{-0.15}$ & $2.98^{+0.10}_{-0.24}$ & $3.13^{+0.08}_{-0.12}$ & -- & $3.07^{+0.09}_{-0.14}$ & $3.92^{+0.01}_{-0.01}$ & $3.88^{+0.01}_{-0.02}$ & 15\\
M17 & $2.66^{+0.08}_{-0.12}$ & $2.62^{+0.11}_{-0.21}$ & $2.33^{+0.09}_{-0.12}$ & -- & $2.56^{+0.10}_{-0.14}$ & $3.95^{+0.01}_{-0.01}$ & $3.90^{+0.01}_{-0.02}$ & 14\\
M20 & $2.50^{+0.06}_{-0.14}$ & $2.31^{+0.12}_{-0.21}$ & $2.52^{+0.03}_{-0.08}$ & -- & $2.44^{+0.08}_{-0.12}$ & $3.92^{+0.01}_{-0.01}$ & $3.89^{+0.01}_{-0.02}$ & 15\\
M42 & $3.80^{+0.35}_{-0.19}$ & $3.69^{+0.16}_{-0.40}$ & $3.91^{+0.20}_{-0.12}$ & $3.83^{+0.59}_{-0.15}$ & $3.86^{+0.31}_{-0.22}$ & $4.00^{+0.02}_{-0.05}$ & $3.92^{+0.01}_{-0.02}$ & 16\\
NGC~3576 & $3.11^{+0.06}_{-0.15}$ & $3.15^{+0.10}_{-0.28}$ & $3.51^{+0.14}_{-0.11}$ & $3.61^{+0.59}_{-0.15}$ & $3.40^{+0.36}_{-0.15}$ & $3.94^{+0.01}_{-0.02}$ & $3.93^{+0.01}_{-0.02}$ & 17\\
NGC~3603 & $3.61^{+0.17}_{-0.16}$ & $3.35^{+0.10}_{-0.29}$ & $3.72^{+0.15}_{-0.11}$ & $3.07^{+0.64}_{-0.11}$ & $3.54^{+0.19}_{-0.18}$ & $4.05^{+0.02}_{-0.01}$ & $3.96^{+0.01}_{-0.02}$ & 15\\
\hline
\end{tabular}
References for the line intensities: (1) \citet{GarciaR12}, (2) \citet{Wang07}, (3) \citet{Liu04}, (4) \citet{Sharpee03}, (5) \citet{Hyung94}, (6) \citet{Hyung01}, (7) \citet{Hyung01b}, (8) \citet{Liu01}, (9) \citet{GarciaR09}, (10) \citet{Liu00}, (11) \citet{Wesson05}, (12) \citet{Fang11}, (13) \citet{PeimbertA03}, (14) \citet{GarciaR07}, (15) \citet{GarciaR06}, (16) \citet{Esteban04}, (17) \citet{GarciaR04}.
\label{tab:physical_conditions}
\end{minipage}
\end{table*}

\begin{table*}
\begin{minipage}{16.5cm}
\renewcommand{\arraystretch}{1.2}
\caption{Median and spread of the distributions of ionic abundances implied by different combinations of atomic data, with $X^{+i}/\mbox{H}^{+}\equiv12+\log(X^{+i}/\mbox{H}^{+})$.}
\begin{tabular}{l c c c c c c c c c c}
\hline					
Object	& O$^{+}$/H$^{+}$ & O$^{++}$/H$^{+}$ & N$^{+}$/H$^{+}$ & Ne$^{++}$/H$^{+}$ & S$^{+}$/H$^{+}$ & S$^{++}$/H$^{+}$ & Cl$^{++}$/H$^{+}$ & Ar$^{++}$/H$^{+}$ & Ar$^{3+}$/H$^{+}$ & $\log\nu^{a}$\\
\hline
\multicolumn{11}{l}{Planetary nebulae}\\
Cn~1-5   & $8.25^{+0.59}_{-0.17}$ & $8.70^{+0.06}_{-0.04}$ & $8.07^{+0.24}_{-0.06}$ & $8.28^{+0.11}_{-0.03}$ & $6.55^{+0.51}_{-0.16}$ & $7.07^{+0.13}_{-0.10}$ & $5.44^{+0.16}_{-0.07}$ & $6.62^{+0.07}_{-0.05}$ & $5.12^{+0.24}_{-0.27}$ & $-2.29$\\
Cn~2-1   & $7.17^{+0.82}_{-0.23}$ & $8.71^{+0.08}_{-0.05}$ & $6.81^{+0.39}_{-0.10}$ & $8.05^{+0.12}_{-0.04}$ & $5.61^{+0.67}_{-0.24}$ & $6.69^{+0.14}_{-0.09}$ & $4.95^{+0.23}_{-0.09}$ & $6.14^{+0.08}_{-0.05}$ & $5.89^{+0.27}_{-0.24}$ & $-1.43$\\
H~1-50   & $7.45^{+0.61}_{-0.27}$ & $8.65^{+0.05}_{-0.04}$ & $7.01^{+0.27}_{-0.11}$ & $8.04^{+0.10}_{-0.03}$ & $5.78^{+0.51}_{-0.25}$ & $6.61^{+0.11}_{-0.09}$ & $4.94^{+0.16}_{-0.09}$ & $6.04^{+0.06}_{-0.05}$ & $6.09^{+0.22}_{-0.24}$ & $-0.95$\\
He~2-86  & $7.77^{+1.32}_{-0.52}$ & $8.80^{+0.19}_{-0.07}$ & $7.54^{+0.75}_{-0.24}$ & $8.27^{+0.20}_{-0.07}$ & $6.12^{+1.02}_{-0.43}$ & $7.09^{+0.20}_{-0.12}$ & $5.41^{+0.43}_{-0.15}$ & $6.42^{+0.13}_{-0.06}$ & $5.66^{+0.36}_{-0.32}$ & --\\
He~2-118 & $7.20^{+0.85}_{-0.29}$ & $8.32^{+0.08}_{-0.05}$ & $6.36^{+0.43}_{-0.13}$ & $7.71^{+0.12}_{-0.04}$ & $5.38^{+0.70}_{-0.28}$ & $6.31^{+0.13}_{-0.08}$ & $4.53^{+0.24}_{-0.09}$ & $5.67^{+0.08}_{-0.06}$ & $5.10^{+0.27}_{-0.21}$ & --\\
Hu~1-2   & $7.18^{+0.22}_{-0.17}$ & $7.64^{+0.04}_{-0.04}$ & $7.20^{+0.09}_{-0.07}$ & $7.10^{+0.06}_{-0.03}$ & $5.55^{+0.23}_{-0.16}$ & $5.93^{+0.08}_{-0.05}$ & $4.15^{+0.08}_{-0.06}$ & $5.40^{+0.05}_{-0.05}$ & $5.55^{+0.20}_{-0.10}$ & $-0.18$\\
IC~418   & $8.49^{+0.48}_{-0.40}$ & $8.10^{+0.07}_{-0.05}$ & $7.68^{+0.21}_{-0.16}$ & $6.79^{+0.11}_{-0.04}$ & $5.85^{+0.35}_{-0.30}$ & $6.59^{+0.13}_{-0.10}$ & $4.86^{+0.16}_{-0.10}$ & $6.02^{+0.07}_{-0.06}$ & $3.24^{+0.24}_{-0.28}$ & --\\
IC~2165  & $6.82^{+0.26}_{-0.14}$ & $8.13^{+0.02}_{-0.03}$ & $6.30^{+0.09}_{-0.07}$ & $7.41^{+0.07}_{-0.02}$ & $4.76^{+0.27}_{-0.14}$ & $5.91^{+0.09}_{-0.06}$ & $4.30^{+0.09}_{-0.06}$ & $5.48^{+0.05}_{-0.05}$ & $5.75^{+0.20}_{-0.16}$ & $-0.22$\\
IC~4846  & $7.10^{+0.44}_{-0.30}$ & $8.52^{+0.05}_{-0.04}$ & $6.41^{+0.19}_{-0.13}$ & $8.05^{+0.10}_{-0.03}$ & $5.30^{+0.41}_{-0.28}$ & $6.41^{+0.12}_{-0.08}$ & $4.86^{+0.15}_{-0.08}$ & $5.96^{+0.06}_{-0.05}$ & $5.55^{+0.23}_{-0.24}$ & $-2.21$\\
IC~5217  & $6.61^{+0.28}_{-0.15}$ & $8.64^{+0.03}_{-0.04}$ & $6.40^{+0.10}_{-0.08}$ & $8.05^{+0.08}_{-0.03}$ & $5.00^{+0.29}_{-0.15}$ & $6.41^{+0.10}_{-0.08}$ & $4.64^{+0.10}_{-0.06}$ & $5.87^{+0.05}_{-0.05}$ & $6.01^{+0.21}_{-0.23}$ & $-0.86$\\
M~1-20   & $7.63^{+4.08}_{-0.43}$ & $8.58^{+0.96}_{-0.06}$ & $6.89^{+2.51}_{-0.17}$ & $7.79^{+0.81}_{-0.05}$ & $5.47^{+3.08}_{-0.35}$ & $6.48^{+0.68}_{-0.11}$ & $4.73^{+1.63}_{-0.12}$ & $5.82^{+0.50}_{-0.06}$ & $5.20^{+1.10}_{-0.35}$ & $-3.34$\\
M~1-42   & $7.64^{+0.07}_{-0.09}$ & $8.40^{+0.04}_{-0.03}$ & $7.85^{+0.05}_{-0.02}$ & $8.03^{+0.09}_{-0.02}$ & $6.18^{+0.09}_{-0.09}$ & $6.73^{+0.12}_{-0.07}$ & $5.05^{+0.08}_{-0.04}$ & $6.32^{+0.06}_{-0.05}$ & $5.82^{+0.21}_{-0.26}$ & $-1.22$\\
M~1-61   & $7.67^{+1.20}_{-0.61}$ & $8.70^{+0.18}_{-0.08}$ & $7.13^{+0.69}_{-0.30}$ & $8.11^{+0.19}_{-0.08}$ & $5.87^{+0.93}_{-0.51}$ & $6.93^{+0.20}_{-0.12}$ & $5.17^{+0.41}_{-0.18}$ & $6.41^{+0.13}_{-0.07}$ & $5.49^{+0.35}_{-0.31}$ & --\\
M~2-4    & $7.84^{+0.53}_{-0.22}$ & $8.70^{+0.06}_{-0.04}$ & $7.34^{+0.22}_{-0.09}$ & $8.15^{+0.11}_{-0.03}$ & $5.93^{+0.46}_{-0.23}$ & $7.37^{+0.13}_{-0.10}$ & $5.27^{+0.15}_{-0.08}$ & $6.46^{+0.07}_{-0.05}$ & $5.29^{+0.24}_{-0.28}$ & --\\
M~2-36   & $7.75^{+0.31}_{-0.16}$ & $8.75^{+0.05}_{-0.04}$ & $7.58^{+0.10}_{-0.07}$ & $8.23^{+0.10}_{-0.03}$ & $6.15^{+0.30}_{-0.16}$ & $6.95^{+0.13}_{-0.09}$ & $5.26^{+0.12}_{-0.06}$ & $6.44^{+0.06}_{-0.05}$ & $5.98^{+0.22}_{-0.28}$ & $-1.72$\\
M~3-15   & $7.08^{+0.55}_{-0.31}$ & $8.80^{+0.07}_{-0.05}$ & $6.63^{+0.24}_{-0.12}$ & $8.09^{+0.12}_{-0.03}$ & $5.44^{+0.48}_{-0.26}$ & $6.91^{+0.14}_{-0.10}$ & $5.24^{+0.17}_{-0.09}$ & $6.41^{+0.07}_{-0.05}$ & $5.69^{+0.24}_{-0.28}$ & --\\
M~3-21   & $7.46^{+1.05}_{-0.29}$ & $8.80^{+0.12}_{-0.05}$ & $7.10^{+0.54}_{-0.13}$ & $8.27^{+0.15}_{-0.04}$ & $5.85^{+0.84}_{-0.26}$ & $6.84^{+0.16}_{-0.10}$ & $5.03^{+0.31}_{-0.10}$ & $6.29^{+0.10}_{-0.06}$ & $6.12^{+0.31}_{-0.25}$ & $-1.24$\\
M~3-32   & $6.22^{+0.12}_{-0.13}$ & $8.60^{+0.04}_{-0.03}$ & $6.04^{+0.08}_{-0.04}$ & $8.09^{+0.10}_{-0.02}$ & $4.76^{+0.14}_{-0.11}$ & $6.58^{+0.12}_{-0.09}$ & $4.94^{+0.10}_{-0.05}$ & $6.17^{+0.06}_{-0.05}$ & $5.87^{+0.21}_{-0.27}$ & $-0.99$\\
NGC~2867 & $7.48^{+0.38}_{-0.15}$ & $8.37^{+0.03}_{-0.04}$ & $6.91^{+0.13}_{-0.07}$ & $7.73^{+0.08}_{-0.03}$ & $5.56^{+0.36}_{-0.15}$ & $6.34^{+0.10}_{-0.08}$ & $4.77^{+0.11}_{-0.07}$ & $5.90^{+0.06}_{-0.05}$ & $5.69^{+0.21}_{-0.20}$ & $-0.51$\\
NGC~5189 & $8.13^{+0.12}_{-0.11}$ & $8.42^{+0.02}_{-0.03}$ & $8.05^{+0.05}_{-0.04}$ & $7.99^{+0.07}_{-0.02}$ & $6.56^{+0.14}_{-0.09}$ & $6.88^{+0.10}_{-0.06}$ & $5.10^{+0.08}_{-0.04}$ & $6.41^{+0.05}_{-0.05}$ & $5.83^{+0.20}_{-0.20}$ & $-0.48$\\
NGC~6153 & $7.18^{+0.22}_{-0.14}$ & $8.64^{+0.04}_{-0.03}$ & $7.03^{+0.07}_{-0.06}$ & $8.19^{+0.10}_{-0.02}$ & $5.59^{+0.24}_{-0.13}$ & $6.71^{+0.13}_{-0.08}$ & $5.08^{+0.10}_{-0.05}$ & $6.32^{+0.06}_{-0.05}$ & $6.03^{+0.22}_{-0.26}$ & $-1.02$\\
NGC~6210 & $7.23^{+0.38}_{-0.15}$ & $8.65^{+0.06}_{-0.04}$ & $6.49^{+0.12}_{-0.07}$ & $8.07^{+0.11}_{-0.03}$ & $5.36^{+0.36}_{-0.15}$ & $6.52^{+0.13}_{-0.09}$ & $4.88^{+0.13}_{-0.07}$ & $6.03^{+0.07}_{-0.05}$ & $5.69^{+0.23}_{-0.26}$ & $-1.86$\\
NGC~6369 & $6.95^{+0.31}_{-0.15}$ & $8.52^{+0.03}_{-0.03}$ & $6.55^{+0.10}_{-0.08}$ & $7.89^{+0.08}_{-0.02}$ & $5.29^{+0.31}_{-0.15}$ & $7.45^{+0.11}_{-0.08}$ & $4.92^{+0.11}_{-0.06}$ & $6.17^{+0.05}_{-0.05}$ & $5.45^{+0.21}_{-0.23}$ & $-2.33$\\
NGC~6439 & $7.73^{+0.41}_{-0.17}$ & $8.61^{+0.04}_{-0.04}$ & $7.53^{+0.15}_{-0.07}$ & $8.15^{+0.09}_{-0.03}$ & $6.11^{+0.38}_{-0.17}$ & $6.85^{+0.11}_{-0.08}$ & $5.15^{+0.12}_{-0.06}$ & $6.40^{+0.06}_{-0.05}$ & $6.13^{+0.22}_{-0.23}$ & $-0.83$\\
NGC~6567 & $7.20^{+0.50}_{-0.25}$ & $8.45^{+0.05}_{-0.04}$ & $6.44^{+0.21}_{-0.10}$ & $7.70^{+0.10}_{-0.03}$ & $5.06^{+0.43}_{-0.25}$ & $6.20^{+0.12}_{-0.08}$ & $4.60^{+0.14}_{-0.08}$ & $5.62^{+0.06}_{-0.05}$ & $5.30^{+0.23}_{-0.24}$ & $-1.99$\\
NGC~6572 & $7.57^{+0.81}_{-0.43}$ & $8.61^{+0.11}_{-0.06}$ & $7.13^{+0.42}_{-0.19}$ & $8.01^{+0.14}_{-0.05}$ & $5.46^{+0.66}_{-0.35}$ & $6.45^{+0.16}_{-0.10}$ & $4.90^{+0.25}_{-0.13}$ & $6.18^{+0.09}_{-0.06}$ & $5.70^{+0.27}_{-0.26}$ & $-2.42$\\
NGC~6620 & $8.22^{+0.20}_{-0.11}$ & $8.72^{+0.04}_{-0.03}$ & $7.98^{+0.06}_{-0.06}$ & $8.16^{+0.09}_{-0.02}$ & $6.55^{+0.21}_{-0.12}$ & $7.03^{+0.12}_{-0.08}$ & $5.30^{+0.10}_{-0.05}$ & $6.52^{+0.06}_{-0.05}$ & $6.03^{+0.21}_{-0.25}$ & $-0.82$\\
NGC~6720 & $8.27^{+0.09}_{-0.10}$ & $8.49^{+0.04}_{-0.04}$ & $7.86^{+0.06}_{-0.02}$ & $8.08^{+0.09}_{-0.02}$ & $6.13^{+0.10}_{-0.08}$ & $6.52^{+0.12}_{-0.06}$ & $4.98^{+0.08}_{-0.04}$ & $6.31^{+0.05}_{-0.05}$ & $5.51^{+0.21}_{-0.23}$ & $-0.79$\\
NGC~6741 & $8.11^{+0.42}_{-0.18}$ & $8.42^{+0.05}_{-0.04}$ & $7.75^{+0.15}_{-0.07}$ & $7.88^{+0.09}_{-0.03}$ & $6.28^{+0.38}_{-0.18}$ & $6.51^{+0.12}_{-0.08}$ & $4.85^{+0.12}_{-0.07}$ & $6.19^{+0.06}_{-0.05}$ & $5.90^{+0.22}_{-0.20}$ & $-0.53$\\
NGC~6803 & $7.36^{+0.41}_{-0.24}$ & $8.68^{+0.05}_{-0.04}$ & $7.23^{+0.16}_{-0.09}$ & $8.23^{+0.10}_{-0.03}$ & $5.83^{+0.39}_{-0.21}$ & $6.79^{+0.13}_{-0.09}$ & $5.23^{+0.14}_{-0.07}$ & $6.38^{+0.07}_{-0.05}$ & $6.07^{+0.23}_{-0.25}$ & $-1.40$\\
NGC~6826 & $7.02^{+0.17}_{-0.11}$ & $8.54^{+0.06}_{-0.04}$ & $6.21^{+0.05}_{-0.06}$ & $7.87^{+0.11}_{-0.02}$ & $4.49^{+0.19}_{-0.10}$ & $6.33^{+0.14}_{-0.08}$ & $4.86^{+0.10}_{-0.05}$ & $6.10^{+0.07}_{-0.05}$ & $5.17^{+0.22}_{-0.26}$ & $-3.64$\\
NGC~6884 & $7.21^{+0.56}_{-0.23}$ & $8.57^{+0.06}_{-0.05}$ & $6.76^{+0.24}_{-0.10}$ & $8.00^{+0.10}_{-0.04}$ & $5.29^{+0.50}_{-0.23}$ & $6.36^{+0.12}_{-0.09}$ & $4.84^{+0.16}_{-0.09}$ & $6.05^{+0.07}_{-0.05}$ & $6.04^{+0.23}_{-0.23}$ & $-0.73$\\
NGC~7009 & $6.81^{+0.30}_{-0.16}$ & $8.64^{+0.04}_{-0.04}$ & $6.34^{+0.10}_{-0.08}$ & $8.14^{+0.10}_{-0.03}$ & $5.06^{+0.29}_{-0.15}$ & $6.57^{+0.12}_{-0.08}$ & $4.90^{+0.11}_{-0.06}$ & $6.15^{+0.06}_{-0.05}$ & $6.09^{+0.22}_{-0.24}$ & $-0.86$\\
NGC~7662 & $6.45^{+0.17}_{-0.12}$ & $8.26^{+0.04}_{-0.03}$ & $5.82^{+0.07}_{-0.05}$ & $7.58^{+0.07}_{-0.02}$ & $4.63^{+0.19}_{-0.12}$ & $6.07^{+0.10}_{-0.06}$ & $4.47^{+0.09}_{-0.05}$ & $5.66^{+0.05}_{-0.05}$ & $5.90^{+0.20}_{-0.18}$ & $-0.40$\\
PC14     & $7.38^{+0.32}_{-0.14}$ & $8.74^{+0.05}_{-0.04}$ & $6.78^{+0.10}_{-0.06}$ & $8.18^{+0.10}_{-0.03}$ & $5.64^{+0.31}_{-0.14}$ & $6.80^{+0.12}_{-0.09}$ & $5.11^{+0.12}_{-0.06}$ & $6.26^{+0.06}_{-0.05}$ & $5.71^{+0.22}_{-0.26}$ & $-1.41$\\
Pe~1-1   & $8.48^{+1.41}_{-0.91}$ & $8.62^{+0.26}_{-0.11}$ & $7.64^{+0.86}_{-0.45}$ & $8.00^{+0.25}_{-0.11}$ & $6.33^{+1.08}_{-0.69}$ & $6.69^{+0.24}_{-0.13}$ & $5.16^{+0.53}_{-0.24}$ & $6.33^{+0.17}_{-0.08}$ & $4.67^{+0.43}_{-0.33}$ & $-3.32$\\
\multicolumn{11}{l}{\ion{H}{ii} regions}\\
30~Doradus & $7.59^{+0.08}_{-0.09}$ & $8.27^{+0.03}_{-0.03}$ & $6.28^{+0.06}_{-0.02}$ & $7.59^{+0.09}_{-0.02}$ & $5.50^{+0.07}_{-0.07}$ & $6.66^{+0.12}_{-0.06}$ & $4.72^{+0.08}_{-0.04}$ & $6.04^{+0.06}_{-0.05}$ & $4.54^{+0.21}_{-0.24}$ & --\\
M8  & $8.38^{+0.18}_{-0.10}$ & $7.89^{+0.05}_{-0.03}$ & $7.54^{+0.05}_{-0.05}$ & $7.05^{+0.10}_{-0.02}$ & $6.06^{+0.19}_{-0.10}$ & $6.98^{+0.13}_{-0.09}$ & $5.06^{+0.10}_{-0.05}$ & $6.20^{+0.06}_{-0.05}$ & $3.97^{+0.22}_{-0.28}$ & --\\
M16 & $8.45^{+0.05}_{-0.08}$ & $7.92^{+0.06}_{-0.03}$ & $7.69^{+0.04}_{-0.02}$ & $7.15^{+0.11}_{-0.02}$ & $6.40^{+0.06}_{-0.07}$ & $6.99^{+0.14}_{-0.09}$ & $5.11^{+0.09}_{-0.04}$ & $6.23^{+0.07}_{-0.04}$ & $4.22^{+0.20}_{-0.31}$ & --\\
M17 & $7.76^{+0.04}_{-0.08}$ & $8.46^{+0.05}_{-0.03}$ & $6.79^{+0.05}_{-0.02}$ & $7.77^{+0.10}_{-0.02}$ & $5.47^{+0.05}_{-0.05}$ & $6.95^{+0.13}_{-0.07}$ & $5.08^{+0.08}_{-0.04}$ & $6.35^{+0.06}_{-0.04}$ & $4.43^{+0.21}_{-0.28}$ & --\\
M20 & $8.46^{+0.04}_{-0.08}$ & $7.74^{+0.05}_{-0.03}$ & $7.55^{+0.05}_{-0.02}$ & $6.68^{+0.11}_{-0.02}$ & $6.26^{+0.05}_{-0.05}$ & $6.90^{+0.14}_{-0.07}$ & $5.05^{+0.09}_{-0.03}$ & $6.20^{+0.06}_{-0.04}$ & $4.33^{+0.21}_{-0.29}$ & --\\
M42 & $7.80^{+0.40}_{-0.20}$ & $8.44^{+0.06}_{-0.04}$ & $6.89^{+0.15}_{-0.08}$ & $7.77^{+0.11}_{-0.03}$ & $5.49^{+0.37}_{-0.18}$ & $7.06^{+0.13}_{-0.10}$ & $5.15^{+0.13}_{-0.07}$ & $6.38^{+0.07}_{-0.05}$ & $4.85^{+0.23}_{-0.28}$ & --\\
NGC~3576 & $8.13^{+0.25}_{-0.11}$ & $8.38^{+0.05}_{-0.03}$ & $7.07^{+0.06}_{-0.06}$ & $7.70^{+0.10}_{-0.02}$ & $5.84^{+0.24}_{-0.11}$ & $6.94^{+0.13}_{-0.09}$ & $4.97^{+0.10}_{-0.05}$ & $6.34^{+0.06}_{-0.05}$ & $4.46^{+0.22}_{-0.27}$ & --\\
NGC~3603 & $7.35^{+0.13}_{-0.12}$ & $8.44^{+0.04}_{-0.03}$ & $6.45^{+0.07}_{-0.04}$ & $7.79^{+0.09}_{-0.02}$ & $5.05^{+0.16}_{-0.11}$ & $6.87^{+0.12}_{-0.09}$ & $5.06^{+0.09}_{-0.05}$ & $6.33^{+0.06}_{-0.05}$ & $5.00^{+0.21}_{-0.26}$ & --\\
\hline
\end{tabular}
$^{a}$ $\nu=\mbox{He}^{++}/(\mbox{He}^++\mbox{He}^{++})$. This parameter is used by the oxygen ICF. The different atomic data introduce variations in $\nu$ that are lower than $\pm0.01$~dex in most objects, and lower than $\pm0.06$~dex in all cases.
\label{tab:ionic abundances}
\end{minipage}
\end{table*}

\begin{table*}
\renewcommand{\arraystretch}{1.25}
\caption{Median and spread of the distributions of total abundances implied by different combinations of atomic data, with $X/\mbox{H}\equiv12+\log(X/\mbox{H})$.}
\begin{tabular}{l l l l l l l}
\hline					
Object	& O/H & N/H & Ne/H & S/H & Cl/H & Ar/H \\
\hline
\multicolumn{7}{l}{Planetary nebulae}\\
Cn 1-5   & $8.84^{+0.26}_{-0.07}$ & $8.65^{+0.11}_{-0.11}$ & $8.46^{+0.25}_{-0.06}$ & $7.23^{+0.18}_{-0.11}$ & $5.57^{+0.14}_{-0.06}$ & $6.67^{+0.10}_{-0.08}$\\
Cn 2-1   & $8.73^{+0.12}_{-0.05}$ & $8.36^{+0.17}_{-0.34}$ & $8.06^{+0.13}_{-0.04}$ & $7.04^{+0.15}_{-0.15}$ & $5.29^{+0.12}_{-0.05}$ & $6.35^{+0.08}_{-0.10}$\\
H 1-50   & $8.70^{+0.11}_{-0.05}$ & $8.24^{+0.16}_{-0.24}$ & $8.06^{+0.10}_{-0.03}$ & $6.93^{+0.14}_{-0.12}$ & $5.22^{+0.10}_{-0.05}$ & $6.24^{+0.09}_{-0.09}$\\
He 2-86  & $8.83^{+0.50}_{-0.10}$ & $8.60^{+0.25}_{-0.18}$ & $8.50^{+0.66}_{-0.14}$ & $7.32^{+0.20}_{-0.15}$ & $5.64^{+0.32}_{-0.08}$ & $6.56^{+0.11}_{-0.13}$\\
He 2-118 & $8.35^{+0.20}_{-0.06}$ & $7.50^{+0.16}_{-0.24}$ & $7.92^{+0.34}_{-0.09}$ & $6.57^{+0.14}_{-0.13}$ & $4.78^{+0.14}_{-0.05}$ & $5.83^{+0.10}_{-0.11}$\\
Hu 1-2   & $8.11^{+0.09}_{-0.06}$ & $8.11^{+0.10}_{-0.09}$ & $7.51^{+0.07}_{-0.05}$ & $6.46^{+0.08}_{-0.07}$ & $4.62^{+0.07}_{-0.06}$ & $5.80^{+0.08}_{-0.07}$\\
IC 418   & $8.64^{+0.39}_{-0.26}$ & $7.84^{+0.12}_{-0.05}$ & $7.68^{+0.40}_{-0.27}$ & $6.65^{+0.18}_{-0.13}$ & $4.99^{+0.20}_{-0.12}$ & $6.09^{+0.18}_{-0.10}$\\
IC 2165  & $8.43^{+0.04}_{-0.04}$ & $7.88^{+0.10}_{-0.17}$ & $7.70^{+0.07}_{-0.03}$ & $6.49^{+0.10}_{-0.10}$ & $4.86^{+0.07}_{-0.05}$ & $5.95^{+0.06}_{-0.07}$\\
IC 4846  & $8.53^{+0.07}_{-0.05}$ & $7.84^{+0.19}_{-0.23}$ & $8.10^{+0.12}_{-0.04}$ & $6.74^{+0.15}_{-0.12}$ & $5.17^{+0.09}_{-0.05}$ & $6.16^{+0.07}_{-0.08}$\\
IC 5217  & $8.68^{+0.03}_{-0.04}$ & $8.45^{+0.11}_{-0.19}$ & $8.08^{+0.08}_{-0.03}$ & $6.92^{+0.18}_{-0.13}$ & $5.12^{+0.08}_{-0.04}$ & $6.14^{+0.05}_{-0.05}$\\
M 1-20   & $8.62^{+3.09}_{-0.08}$ & $7.99^{+1.42}_{-0.25}$ & $7.89^{+2.71}_{-0.10}$ & $6.73^{+0.46}_{-0.18}$ & $4.95^{+0.42}_{-0.06}$ & $6.13^{+0.58}_{-0.24}$\\
M 1-42   & $8.48^{+0.04}_{-0.04}$ & $8.71^{+0.08}_{-0.08}$ & $8.05^{+0.09}_{-0.02}$ & $6.97^{+0.13}_{-0.08}$ & $5.25^{+0.10}_{-0.05}$ & $6.46^{+0.08}_{-0.06}$\\
M 1-61   & $8.73^{+0.43}_{-0.11}$ & $8.20^{+0.28}_{-0.16}$ & $8.33^{+0.59}_{-0.15}$ & $7.16^{+0.20}_{-0.16}$ & $5.41^{+0.30}_{-0.09}$ & $6.56^{+0.10}_{-0.12}$\\
M 2-4    & $8.77^{+0.14}_{-0.07}$ & $8.26^{+0.15}_{-0.20}$ & $8.45^{+0.23}_{-0.12}$ & $7.51^{+0.20}_{-0.16}$ & $5.45^{+0.11}_{-0.05}$ & $6.58^{+0.11}_{-0.10}$\\
M 2-36   & $8.80^{+0.08}_{-0.05}$ & $8.61^{+0.13}_{-0.17}$ & $8.28^{+0.12}_{-0.03}$ & $7.20^{+0.16}_{-0.12}$ & $5.48^{+0.10}_{-0.04}$ & $6.60^{+0.09}_{-0.08}$\\
M 3-15   & $8.81^{+0.09}_{-0.05}$ & $8.37^{+0.21}_{-0.28}$ & $8.23^{+0.16}_{-0.05}$ & $7.29^{+0.22}_{-0.15}$ & $5.61^{+0.10}_{-0.05}$ & $6.63^{+0.07}_{-0.07}$\\
M 3-21   & $8.84^{+0.23}_{-0.06}$ & $8.46^{+0.17}_{-0.31}$ & $8.28^{+0.16}_{-0.04}$ & $7.16^{+0.16}_{-0.14}$ & $5.33^{+0.18}_{-0.06}$ & $6.48^{+0.09}_{-0.12}$\\
M 3-32   & $8.63^{+0.04}_{-0.04}$ & $8.46^{+0.11}_{-0.12}$ & $8.12^{+0.10}_{-0.03}$ & $7.27^{+0.23}_{-0.17}$ & $5.50^{+0.10}_{-0.05}$ & $6.44^{+0.06}_{-0.05}$\\
NGC 2867 & $8.53^{+0.08}_{-0.05}$ & $7.93^{+0.10}_{-0.18}$ & $7.83^{+0.08}_{-0.03}$ & $6.66^{+0.12}_{-0.10}$ & $5.05^{+0.08}_{-0.05}$ & $6.13^{+0.09}_{-0.08}$\\
NGC 5189 & $8.71^{+0.06}_{-0.05}$ & $8.62^{+0.07}_{-0.07}$ & $8.10^{+0.08}_{-0.02}$ & $7.16^{+0.10}_{-0.06}$ & $5.33^{+0.08}_{-0.04}$ & $6.56^{+0.07}_{-0.06}$\\
NGC 6153 & $8.68^{+0.05}_{-0.04}$ & $8.51^{+0.13}_{-0.16}$ & $8.22^{+0.10}_{-0.02}$ & $7.07^{+0.15}_{-0.11}$ & $5.43^{+0.06}_{-0.05}$ & $6.55^{+0.07}_{-0.06}$\\
NGC 6210 & $8.68^{+0.07}_{-0.05}$ & $7.91^{+0.14}_{-0.24}$ & $8.13^{+0.12}_{-0.04}$ & $6.84^{+0.16}_{-0.13}$ & $5.18^{+0.11}_{-0.05}$ & $6.23^{+0.08}_{-0.07}$\\
NGC 6369 & $8.54^{+0.04}_{-0.04}$ & $8.10^{+0.11}_{-0.20}$ & $7.94^{+0.09}_{-0.03}$ & $7.77^{+0.14}_{-0.12}$ & $5.26^{+0.09}_{-0.04}$ & $6.38^{+0.06}_{-0.06}$\\
NGC 6439 & $8.71^{+0.09}_{-0.05}$ & $8.49^{+0.13}_{-0.18}$ & $8.19^{+0.09}_{-0.03}$ & $7.10^{+0.14}_{-0.11}$ & $5.38^{+0.09}_{-0.04}$ & $6.56^{+0.09}_{-0.09}$\\
NGC 6567 & $8.48^{+0.08}_{-0.05}$ & $7.71^{+0.16}_{-0.23}$ & $7.77^{+0.13}_{-0.05}$ & $6.48^{+0.15}_{-0.13}$ & $4.87^{+0.09}_{-0.04}$ & $5.80^{+0.08}_{-0.09}$\\
NGC 6572 & $8.65^{+0.23}_{-0.09}$ & $8.20^{+0.22}_{-0.21}$ & $8.10^{+0.23}_{-0.09}$ & $6.69^{+0.18}_{-0.15}$ & $5.13^{+0.18}_{-0.06}$ & $6.34^{+0.10}_{-0.12}$\\
NGC 6620 & $8.89^{+0.08}_{-0.06}$ & $8.62^{+0.09}_{-0.10}$ & $8.20^{+0.09}_{-0.02}$ & $7.24^{+0.12}_{-0.08}$ & $5.48^{+0.09}_{-0.05}$ & $6.63^{+0.09}_{-0.07}$\\
NGC 6720 & $8.74^{+0.06}_{-0.06}$ & $8.34^{+0.07}_{-0.07}$ & $8.13^{+0.09}_{-0.02}$ & $6.71^{+0.10}_{-0.06}$ & $5.14^{+0.09}_{-0.04}$ & $6.39^{+0.08}_{-0.06}$\\
NGC 6741 & $8.69^{+0.19}_{-0.08}$ & $8.31^{+0.10}_{-0.08}$ & $7.97^{+0.09}_{-0.03}$ & $6.82^{+0.18}_{-0.11}$ & $5.06^{+0.11}_{-0.06}$ & $6.32^{+0.09}_{-0.07}$\\
NGC 6803 & $8.71^{+0.08}_{-0.05}$ & $8.58^{+0.18}_{-0.23}$ & $8.25^{+0.11}_{-0.03}$ & $7.12^{+0.14}_{-0.11}$ & $5.53^{+0.10}_{-0.05}$ & $6.59^{+0.07}_{-0.08}$\\
NGC 6826 & $8.55^{+0.06}_{-0.04}$ & $7.73^{+0.12}_{-0.15}$ & $7.92^{+0.11}_{-0.03}$ & $6.65^{+0.16}_{-0.11}$ & $5.20^{+0.11}_{-0.05}$ & $6.31^{+0.07}_{-0.06}$\\
NGC 6884 & $8.64^{+0.09}_{-0.05}$ & $8.18^{+0.18}_{-0.28}$ & $8.05^{+0.10}_{-0.04}$ & $6.73^{+0.15}_{-0.13}$ & $5.18^{+0.09}_{-0.05}$ & $6.30^{+0.08}_{-0.10}$\\
NGC 7009 & $8.68^{+0.04}_{-0.04}$ & $8.18^{+0.12}_{-0.20}$ & $8.17^{+0.09}_{-0.03}$ & $7.00^{+0.18}_{-0.12}$ & $5.33^{+0.10}_{-0.05}$ & $6.41^{+0.06}_{-0.06}$\\
NGC 7662 & $8.41^{+0.04}_{-0.04}$ & $7.77^{+0.11}_{-0.13}$ & $7.72^{+0.08}_{-0.03}$ & $6.62^{+0.13}_{-0.09}$ & $5.01^{+0.08}_{-0.05}$ & $6.03^{+0.06}_{-0.05}$\\
PC14     & $8.77^{+0.06}_{-0.04}$ & $8.15^{+0.13}_{-0.20}$ & $8.19^{+0.10}_{-0.03}$ & $7.11^{+0.15}_{-0.12}$ & $5.41^{+0.10}_{-0.05}$ & $6.47^{+0.07}_{-0.07}$\\
Pe 1-1   & $8.86^{+1.07}_{-0.30}$ & $8.07^{+0.47}_{-0.08}$ & $8.24^{+0.88}_{-0.26}$ & $6.87^{+0.66}_{-0.16}$ & $5.28^{+0.62}_{-0.17}$ & $6.40^{+0.36}_{-0.10}$\\
\multicolumn{7}{l}{\ion{H}{ii} regions}\\
30 Doradus & $8.36^{+0.04}_{-0.04}$ & $7.06^{+0.08}_{-0.08}$ & $7.91^{+0.10}_{-0.04}$ & $6.79^{+0.14}_{-0.08}$ & $4.89^{+0.09}_{-0.05}$ & $6.15^{+0.08}_{-0.06}$\\
M8 & $8.50^{+0.15}_{-0.08}$ & $7.65^{+0.04}_{-0.03}$ & $8.02^{+0.18}_{-0.10}$ & $7.03^{+0.14}_{-0.09}$ & $5.21^{+0.11}_{-0.05}$ & $6.30^{+0.09}_{-0.06}$\\
M16 & $8.57^{+0.05}_{-0.07}$ & $7.81^{+0.05}_{-0.02}$ & $8.14^{+0.09}_{-0.07}$ & $7.09^{+0.13}_{-0.08}$ & $5.26^{+0.09}_{-0.04}$ & $6.34^{+0.07}_{-0.05}$\\
M17 & $8.53^{+0.05}_{-0.04}$ & $7.58^{+0.08}_{-0.06}$ & $8.08^{+0.10}_{-0.04}$ & $7.06^{+0.17}_{-0.09}$ & $5.25^{+0.10}_{-0.04}$ & $6.46^{+0.09}_{-0.05}$\\
M20 & $8.54^{+0.04}_{-0.07}$ & $7.63^{+0.06}_{-0.02}$ & $7.80^{+0.08}_{-0.07}$ & $6.99^{+0.12}_{-0.06}$ & $5.22^{+0.08}_{-0.04}$ & $6.35^{+0.06}_{-0.05}$\\
M42 & $8.54^{+0.12}_{-0.07}$ & $7.62^{+0.14}_{-0.14}$ & $8.12^{+0.21}_{-0.10}$ & $7.14^{+0.19}_{-0.14}$ & $5.31^{+0.11}_{-0.05}$ & $6.47^{+0.11}_{-0.09}$\\
NGC 3576 & $8.58^{+0.12}_{-0.07}$ & $7.48^{+0.09}_{-0.08}$ & $8.25^{+0.16}_{-0.10}$ & $6.97^{+0.13}_{-0.09}$ & $5.09^{+0.10}_{-0.05}$ & $6.37^{+0.09}_{-0.07}$\\
NGC 3603 & $8.47^{+0.05}_{-0.04}$ & $7.58^{+0.11}_{-0.11}$ & $8.00^{+0.11}_{-0.04}$ & $7.09^{+0.16}_{-0.11}$ & $5.30^{+0.10}_{-0.04}$ & $6.51^{+0.08}_{-0.06}$\\
\hline
\end{tabular}
\label{tab:total_abundances}
\end{table*}

\begin{table*}
\renewcommand{\arraystretch}{1.25}
\caption{Median and spread of the distributions of total abundances implied by different combinations of atomic data, with $X/\mbox{O}\equiv\log(X/\mbox{O})$.}
\begin{tabular}{l l l l l l}
\hline					
Object	& N/O & Ne/O & S/O & Cl/O & Ar/O \\
\hline
\multicolumn{6}{l}{Planetary nebulae}\\
Cn 1-5   & $-0.18^{+0.12}_{-0.35}$ & $-0.38^{+0.05}_{-0.07}$ & $-1.63^{+0.14}_{-0.18}$ & $-3.27^{+0.09}_{-0.19}$ & $-2.15^{+0.14}_{-0.32}$  \\
Cn 2-1   & $-0.36^{+0.16}_{-0.43}$ & $-0.66^{+0.05}_{-0.07}$ & $-1.68^{+0.15}_{-0.22}$ & $-3.44^{+0.07}_{-0.12}$ & $-2.38^{+0.07}_{-0.20}$  \\
H 1-50   & $-0.45^{+0.17}_{-0.33}$ & $-0.63^{+0.06}_{-0.08}$ & $-1.76^{+0.14}_{-0.18}$ & $-3.48^{+0.07}_{-0.09}$ & $-2.46^{+0.12}_{-0.19}$  \\
He 2-86  & $-0.23^{+0.28}_{-0.57}$ & $-0.33^{+0.16}_{-0.07}$ & $-1.51^{+0.24}_{-0.41}$ & $-3.20^{+0.10}_{-0.25}$ & $-2.26^{+0.18}_{-0.60}$  \\
He 2-118 & $-0.84^{+0.17}_{-0.43}$ & $-0.43^{+0.14}_{-0.06}$ & $-1.78^{+0.15}_{-0.29}$ & $-3.59^{+0.08}_{-0.14}$ & $-2.52^{+0.13}_{-0.31}$  \\
Hu 1-2   & $+0.01^{+0.11}_{-0.15}$ & $-0.61^{+0.07}_{-0.05}$ & $-1.65^{+0.09}_{-0.07}$ & $-3.49^{+0.08}_{-0.06}$ & $-2.31^{+0.13}_{-0.14}$  \\
IC 418   & $-0.81^{+0.26}_{-0.27}$ & $-0.96^{+0.05}_{-0.04}$ & $-1.99^{+0.28}_{-0.33}$ & $-3.64^{+0.19}_{-0.25}$ & $-2.55^{+0.24}_{-0.34}$  \\
IC 2165  & $-0.54^{+0.09}_{-0.18}$ & $-0.73^{+0.05}_{-0.03}$ & $-1.94^{+0.10}_{-0.11}$ & $-3.57^{+0.07}_{-0.05}$ & $-2.49^{+0.07}_{-0.08}$  \\
IC 4846  & $-0.69^{+0.19}_{-0.26}$ & $-0.42^{+0.05}_{-0.02}$ & $-1.79^{+0.14}_{-0.15}$ & $-3.36^{+0.07}_{-0.08}$ & $-2.37^{+0.11}_{-0.12}$  \\
IC 5217  & $-0.23^{+0.10}_{-0.19}$ & $-0.59^{+0.05}_{-0.02}$ & $-1.75^{+0.15}_{-0.14}$ & $-3.56^{+0.06}_{-0.05}$ & $-2.55^{+0.08}_{-0.07}$  \\
M 1-20   & $-0.73^{+0.27}_{-1.37}$ & $-0.48^{+0.07}_{-0.38}$ & $-1.16^{+0.48}_{-0.85}$ & $-2.75^{+0.45}_{-0.82}$ & $-2.57^{+0.69}_{-0.87}$  \\
M 1-42   & $+0.23^{+0.06}_{-0.07}$ & $-0.43^{+0.06}_{-0.02}$ & $-1.51^{+0.10}_{-0.07}$ & $-3.23^{+0.07}_{-0.04}$ & $-2.03^{+0.11}_{-0.09}$  \\
M 1-61   & $-0.54^{+0.33}_{-0.51}$ & $-0.40^{+0.15}_{-0.07}$ & $-1.57^{+0.25}_{-0.42}$ & $-3.33^{+0.10}_{-0.23}$ & $-2.16^{+0.19}_{-0.53}$  \\
M 2-4    & $-0.49^{+0.14}_{-0.31}$ & $-0.32^{+0.10}_{-0.06}$ & $-1.24^{+0.18}_{-0.27}$ & $-3.31^{+0.08}_{-0.12}$ & $-2.17^{+0.14}_{-0.23}$  \\
M 2-36   & $-0.18^{+0.11}_{-0.20}$ & $-0.51^{+0.05}_{-0.03}$ & $-1.60^{+0.14}_{-0.14}$ & $-3.31^{+0.07}_{-0.07}$ & $-2.21^{+0.13}_{-0.14}$  \\
M 3-15   & $-0.44^{+0.19}_{-0.31}$ & $-0.58^{+0.07}_{-0.02}$ & $-1.52^{+0.20}_{-0.18}$ & $-3.19^{+0.07}_{-0.09}$ & $-2.18^{+0.11}_{-0.13}$  \\
M 3-21   & $-0.36^{+0.17}_{-0.52}$ & $-0.54^{+0.06}_{-0.15}$ & $-1.67^{+0.15}_{-0.28}$ & $-3.50^{+0.08}_{-0.17}$ & $-2.34^{+0.13}_{-0.34}$  \\
M 3-32   & $-0.16^{+0.08}_{-0.09}$ & $-0.50^{+0.04}_{-0.02}$ & $-1.36^{+0.19}_{-0.14}$ & $-3.12^{+0.05}_{-0.04}$ & $-2.20^{+0.09}_{-0.08}$  \\
NGC 2867 & $-0.59^{+0.09}_{-0.24}$ & $-0.68^{+0.06}_{-0.07}$ & $-1.86^{+0.12}_{-0.16}$ & $-3.47^{+0.08}_{-0.08}$ & $-2.39^{+0.12}_{-0.15}$  \\
NGC 5189 & $-0.08^{+0.07}_{-0.10}$ & $-0.60^{+0.08}_{-0.05}$ & $-1.55^{+0.09}_{-0.07}$ & $-3.38^{+0.08}_{-0.06}$ & $-2.15^{+0.12}_{-0.10}$  \\
NGC 6153 & $-0.16^{+0.10}_{-0.15}$ & $-0.45^{+0.05}_{-0.02}$ & $-1.61^{+0.11}_{-0.10}$ & $-3.25^{+0.07}_{-0.05}$ & $-2.13^{+0.05}_{-0.06}$  \\
NGC 6210 & $-0.75^{+0.10}_{-0.25}$ & $-0.54^{+0.05}_{-0.02}$ & $-1.82^{+0.12}_{-0.14}$ & $-3.48^{+0.07}_{-0.07}$ & $-2.44^{+0.11}_{-0.12}$  \\
NGC 6369 & $-0.42^{+0.10}_{-0.21}$ & $-0.59^{+0.05}_{-0.02}$ & $-0.76^{+0.13}_{-0.13}$ & $-3.27^{+0.07}_{-0.05}$ & $-2.16^{+0.09}_{-0.08}$  \\
NGC 6439 & $-0.21^{+0.12}_{-0.25}$ & $-0.50^{+0.06}_{-0.08}$ & $-1.60^{+0.14}_{-0.16}$ & $-3.32^{+0.07}_{-0.09}$ & $-2.14^{+0.12}_{-0.17}$  \\
NGC 6567 & $-0.77^{+0.17}_{-0.28}$ & $-0.71^{+0.05}_{-0.02}$ & $-1.99^{+0.14}_{-0.18}$ & $-3.61^{+0.07}_{-0.09}$ & $-2.68^{+0.11}_{-0.15}$  \\
NGC 6572 & $-0.45^{+0.26}_{-0.39}$ & $-0.56^{+0.05}_{-0.04}$ & $-1.96^{+0.19}_{-0.28}$ & $-3.51^{+0.09}_{-0.13}$ & $-2.31^{+0.16}_{-0.31}$  \\
NGC 6620 & $-0.26^{+0.08}_{-0.15}$ & $-0.67^{+0.08}_{-0.06}$ & $-1.64^{+0.11}_{-0.10}$ & $-3.40^{+0.08}_{-0.07}$ & $-2.26^{+0.13}_{-0.13}$  \\
NGC 6720 & $-0.39^{+0.06}_{-0.08}$ & $-0.60^{+0.08}_{-0.05}$ & $-2.02^{+0.09}_{-0.07}$ & $-3.59^{+0.08}_{-0.06}$ & $-2.35^{+0.11}_{-0.10}$  \\
NGC 6741 & $-0.37^{+0.12}_{-0.26}$ & $-0.72^{+0.11}_{-0.19}$ & $-1.89^{+0.10}_{-0.10}$ & $-3.64^{+0.09}_{-0.13}$ & $-2.37^{+0.14}_{-0.23}$  \\
NGC 6803 & $-0.13^{+0.16}_{-0.25}$ & $-0.46^{+0.05}_{-0.04}$ & $-1.59^{+0.12}_{-0.13}$ & $-3.18^{+0.07}_{-0.07}$ & $-2.12^{+0.06}_{-0.12}$  \\
NGC 6826 & $-0.81^{+0.07}_{-0.14}$ & $-0.62^{+0.05}_{-0.02}$ & $-1.89^{+0.11}_{-0.10}$ & $-3.35^{+0.06}_{-0.05}$ & $-2.24^{+0.10}_{-0.09}$  \\
NGC 6884 & $-0.46^{+0.16}_{-0.32}$ & $-0.59^{+0.06}_{-0.06}$ & $-1.91^{+0.13}_{-0.17}$ & $-3.45^{+0.08}_{-0.09}$ & $-2.34^{+0.11}_{-0.16}$  \\
NGC 7009 & $-0.49^{+0.10}_{-0.20}$ & $-0.50^{+0.05}_{-0.02}$ & $-1.67^{+0.14}_{-0.12}$ & $-3.34^{+0.07}_{-0.05}$ & $-2.28^{+0.10}_{-0.09}$  \\
NGC 7662 & $-0.63^{+0.08}_{-0.12}$ & $-0.68^{+0.05}_{-0.02}$ & $-1.78^{+0.11}_{-0.09}$ & $-3.39^{+0.07}_{-0.04}$ & $-2.38^{+0.08}_{-0.07}$  \\
PC14     & $-0.61^{+0.10}_{-0.21}$ & $-0.57^{+0.05}_{-0.03}$ & $-1.66^{+0.13}_{-0.12}$ & $-3.36^{+0.07}_{-0.07}$ & $-2.31^{+0.11}_{-0.11}$  \\
Pe 1-1   & $-0.83^{+0.46}_{-0.55}$ & $-0.61^{+0.07}_{-0.25}$ & $-2.01^{+0.36}_{-0.42}$ & $-3.55^{+0.17}_{-0.52}$ & $-2.49^{+0.42}_{-0.94}$  \\
\multicolumn{6}{l}{\ion{H}{ii} regions}\\
30 Doradus & $-1.30^{+0.05}_{-0.06}$ & $-0.44^{+0.05}_{-0.03}$ & $-1.56^{+0.11}_{-0.08}$ & $-3.46^{+0.07}_{-0.05}$ & $-2.20^{+0.11}_{-0.09}$  \\
M8 & $-0.85^{+0.07}_{-0.14}$ & $-0.50^{+0.05}_{-0.02}$ & $-1.49^{+0.17}_{-0.19}$ & $-3.30^{+0.11}_{-0.12}$ & $-2.22^{+0.13}_{-0.15}$  \\
M16 & $-0.75^{+0.05}_{-0.05}$ & $-0.43^{+0.05}_{-0.01}$ & $-1.48^{+0.15}_{-0.09}$ & $-3.31^{+0.11}_{-0.05}$ & $-2.23^{+0.10}_{-0.06}$  \\
M17 & $-0.96^{+0.05}_{-0.05}$ & $-0.45^{+0.05}_{-0.03}$ & $-1.47^{+0.13}_{-0.08}$ & $-3.28^{+0.07}_{-0.04}$ & $-2.08^{+0.12}_{-0.10}$  \\
M20 & $-0.89^{+0.04}_{-0.04}$ & $-0.73^{+0.05}_{-0.02}$ & $-1.54^{+0.15}_{-0.09}$ & $-3.31^{+0.11}_{-0.06}$ & $-2.18^{+0.09}_{-0.07}$  \\
M42 & $-0.91^{+0.13}_{-0.24}$ & $-0.42^{+0.09}_{-0.06}$ & $-1.39^{+0.19}_{-0.23}$ & $-3.22^{+0.08}_{-0.10}$ & $-2.06^{+0.14}_{-0.20}$  \\
NGC 3576 & $-1.08^{+0.08}_{-0.17}$ & $-0.35^{+0.06}_{-0.04}$ & $-1.61^{+0.15}_{-0.15}$ & $-3.49^{+0.08}_{-0.09}$ & $-2.21^{+0.14}_{-0.16}$  \\
NGC 3603 & $-0.90^{+0.08}_{-0.10}$ & $-0.47^{+0.05}_{-0.03}$ & $-1.38^{+0.13}_{-0.11}$ & $-3.16^{+0.07}_{-0.04}$ & $-1.97^{+0.10}_{-0.09}$  \\
\hline
\end{tabular}
\label{tab:total_abundances2}
\end{table*}

\begin{figure}
	\includegraphics[width=\columnwidth]{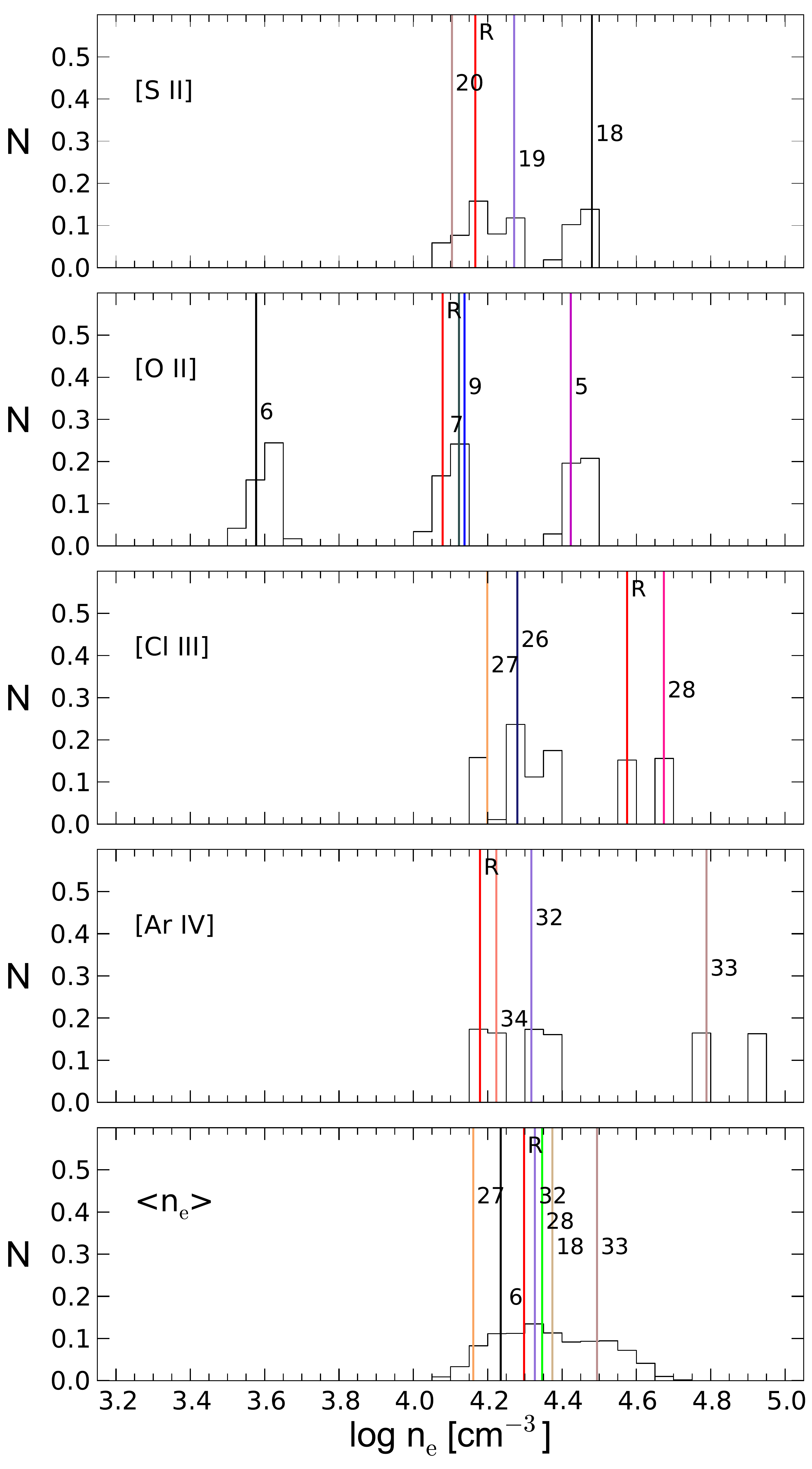}
	\vspace{-0.5cm}
    \caption{Electron densities derived for the PN NGC~6572 with the diagnostics identified in each panel (upper four panels), and the average density $\langle{n_{\rm{e}}}\rangle$ (lower panel), using different combinations of atomic data. The numbers and lines identify the results implied by the set of atomic data changed in the default combination following the notation of Table~\ref{tab:atomic_data}. The line identified with the letter R shows the results that do not differ significantly from those obtained with the default combination. A color version of this plot is available in the online article.}
    \label{fig:densities1}
\end{figure}

\begin{figure}
	\includegraphics[width=\columnwidth]{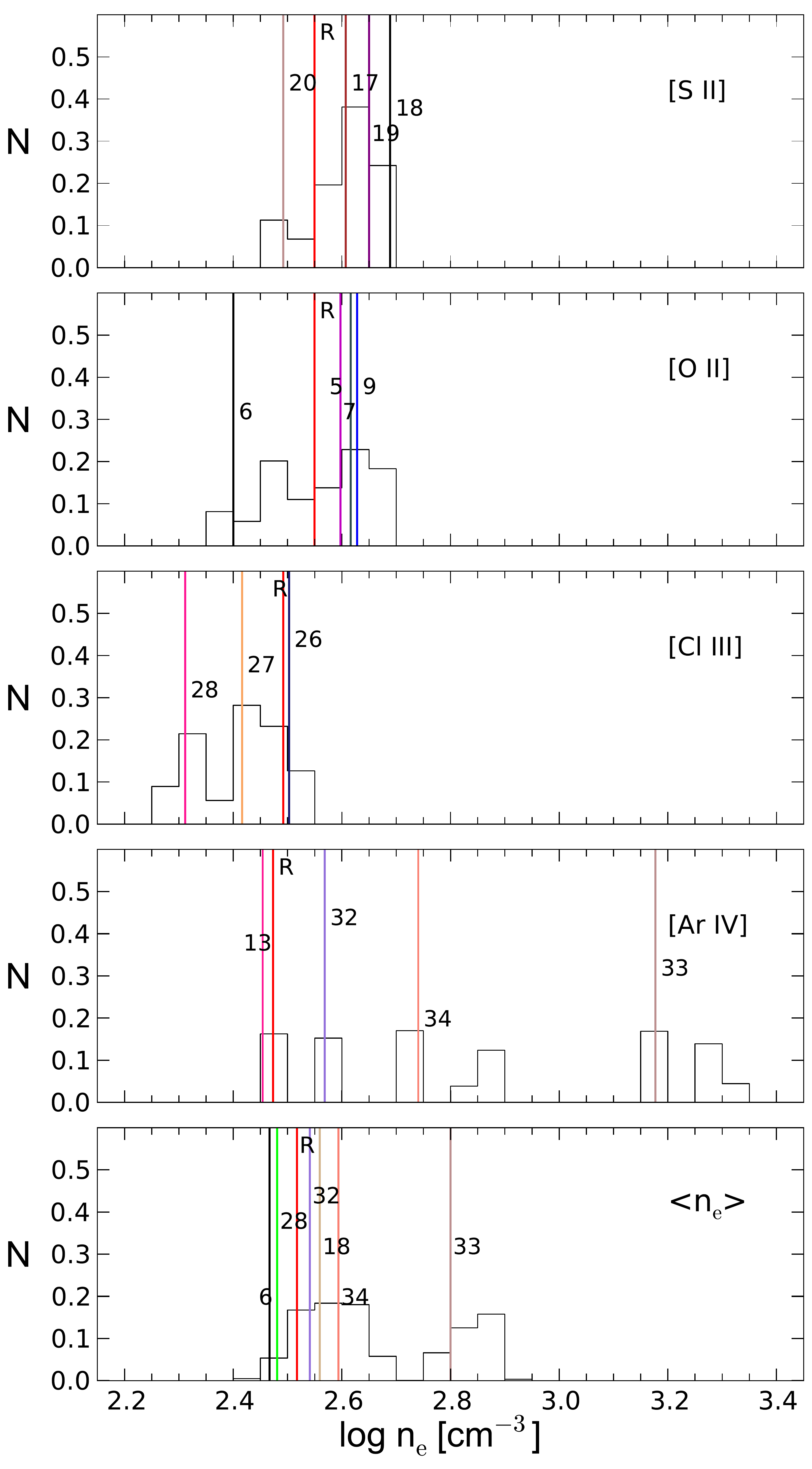}
	\vspace{-0.5cm}
    \caption{Electron densities derived for the \ion{H}{ii} region 30~Doradus with the diagnostics identified in each panel (upper four panels), and the average density $\langle{n_{\rm{e}}}\rangle$ (lower panel), using different combinations of atomic data. The numbers and lines identify the results implied by the set of atomic data changed in the default combination following the notation of Table~\ref{tab:atomic_data}. The line identified with the letter R shows the results that do not differ significantly from those obtained with the default combination. A color version of this plot is available in the online article.}
    \label{fig:densities2}
\end{figure}

\begin{figure}
	\includegraphics[width=\columnwidth, height=7.0cm]{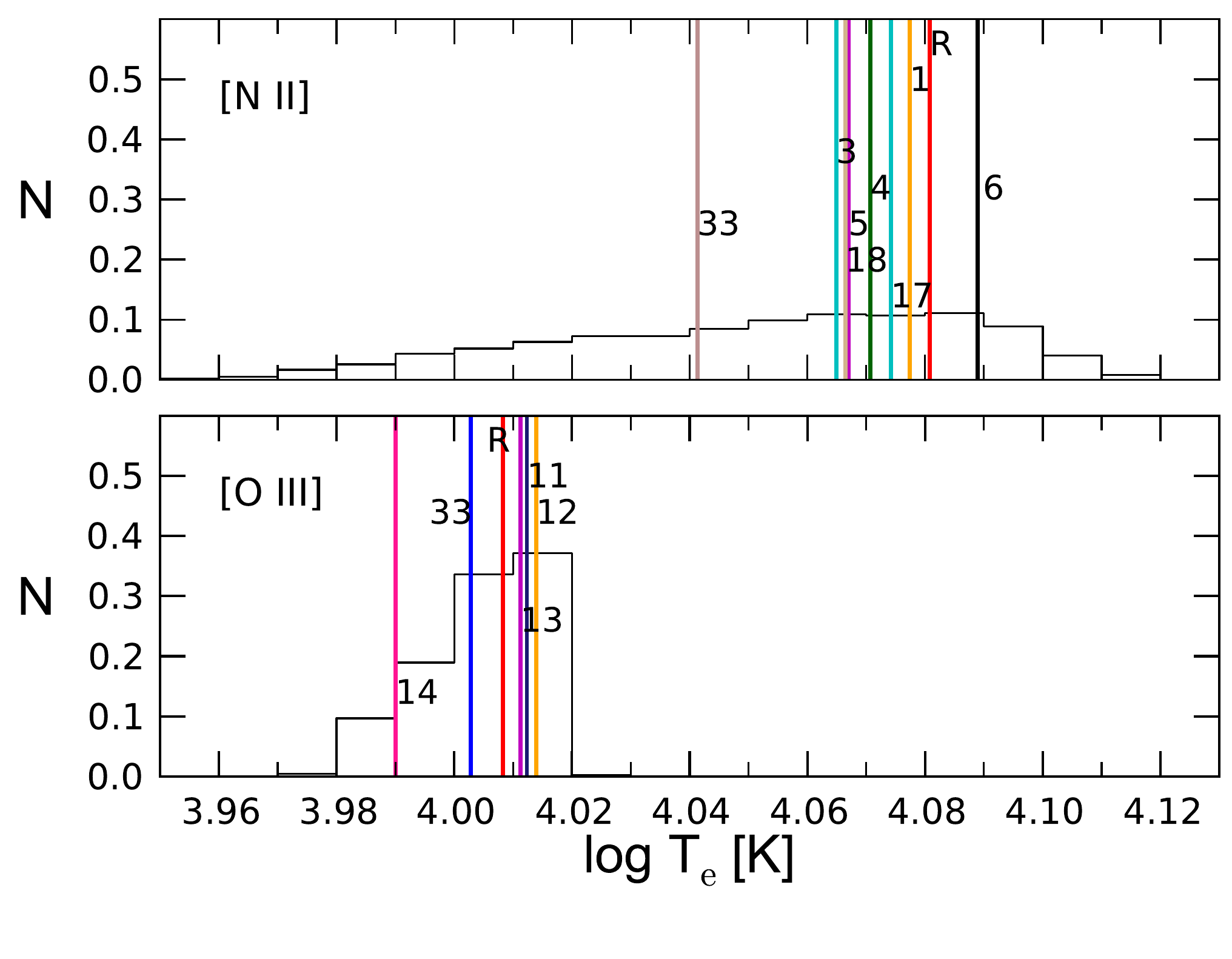}
	\vspace{-0.9cm}
    \caption{Electron temperatures derived for the PN NGC~6572 with the diagnostics identified in each panel and using different combinations of atomic data. The numbers and lines identify the results implied by the set of atomic data changed in the default combination following the notation of Table~\ref{tab:atomic_data}. The line identified with the letter R shows the results that do not differ significantly from those obtained with the default combination. A color version of this plot is available in the online article.}
    \label{fig:temperatures1}
\end{figure}

\begin{figure}
	\includegraphics[width=\columnwidth, height=7.0cm]{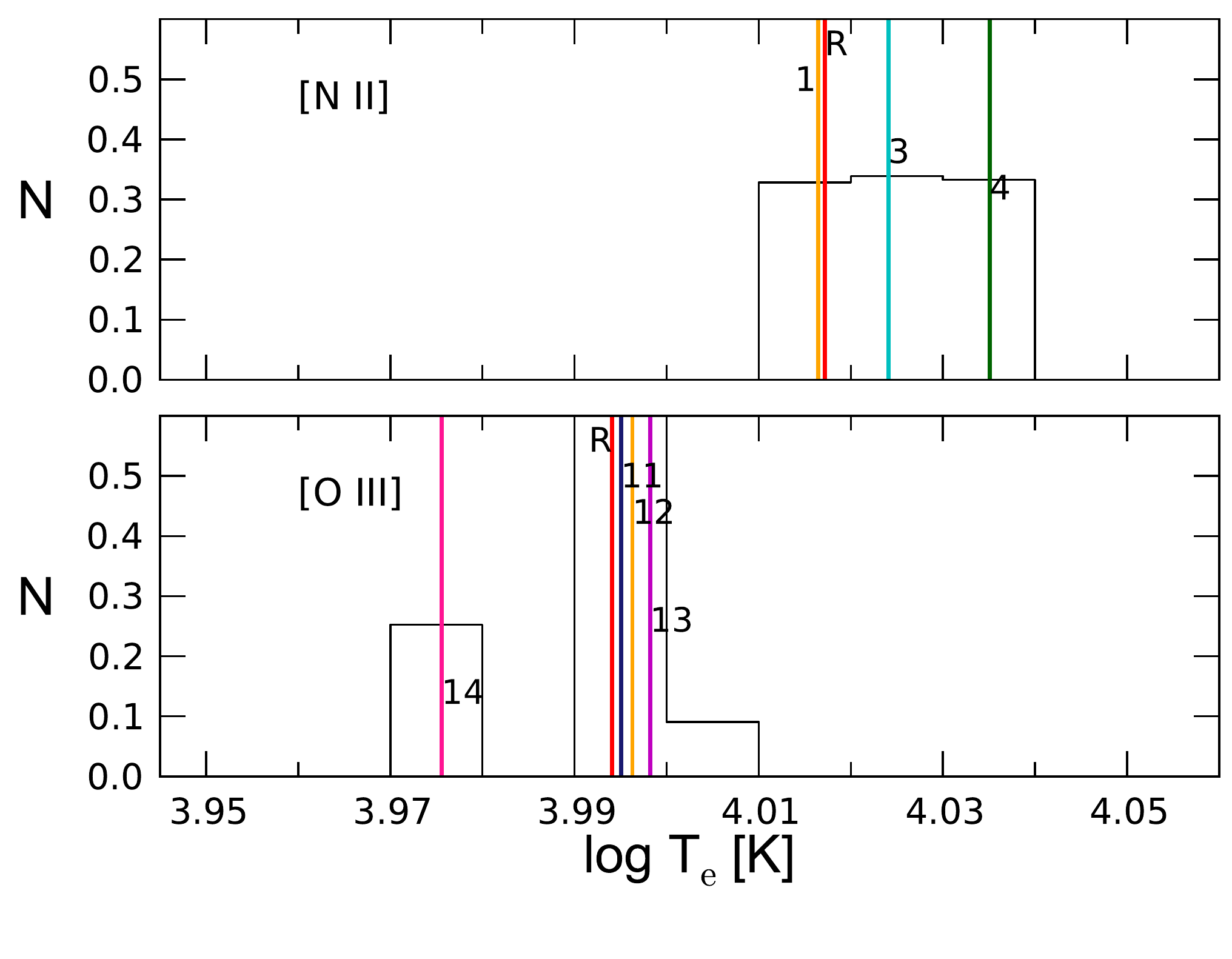}
	\vspace{-0.9cm}
    \caption{Electron temperatures derived for the \ion{H}{ii} region 30~Doradus with the diagnostics identified in each panel and using different combinations of atomic data. The numbers and lines identify the results implied by the set of atomic data changed in the default combination following the notation of Table~\ref{tab:atomic_data}. The line identified with the letter R shows the results that do not differ significantly from those obtained with the default combination. A color version of this plot is available in the online article.}
    \label{fig:temperatures2}
\end{figure}

\begin{figure*}
	\includegraphics[width=\textwidth]{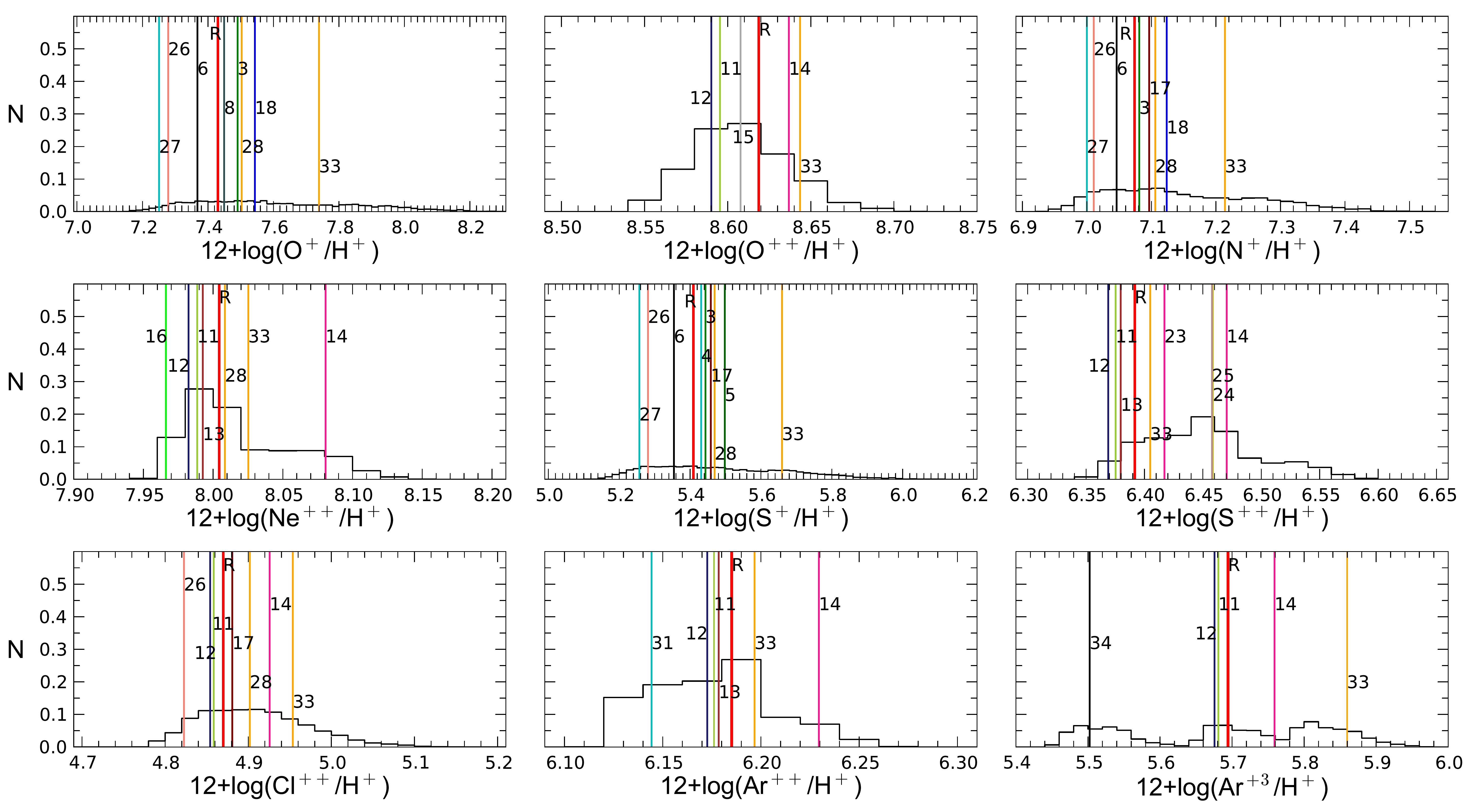}
	\vspace{-0.5cm}
    \caption{Ionic abundances derived for the PN NGC~6572 using different combinations of atomic data. The numbers and lines identify the results implied by the set of atomic data changed in the default combination following the notation of Table~\ref{tab:atomic_data}. The line identified with the letter R shows the results that do not differ significantly from those obtained with the default combination. A color version of this plot is available in the online article.}
    \label{fig:Ionic_abundances1}
\end{figure*}

\begin{figure*}
	\includegraphics[width=\textwidth]{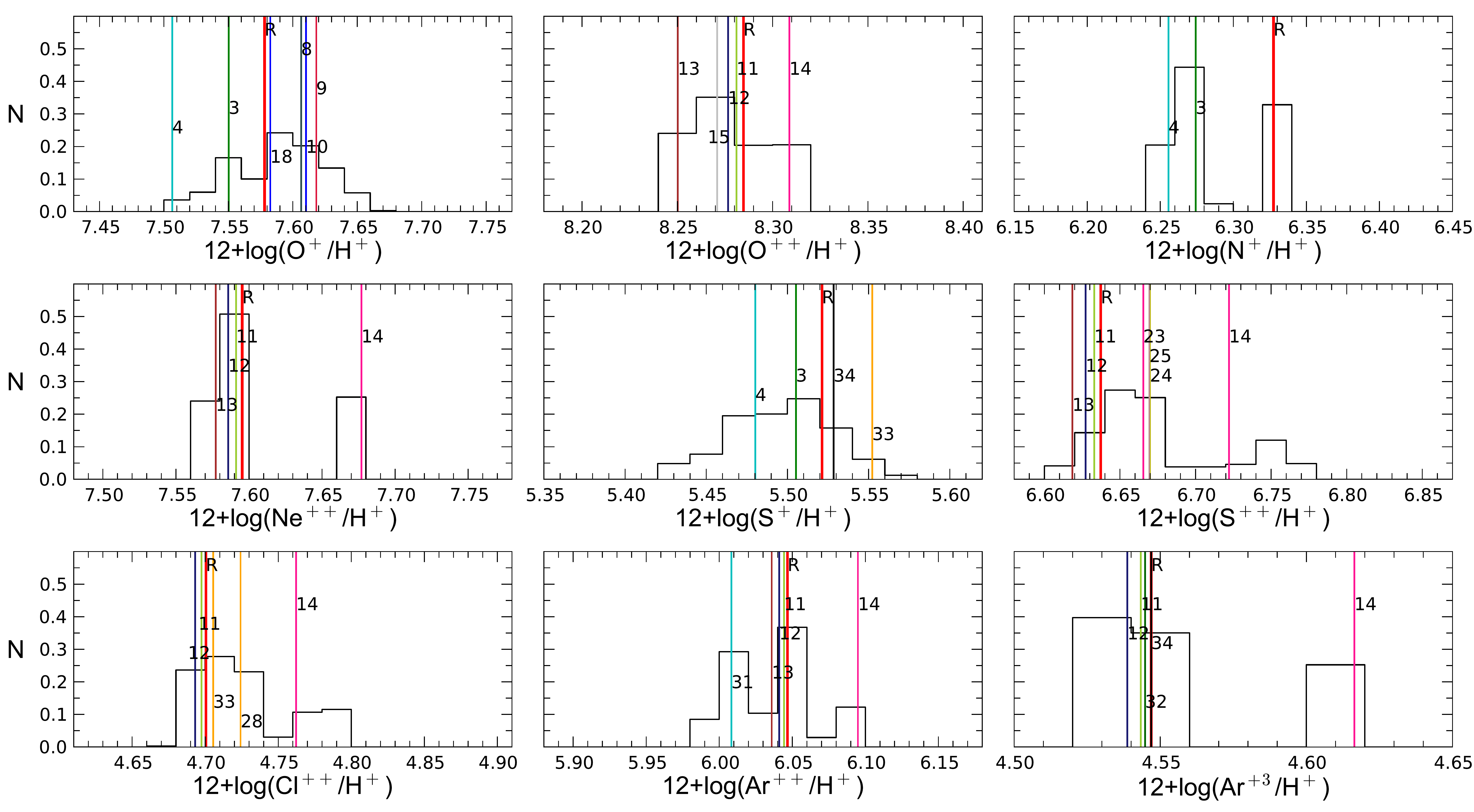}
	\vspace{-0.5cm}
    \caption{Ionic abundances derived for the \ion{H}{ii} region 30~Doradus using different combinations of atomic data. The numbers and lines identify the results implied by the set of atomic data changed in the default combination following the notation of Table~\ref{tab:atomic_data}. The line identified with the letter R shows the results that do not differ significantly from those obtained with the default combination. A color version of this plot is available in the online article.}
    \label{fig:Ionic_abundances2}
\end{figure*}

\begin{figure*}
	\includegraphics[width=\textwidth]{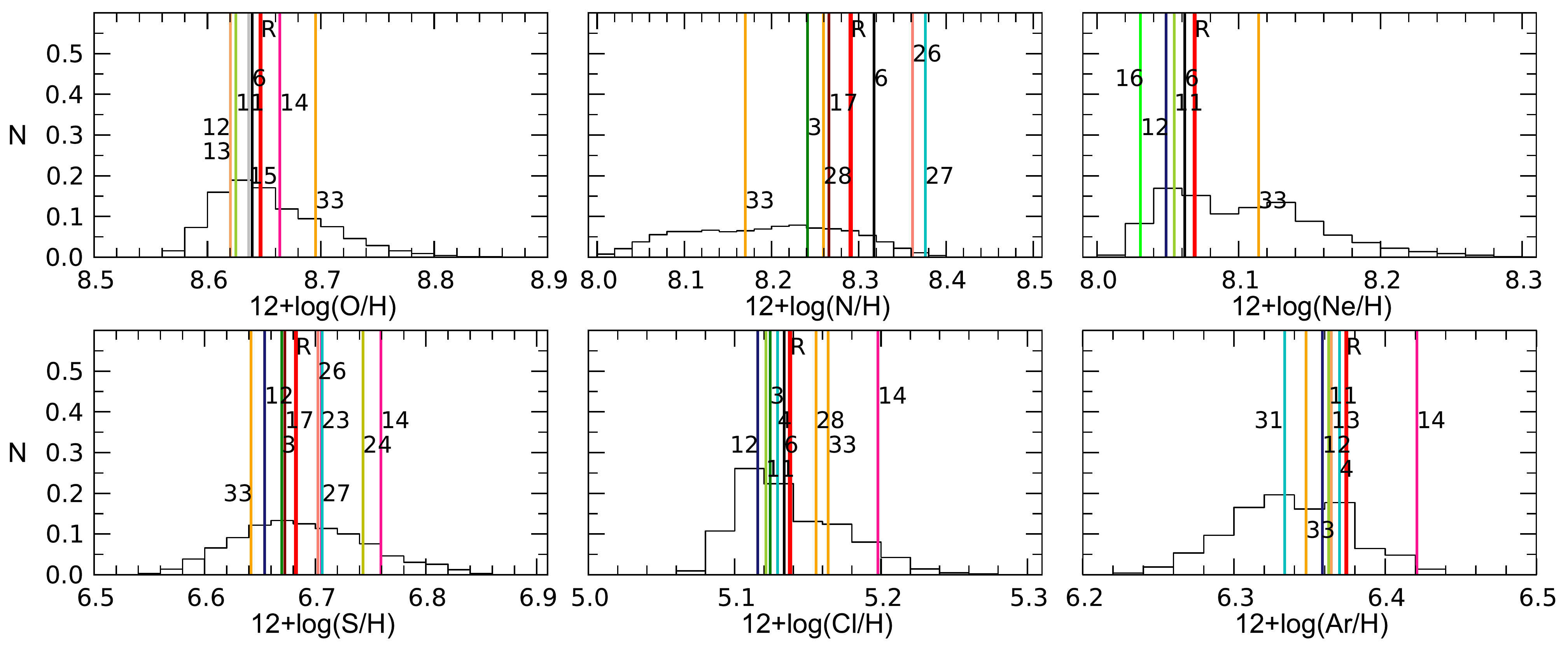}
	\vspace{-0.5cm}
    \caption{Total abundances derived for the PN NGC~6572 using different combinations of atomic data. The numbers and lines identify the results implied by the set of atomic data changed in the default combination following the notation of Table~\ref{tab:atomic_data}. The line identified with the letter R shows the results that do not differ significantly from those obtained with the default combination. A color version of this plot is available in the online article.}
    \label{fig:total_abundances1}
    
\end{figure*}

\begin{figure*}
	\includegraphics[width=\textwidth]{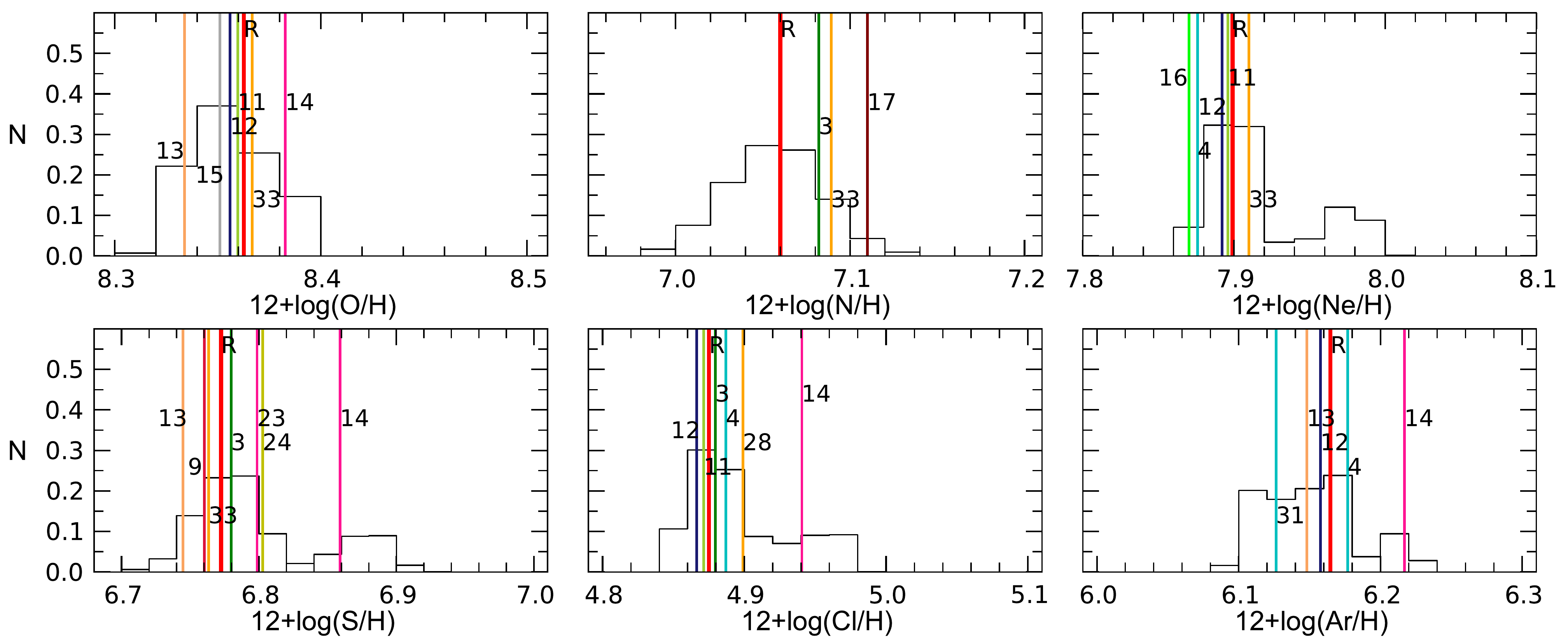}
	\vspace{-0.5cm}
    \caption{Total abundances derived for the \ion{H}{ii} region 30~Doradus using different combinations of atomic data. The numbers and lines identify the results implied by the set of atomic data changed in the default combination following the notation of Table~\ref{tab:atomic_data}. The line identified with the letter R shows the results that do not differ significantly from those obtained with the default combination. A color version of this plot is available in the online article.}
    \label{fig:total_abundances2}
\end{figure*}

We show in Figs.~\ref{fig:densities1}--\ref{fig:total_abundances2} the shapes of these distributions for two objects, the high-density PN NGC~6572 and the low-density \ion{H}{ii} region 30~Doradus. We use in the plots normalized histograms with bin sizes of 0.05 dex in density, 0.01 dex in temperature, and 0.02 dex in the ionic and total abundances. These bin sizes are smaller than the observational uncertainties typically reported for these quantities, which result from the propagation of the errors associated to the measurement of the line intensity ratios.

The vertical lines in Figs.~\ref{fig:densities1}--\ref{fig:total_abundances2} identify the results obtained when only one set of atomic data is changed in the dataset used by default in {\sc pyneb}. The numbers are those listed in Table~\ref{tab:atomic_data}. When these single changes of atomic data do not affect a particular result, their lines are located at the position corresponding to the default calculations, which are identified with the letter R. These results and the trends we find for the full sample of objects are discussed below.


\subsection{Densities}
\label{sec:densities}

\subsubsection{Density differences among the four diagnostics}

The density distributions found for NGC~6572 and 30~Doradus, plotted in Figs.~\ref{fig:densities1} and \ref{fig:densities2}, and the results presented in Table~\ref{tab:physical_conditions}, show that in many objects the four density diagnostics are in broad agreement. However, an inspection of Table~\ref{tab:physical_conditions} shows that some nebulae might have density gradients or density fluctuations. For example, in Cn~2-1, He~2-118, and NGC~3576, most values of $n_{\rm{e}}$[\ion{Cl}{iii}] and $n_{\rm{e}}$[\ion{Ar}{iv}] are higher than those of $n_{\rm{e}}$[\ion{S}{ii}] and $n_{\rm{e}}$[\ion{O}{ii}], whereas M~3-32 shows the opposite behaviour. Establishing whether these variations are real or not will require more investigation, since they could be an artefact introduced by unreliable atomic data or by errors in the line intensity ratios. In the absence of more information, we use the average density from the available diagnostics in our calculations for all the objects in the sample.

\subsubsection{Atomic data that lead to divergent results}

Three sets of atomic data lead to extreme values of density that do not agree in the majority of objects with most of the other density results: the transition probabilities calculated by \citet{TZ10} for S$^+$, the transition probabilities of \citet{WFD96} for O$^+$, and the collision strengths of \citet{M83} for Ar$^{+3}$.

The set of transition probabilities for S$^+$ labelled with the number 17 in Table~\ref{tab:atomic_data} cannot reproduce the observed values of the [\ion{S}{ii}]~$\lambda6716/\lambda6731$ intensity ratio in most of the high-density objects. This dataset uses the default atomic data \citep{PKW09} but includes the transition probabilities of \citet{TZ10} for some transitions, in particular [\ion{S}{ii}]~$\lambda\lambda6716,6731$. The data calculated by \citet{TZ10} imply that the [\ion{S}{ii}]~$\lambda6716/\lambda6731$ intensity ratio reaches at high densities the value 0.503 (for $T_{\rm{e}}=10^4$~K), which is higher than the limiting values implied by the other datasets.  Hence, as previously reported by \citet{Stasinska13}, this dataset fails for nebulae with lower observed values for this ratio: He~2-86, IC~418, NGC~6572, M~1-61, M~3-21, and Pe~1-1, all of them with high density. In M~1-20, where this ratio has an observed value of 0.504, these transition probabilities lead to very high values of $n_{\rm{e}}$[\ion{S}{ii}] (see Table~\ref{tab:physical_conditions}). These extreme densities introduce large uncertainties in all the calculations, and the object does not appear in some of the figures presented below because it falls outside the plotted ranges. Note that in the case of NGC~6572, where these atomic data fail to produce a value for $n_{\rm{e}}$[\ion{S}{ii}], line 17 is missing in the top panel of Fig.~\ref{fig:densities1} showing the distribution of values of $n_{\rm{e}}$[\ion{S}{ii}], but appears in some of the panels of Figs.~\ref{fig:temperatures1}, \ref{fig:Ionic_abundances1}, and \ref{fig:total_abundances1}, showing the results obtained when the other three density diagnostics are used to derive the average density.

In the case of $n_{\rm{e}}$[\ion{O}{ii}], for most of our sample objects the transition probabilities of \citet{WFD96} lead to densities that are lower than those implied by the other atomic datasets for this ion and for the other diagnostics, in particular $n_{\rm{e}}$[\ion{S}{ii}], which is expected to sample similar regions in the nebula. The density results for NGC~6572 in Fig.~\ref{fig:densities1} are a good example of this behaviour: the values of $n_{\rm{e}}$[\ion{O}{ii}] implied by the transition probabilities of \citet{WFD96}, which correspond to the position of line 6 and the results around this line in the second panel of this figure, are much lower than all the other density results. Previous reports of problems with this dataset were given by \citet{Copetti2002}.
 
On the other hand, the collision strengths of \citet{M83} for Ar$^{+3}$ show a similar but opposite behaviour for $n_{\rm{e}}$[\ion{Ar}{iv}]: these collision strengths, which are identified by line 33 and the surrounding results in Figs.~\ref{fig:densities1} and \ref{fig:densities2}, lead to densities that are generally much higher than those implied by the other diagnostics.

The density jumps implied by these three sets of data seem unlikely so that it is plausible to think that there are problems with them. As we show below, the density spreads introduced by the different atomic data introduce important uncertainties in the derived chemical abundances, and these three atomic datasets introduce the greatest variations. Hence, we have repeated our calculations excluding these atomic datasets. We also exclude the collision strengths of \citet{Pal12} for O$^{++}$ since they lead to systematically lower temperatures of $T_{\rm{e}}$[\ion{O}{iii}] in all the objects (see Figs.~\ref{fig:temperatures1} and \ref{fig:temperatures2}), and the results could not be reproduced by \citet{SSB14}, who performed similar calculations.

\subsubsection{The density spreads as a function of density}

\begin{figure}
	\includegraphics[width=\columnwidth]{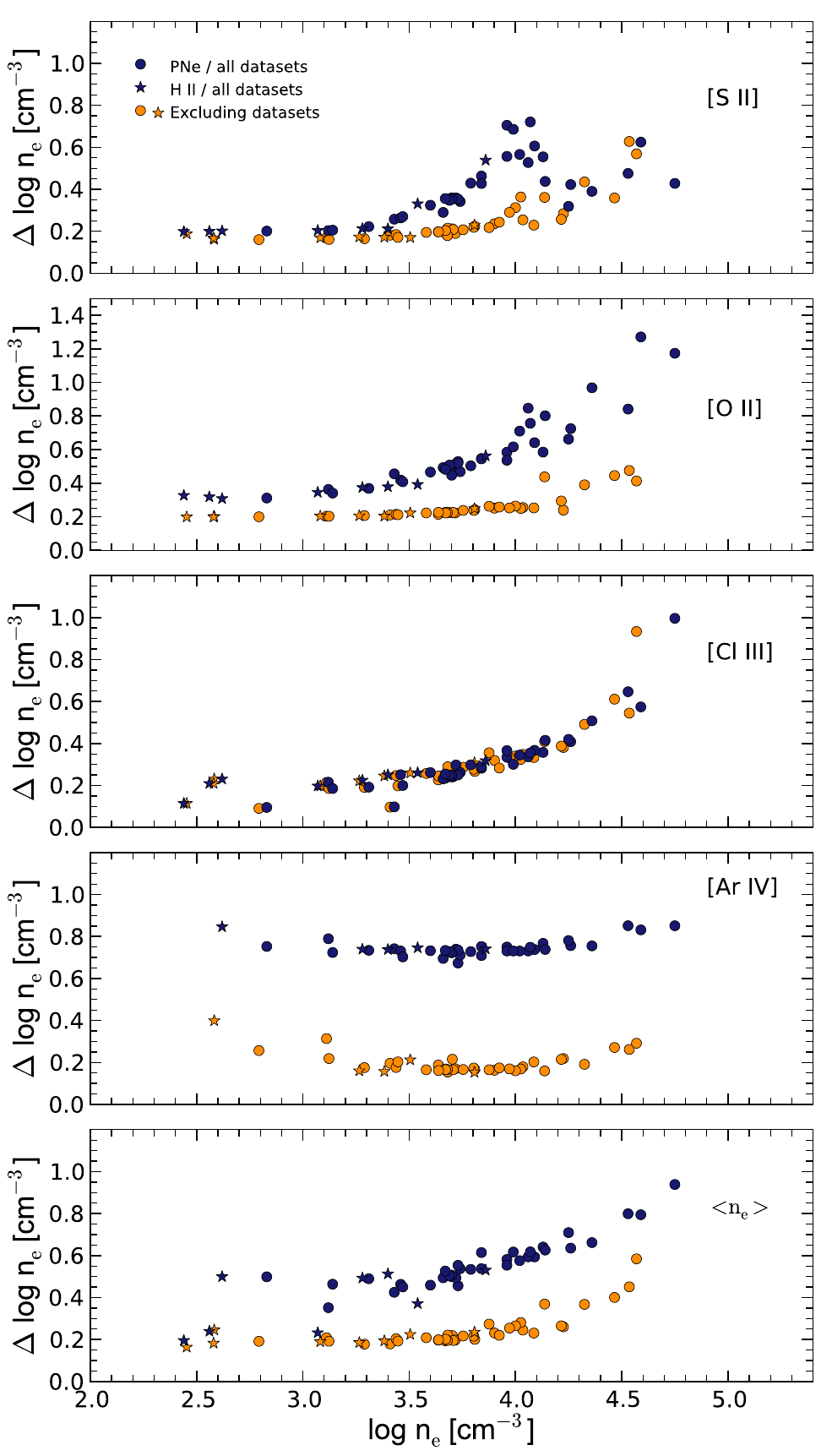}
	\vspace{-0.5cm}
    \caption{Widths of the density distributions for all the sample objects plotted against the medians of the distributions of average density. The dark (blue) symbols show the results implied by all the available 52 datasets; the light (orange) symbols show the results when we exclude the four datasets that lead to discordant values of density or temperature. A color version of this plot is available in the online article.}
    \label{fig:Comparacion_densidad}
\end{figure}

Fig.~\ref{fig:Comparacion_densidad} shows the total widths of the density distributions of $n_{\rm{e}}$[\ion{S}{ii}], $n_{\rm{e}}$[\ion{O}{ii}], $n_{\rm{e}}$[\ion{Cl}{iii}], and $n_{\rm{e}}$[\ion{Ar}{iv}] for all the objects, plotted as a function of the median of the distribution of the averages of the previous values, $\langle{n_{\rm{e}}}\rangle$ (shown in the 6th column of Table~\ref{tab:physical_conditions}). The distributions that we are finding do not resemble the Gaussian distribution in most cases but the full widths are representative of the uncertainties involved in the calculations. Dark symbols show the results implied by all the 52 atomic datasets; the light symbols show the results obtained when we exclude the four datasets that lead to discordant values of density or temperature. 

The distributions of $n_{\rm{e}}$[\ion{S}{ii}], $n_{\rm{e}}$[\ion{O}{ii}], and $n_{\rm{e}}$[\ion{Cl}{iii}] in the sample objects have nearly constant widths up to $n_{\rm{e}}\sim2000$~cm$^{-3}$, as shown in Fig.~\ref{fig:Comparacion_densidad}. Beyond this density, the widths of these distributions generally increase. On the other hand, the width of the $n_{\rm{e}}$[\ion{Ar}{iv}] distribution remains nearly constant at all densities, in contrast to the behaviour shown by the other diagnostics. Note also that the two \ion{H}{ii} regions where $n_{\rm{e}}$[\ion{Ar}{iv}] is not available, M16 and M20, have less variations in $\langle{n_{\rm{e}}}\rangle$ because they do not have the large differences introduced by this diagnostic. Another noticeable feature in Fig.~\ref{fig:Comparacion_densidad} is that for densities above $10^4$~cm$^{-3}$, the spread in $n_{\rm{e}}$[\ion{S}{ii}] decreases due to the fact that \citet{TZ10} cannot reproduce some of the observed values of the $\lambda6716/\lambda6731$ intensity ratio, as mentioned above, and hence stops contributing in full to this density distribution spread. A comparison between the results shown with dark and light symbols in Fig.~\ref{fig:Comparacion_densidad} shows the significant reduction in the widths of the distributions attained by the exclusion of just 4 of the 52 atomic datasets.

\subsubsection{Dependence of the diagnostics on atomic data}

The behaviour of the four density diagnostics as a function of density can be understood in terms of their density-dependent sensitivity to changes in the atomic data involved in the calculations. The four ions used as density diagnostics have the same ground electron configuration and thus the same level structure. As the four diagnostics are based on transitions from levels 2 and 3 to the ground level, we can illustrate their dependence on atomic data using a three-level approximation. We have solved analytically the equations of statistical equilibrium for these three levels assuming that level 1 is significantly more populated than levels 2 and 3. This is a good approximation for these ions, since they keep the population of these upper levels below or around 10 per cent for densities lower than $10^5$~cm$^{-3}$. Levels 2 and 3 have very similar energies and we have also considered negligible their energy difference. Besides, an inspection of the transition probabilities implies that radiative transitions from level 3 to level 2 are not important and can be neglected. With these approximations we find an expression for the intensity ratio between the lines produced by the transitions from levels 3 and 2 to level 1:

\begin{equation}
\frac{I(\lambda_{31})}{I(\lambda_{21})}=\frac{\lambda_{21}}{\lambda_{31}}\frac{A_{31}}{A_{21}}\frac{(1+n_{\rm{c}}(2)/n_{\rm{e}})(1+\Upsilon_{12}/\Upsilon_{23})(\omega_3/\omega_2)+1}{(1+n_{\rm{c}}(3)/n_{\rm{e}})(1+\Upsilon_{13}/\Upsilon_{23})(\omega_2/\omega_3)+1},
\label{eq1}
\end{equation}

\noindent
where $\omega_2$ and $\omega_3$ are the statistical weights of levels 2 and 3, and $\Upsilon_{13}/\Upsilon_{12}=\omega_3/\omega_2$, since levels 2 and 3 belong to the same term (this is a good approximation for low-lying states, where relativistic effects are usually negligible). The critical densities for levels 2 and 3, $n_{\rm{c}}(2)$ and $n_{\rm{c}}(3)$, are the densities at which radiative and collisional transitions are equally important for depopulating these levels. Critical densities can be calculated with the expression:

\begin{equation}
n_{\rm{c}}(i)=\sum_{j<i} A_{ij}\Bigg/\sum_{j\neq i}q_{ij},
\end{equation}

\noindent
with $n_{\rm{c}}(2)=A_{21}/(q_{21}+q_{23})$ and $n_{\rm{c}}(3)=(A_{31}+A_{32})/(q_{31}+q_{32})$ for a three-level ion, where $q_{ij}$ is the collisional transition rate from level $i$ to level $j$ per ion in level $i$ per colliding electron per unit volume and per unit time. The value of $q_{ij}$ is proportional to the collision strength of the transition, $\Upsilon_{ij}$, and inversely proportional to the square root of the electron temperature \citep{Osterbrock}.

At densities much lower than the critical density of an energy level, the population of this level tends to zero; at much higher densities the level tends to the Boltzmann population. Table~\ref{tab:Critical_densities} lists the critical densities for all the upper levels of the transitions that we are using in our calculations. We supply in this table the values for $T_{\rm{e}}=10^4$~K derived with the default atomic datasets, and the range of values implied by all the combinations of atomic data for each ion. Besides their dependence on $T_{\rm{e}}$, the critical densities depend on the atomic data used in their calculation, but an inspection of Table~\ref{tab:Critical_densities} shows that in most cases the values calculated with different atomic data are in broad agreement. Therefore, the values implied by the default dataset can be used as crude estimates of the density regions where the level populations of each ion start to be more sensitive to collisional effects and hence to density.

\begin{table}
\renewcommand{\arraystretch}{0.63}
\caption{Critical densities for the lines used in our analysis.}
\begin{tabular}{l c c c} 
\hline
Line & Upper & $\log n_{\rm{c}}$(D) & $\log n_{\rm{c}}$(all) \\
     & level & [cm$ ^{-3}$]      & [cm$ ^{-3}$] \\
\hline
$[$\ion{N}{ii}$]$ $\lambda5755$               & 5 & 7.2 & 7.0--7.2 \\
$[$\ion{N}{ii}$]$ $\lambda\lambda6548,6583$   & 4 & 5.0	& 4.9--5.0 \\
$[$\ion{O}{ii}$]$ $\lambda3726$               & 3 & 3.6	& 3.5--3.7 \\
$[$\ion{O}{ii}$]$ $\lambda3729$               & 2 & 3.0 & 2.9--3.1 \\
$[$\ion{O}{iii}$]$ $\lambda4363$              & 5 & 7.4 & 7.3--7.4 \\
$[$\ion{O}{iii}$]$ $\lambda\lambda4959,5007$  & 4 & 5.9 & 5.8--5.9 \\
$[$\ion{Ne}{iii}$]$ $\lambda\lambda3869,3968$ & 4 & 7.0 & 7.0--7.0 \\
$[$\ion{S}{ii}$]$ $\lambda6716$               & 3 & 3.1 & 3.0--3.2 \\
$[$\ion{S}{ii}$]$ $\lambda6731$               & 2 & 3.5 & 3.4--3.6 \\
$[$\ion{S}{iii}$]$ $\lambda6312$              & 5 & 7.1 & 7.1--7.2 \\
$[$\ion{Cl}{iii}$]$ $\lambda5518$             & 3 & 3.9 & 3.8--4.0 \\
$[$\ion{Cl}{iii}$]$ $\lambda5538$             & 2 & 4.4 & 4.4--4.7 \\
$[$\ion{Ar}{iii}$]$ $\lambda7136$             & 4 & 6.7 & 6.6--6.7 \\
$[$\ion{Ar}{iv}$]$ $\lambda4711$              & 3 & 4.1 & 4.1--4.7 \\
$[$\ion{Ar}{iv}$]$ $\lambda4740$              & 2 & 5.1 & 5.0--5.7 \\
\hline
\end{tabular}
\label{tab:Critical_densities}
\end{table}

Equation~\ref{eq1} can be used to illustrate the behaviour of the density diagnostics as a function of the critical densities of the diagnostic lines. When $n_{\rm{e}}<<n_{\rm{c}}$, the line intensity ratio tends to $(\lambda_{21}/\lambda_{31})(\omega_3/\omega_2)$; at $n_{\rm{e}}\sim n_{\rm{c}}$, the ratio depends strongly on density and on the relative values of the radiative and  collisional atomic data, especially on the ratio $A_{31}/A_{21}$; and when $n_{\rm{e}}>>n_{\rm{c}}$, it tends to $(\lambda_{21}/\lambda_{31})(\omega_3/\omega_2)(A_{31}/A_{21})$. Therefore, the density distributions of the four diagnostics increase their widths at $n_{\rm{e}}\ga n_{\rm{c}}$ because of their growing sensitivity to changes in the values of $A_{31}/A_{21}$ implied by different atomic data. The [\ion{Ar}{iv}] diagnostic lines have the highest values of $n_{\rm{c}}$ and the width of the $n_{\rm{e}}$[\ion{Ar}{iv}] distributions only increases near our high density limit.

On the other hand, the large range of values of $n_{\rm{c}}$ for Ar$^{3+}$ listed in Table~\ref{tab:Critical_densities} and the comparison between atomic datasets in Fig.~\ref{fig:Histogram_atomic_dif} show the important differences and hence uncertainties in the atomic data for this ion. The large width of $\sim0.7$~dex displayed by the $n_{\rm{e}}$[\ion{Ar}{iv}] distribution at all densities is mostly due to the small value found by \citet{M83} for the collision strength between levels 2 and 3, which implies $\Upsilon_{13}/\Upsilon_{23}=0.63$ for $T_{\rm{e}}=10^4$~K. This ratio is very different from those derived by \citet{RB97} and \citet{ZBL87}, $\Upsilon_{13}/\Upsilon_{23}=0.16$ and $0.32$, respectively. Because of this difference, the collision strengths of \citet{M83} lead to very high densities in all the objects (see the values identified by line 33 in Figs.~\ref{fig:densities1} and \ref{fig:densities2}).

The approximations used to derive Equation~\ref{eq1} are only good at a level of $\sim10$ per cent and cannot explain the rise in the width of the $n_{\rm{e}}$[\ion{Ar}{iv}] distribution at low densities. Levels 4 and 5 of the ions used as density diagnostics, which belong to the same spectral term, are collisionally excited at all the densities that we are considering, and radiative transitions from these levels help populate levels 2 and 3 with contributions of a few per cent. This implies that the line intensity ratios do not reach the value $(\lambda_{21}/\lambda_{31})(\omega_3/\omega_2)$ at low densities but have slightly different values depending on the atomic data. Test calculations with the statistical equilibrium equations for 5-level ions show that the large differences in $n_{\rm{e}}$[\ion{Ar}{iv}] at $n_{\rm{e}}\simeq400$~cm$^{-3}$ arise from the low value of $\Upsilon_{14}$ and $\Upsilon_{15}$ derived by \citet{ZBL87}, which implies $\Upsilon_{15}/\Upsilon_{13}=0.15$ for $T_{\rm{e}}=10^4$~K, when \citet{M83} and \citet{RB97} find $\Upsilon_{15}/\Upsilon_{13}=0.50$ and $0.69$, respectively.

\subsubsection{The influence of other atomic data}

Part of the spread in the density distributions must arise from the atomic data used to calculate electron temperature. We have constrained this effect by fixing the temperatures to the values obtained when using the default combination of atomic data for all ions and repeating the caculations of electron densities. We find that the spread decreases but not by a large amount. The remaining spread goes up to 1.2~dex in $n_{\rm{e}}$[\ion{O}{ii}], 0.8~dex in $n_{\rm{e}}$[\ion{Cl}{iii}], and 0.6~dex in $n_{\rm{e}}$[\ion{S}{ii}] and in the average density. The spread of $n_{\rm{e}}$[\ion{Ar}{iv}] is not much affected.

\subsubsection{Final comments on density}

The spreads of the density distributions for $n_{\rm{e}}$[\ion{S}{ii}], $n_{\rm{e}}$[\ion{O}{ii}], $n_{\rm{e}}$[\ion{Cl}{iii}], and $n_{\rm{e}}$[\ion{Ar}{iv}] in all the objects produce the spreads in the distributions of average densities shown in the lower panel of Fig.~\ref{fig:Comparacion_densidad}. The average densities are used in the calculation of electron temperatures and ionic abundances and the large spreads introduce important uncertainties in these quantities, as we show below. Note that the use of a single density diagnostic would not reduce in a significant way the density uncertainties and would increase the effect of observational errors. On the other hand, although our interest here was to explore the behaviour of the four diagnostics in the full density range covered by the objects, one should always consider the sensitivity of the diagnostics to density, avoiding, whenever it is possible, the use of the [\ion{Cl}{iii}] and [\ion{Ar}{iv}] diagnostics at low density, and the [\ion{S}{ii}] and [\ion{O}{ii}] diagnostics at high density.

\subsection{Temperatures}
\label{sec:temperatures}

The distributions of values of $T_{\rm{e}}$[\ion{N}{ii}] and $T_{\rm{e}}$[\ion{O}{iii}] for NGC~6572 and 30~Doradus are shown in Figs.~\ref{fig:temperatures1} and \ref{fig:temperatures2}. Fig.~\ref{fig:temperatures2} shows that at low densities both temperatures show narrow distributions, with differences lower than 10 per cent, which are mainly introduced by the atomic data of the corresponding ions: an inspection of Table~\ref{tab:atomic_data} shows that lines 1, 3, and 4 in the upper panel of Fig.~\ref{fig:temperatures2} identify changes of sets of N$^+$ atomic data in the default combination, whereas lines 11--14 in the lower panel all correspond to changes in datasets for O$^{++}$. Most of the spread in the values of $T_{\rm{e}}$[\ion{O}{iii}] is introduced by the collision strengths of \citet{Pal12}, that lead to systematically lower temperatures in all objects as commented before.

\begin{figure}
	\includegraphics[width=\columnwidth, height=7.0cm]{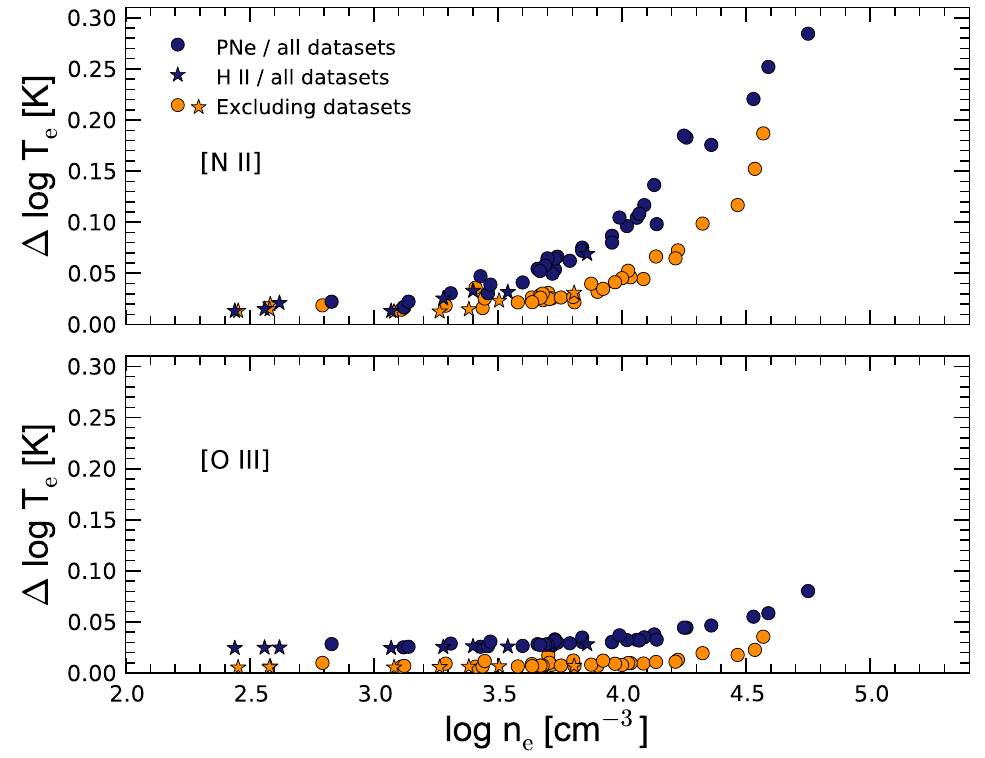}
	\vspace{-0.5cm}
    \caption{Widths of the temperature distributions for all the sample objects plotted against the medians of the distributions of average density. The dark (blue) symbols show the results implied by all the available 52 datasets; the light (orange) symbols show the results when we exclude the four datasets that lead to discordant values of density or temperature. A color version of this plot is available in the online article.}
    \label{fig:Comparacion_Temperaturas}
\end{figure}

A comparison with the results for NGC~6572 in Fig.~\ref{fig:temperatures1} shows that at higher densities the uncertainties introduced by the atomic data used in the diagnostic of density start playing a part in the spread of temperature. In the case of $T_{\rm{e}}$[\ion{O}{iii}], the effect is small and besides lines 11--14, only line 33, corresponding to the collision strengths of \citet{M83} for Ar$^{3+}$, departs significantly from the results of the default combination of atomic data. However, the values calculated for $T_{\rm{e}}$[\ion{N}{ii}] have a stronger dependence on density (note that the values of $n_{\rm{c}}$ in Table~\ref{tab:Critical_densities} are lower for this ion), and the distributions of values for this quantity have a much larger spread. These trends can be seen in Fig.~\ref{fig:Comparacion_Temperaturas}, where the widths of the $T_{\rm{e}}$[\ion{N}{ii}] and $T_{\rm{e}}$[\ion{O}{iii}] distributions are plotted as a function of the median of the average densities for each object. This figure shows that the spreads in density create uncertainties in temperature that can reach 20 per cent in $T_{\rm{e}}$[\ion{O}{iii}] and nearly a factor of 2 in $T_{\rm{e}}$[\ion{N}{ii}]. These uncertainties are reduced to 7 per cent in $T_{\rm{e}}$[\ion{O}{iii}] and 60 per cent in $T_{\rm{e}}$[\ion{N}{ii}] when the three datasets that lead to discordant densities and the collision strengths of \citet{Pal12} are excluded from the calculations (see the light/orange symbols in Fig.~\ref{fig:Comparacion_Temperaturas}).

The spread introduced solely by the atomic data used for the temperature determinations can be obtained by repeating the calculations keeping fixed the density of each object to the value obtained using the default combination of atomic data. In this way we find that the O$^{++}$ atomic data create uncertainties smaller than 0.03~dex in $T_{\rm{e}}$[\ion{O}{iii}] at all densities, whereas the N$^+$ data create even smaller uncertainties for $T_{\rm{e}}$[\ion{N}{ii}] at low density, that only increase up to 0.04~dex at the highest densities.

\subsection{Ionic abundances}
\label{sec:iabundances}

All the ionic abundances have been calculated using the average density obtained from the four density diagnostics. The abundances of O$^{++}$, Cl$^{++}$, Ar$^{++}$, Ar$^{+3}$, Ne$^{++}$, and S$^{++}$ are also based on the values of $T_{\rm{e}}$[\ion{O}{iii}]; whereas the calculations of the abundances of O$^{+}$, N$^{+}$, and S$^{+}$ use the values of $T_{\rm{e}}$[\ion{N}{ii}]. Figs.~\ref{fig:Ionic_abundances1} and \ref{fig:Ionic_abundances2} show the distributions of ionic abundances for NGC~6572 and 30~Doradus implied by the different combinations of atomic data. Since the ionic abundances are calculated using the previous results for the electron density and electron temperature, these figures illustrate the complex interplay between all the atomic data involved in the calculations. For example, note the distributions of the O$^+$, N$^+$, and S$^+$ abundances for NGC~6572 in Fig.~\ref{fig:Ionic_abundances1}. The vertical lines in the figure, which identify the abundances obtained using the default combination of atomic data (line R) and those obtained when a single dataset is changed in this combination (lines 1--34, see Table~\ref{tab:atomic_data}), show how all these ionic abundances depend on the atomic data used for O$^+$, N$^+$, S$^+$, Cl$^{++}$, and Ar$^{3+}$, the ions used in the derivation of physical conditions. The spreads shown by these distributions also illustrate the large uncertainties introduced by atomic data at high density.

\begin{figure*}
	\includegraphics[width=1\textwidth]{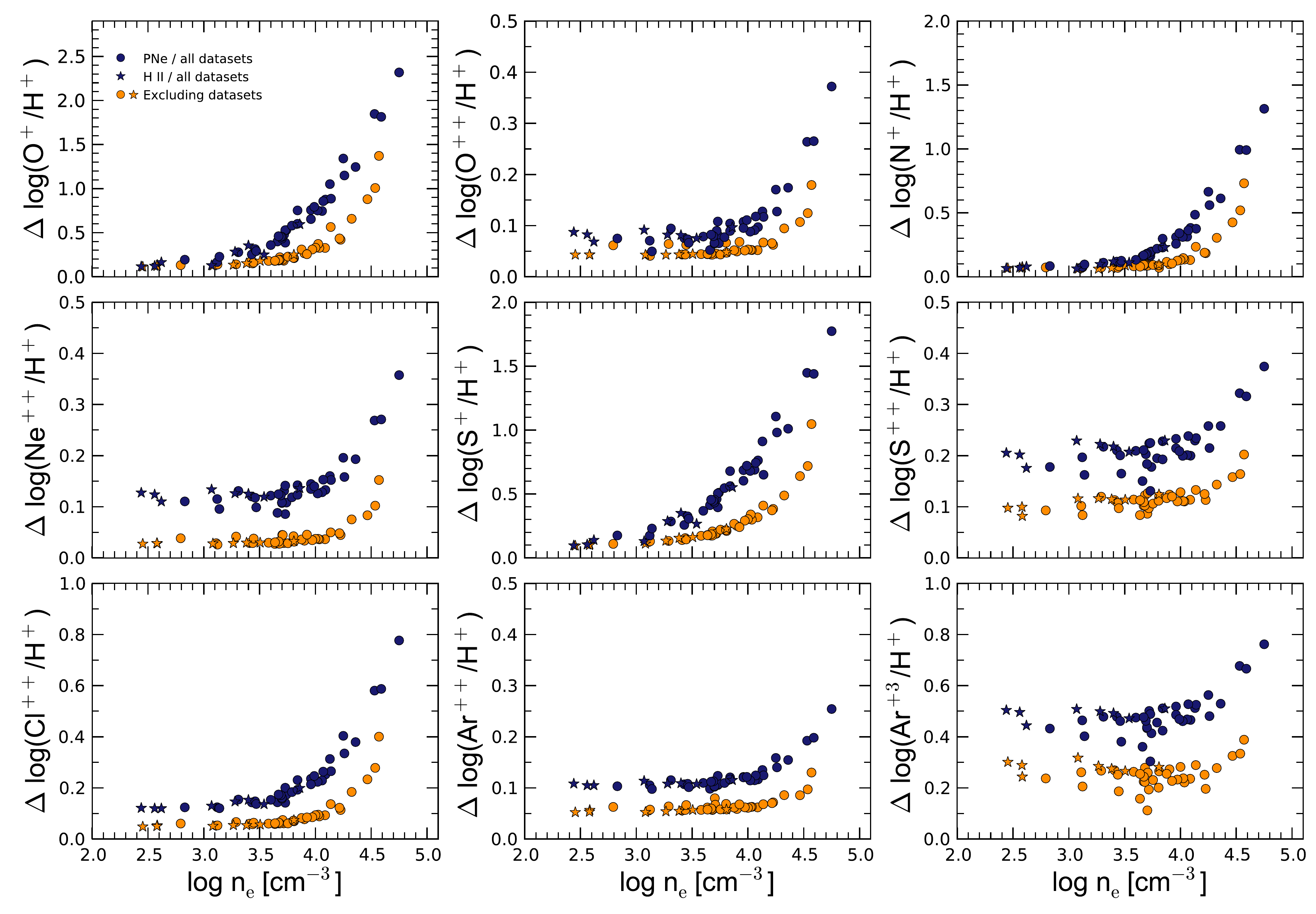}
	\vspace{-0.9cm}
    \caption{Widths of the distributions of ionic abundances in all the objects plotted as a function of the median density. The dark (blue) symbols show the results implied by all the available 52 datasets; the light (orange) symbols show the results when we exclude the four datasets that lead to discordant values of density or temperature. A color version of this plot is available in the online article.}
    \label{fig:Comparacion_ab_ionicas}
\end{figure*}

Fig.~\ref{fig:Comparacion_ab_ionicas} shows the widths of the distributions of ionic abundances for all the objects, plotted against the median of their average densities. It can be seen in Fig.~\ref{fig:Comparacion_ab_ionicas} that the uncertainties introduced by the choice of atomic data at low densities are small, around 0.1 dex, for most ions. One exception is the S$^{++}$ abundance, whose spread reaches $\sim0.2$~dex at low densities because of the high sensitivity of the [\ion{S}{iii}]~$\lambda6312$ line to small changes in $T_{\rm{e}}$. The second and main exception is the Ar$^{3+}$ abundance, with a total width of $\sim0.5$~dex in most objects, which is due to the large differences in the available atomic data for this ion. 
 
As the density increases the situation changes drastically, with some ions, like S$^+$ and O$^+$, reaching uncertainties of a factor of 5 at $n_{\rm{e}}=10^4$~cm$^{-3}$, which continue to increase at higher densities, reaching factors of 60 and 200, respectively, at $n_{\rm{e}}\simeq5\times10^4$~cm$^{-3}$. For other ions, like O$^{++}$, Ne$^{++}$, S$^{++}$, and Ar$^{++}$, the uncertainties are lower, only approaching factors of 2 at the highest densities. The light/orange symbols in Fig.~\ref{fig:Comparacion_ab_ionicas} show that the uncertainties are much lower when the four datasets discussed in Section~\ref{sec:densities} are excluded, although they still reach factors $\sim25$ for O$^+$ at high density.

The behaviour of the uncertainties in the ionic abundances is the result of three causes. The first and main cause is the uncertainty or spread in density, which increases at higher densities. This spread affects mainly those ionic abundances based on lines whose upper levels have low critical densities, that is O$^+$ and S$^+$, followed by Cl$^{++}$, since their emissivities are very sensitive to changes in density. The second cause is the spread in temperature introduced by the spread in density, which is more important for $T_{\rm{e}}$[\ion{N}{ii}] and hence affects mostly those ionic abundances where this temperature is used for the calculations: O$^+$, S$^+$, and N$^+$. The spreads in ionic abundances introduced by $T_{\rm{e}}$[\ion{O}{iii}] for the other ions are smaller and due to changes in all the atomic data for O$^{++}$. The third and least important cause of the spreads in the ionic abundances shown in Fig.~\ref{fig:Comparacion_ab_ionicas} is the uncertainties arising from the use of different atomic data for each ion, independently of the uncertainties that they might introduce in the calculation of the physical conditions. We have checked for the effect of this third source of uncertainty by calculating the abundance spreads when the physical conditions are fixed to the values implied by the default datasets. We find that in this case the spread in ionic abundances remains below or around 0.1 dex for all ions except Ar$^{3+}$, that reaches spreads of a factor of two because of the large differences in its available atomic data.

\subsection{Total abundances}
\label{sec:tabundances}

\begin{figure*}
	\includegraphics[width=1\textwidth]{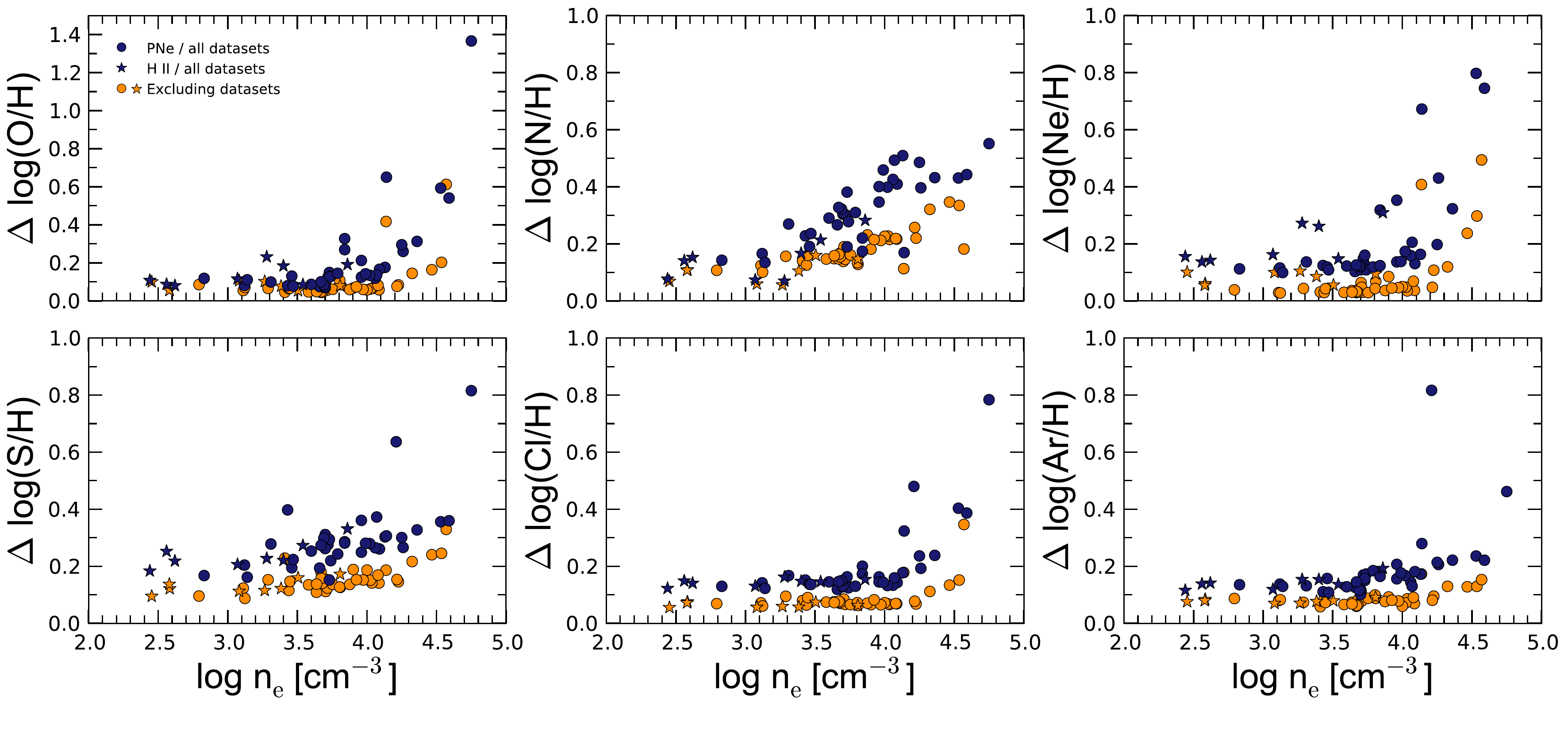}
	\vspace{-0.9cm}
    \caption{Widths of the distributions of total abundances in all the objects as a function of the median density. The dark (blue) symbols show the results implied by all the available 52 datasets; the light (orange) symbols show the results when we exclude the four datasets that lead to discordant values of density or temperature. A color version of this plot is available in the online article.}
    \label{fig:comparacion_ab_totales}
\end{figure*}

The total abundances are based on the ionic abundances of O$^{+}$, O$^{++}$, N$^{+}$, Cl$^{++}$, Ar$^{++}$, Ne$^{++}$, S$^{+}$, and S$^{++}$ along with ICFs that depend on $\mbox{He}^{++}/(\mbox{He}^{+}+\mbox{He}^{++})$ for oxygen and on  $\mbox{O}^{++}/(\mbox{O}^{+}+\mbox{O}^{++})$ for the other elements \citep{DelIng14}. Note that the ICF for argon is based on the Ar$^{++}$ so that the argon total abundances do not reflect the uncertainties introduced by the atomic data in the Ar$^{+3}$ abundance. The resulting distributions for NGC~6572 and 30~Doradus are shown in Figs.~\ref{fig:total_abundances1} and \ref{fig:total_abundances2}, whereas Fig.~\ref{fig:comparacion_ab_totales} shows the widths of the distributions as a function of density. Since the procedure used to derive total abundances depends on the degree of ionization of each object, the widths of the distributions have a larger dispersion for a given density than the ionic abundances, but the figures show that the uncertainties introduced by atomic data remain low, below or around 0.2 dex, at low densities, $n_{\rm{e}}\la10^3$~cm$^{-3}$, but increase to factors of 2--3 at $n_{\rm{e}}\sim10^4$~cm$^{-3}$, reaching factors of 4--6 at the highest densities. At a given density, the uncertainties in O/H, Ne/H, Cl/H, and Ar/H are larger for objects with higher values of O$^+$/O; but N/H and S/H are less uncertain when O$^+$/O is higher.

\begin{figure*}
	\includegraphics[width=1\textwidth]{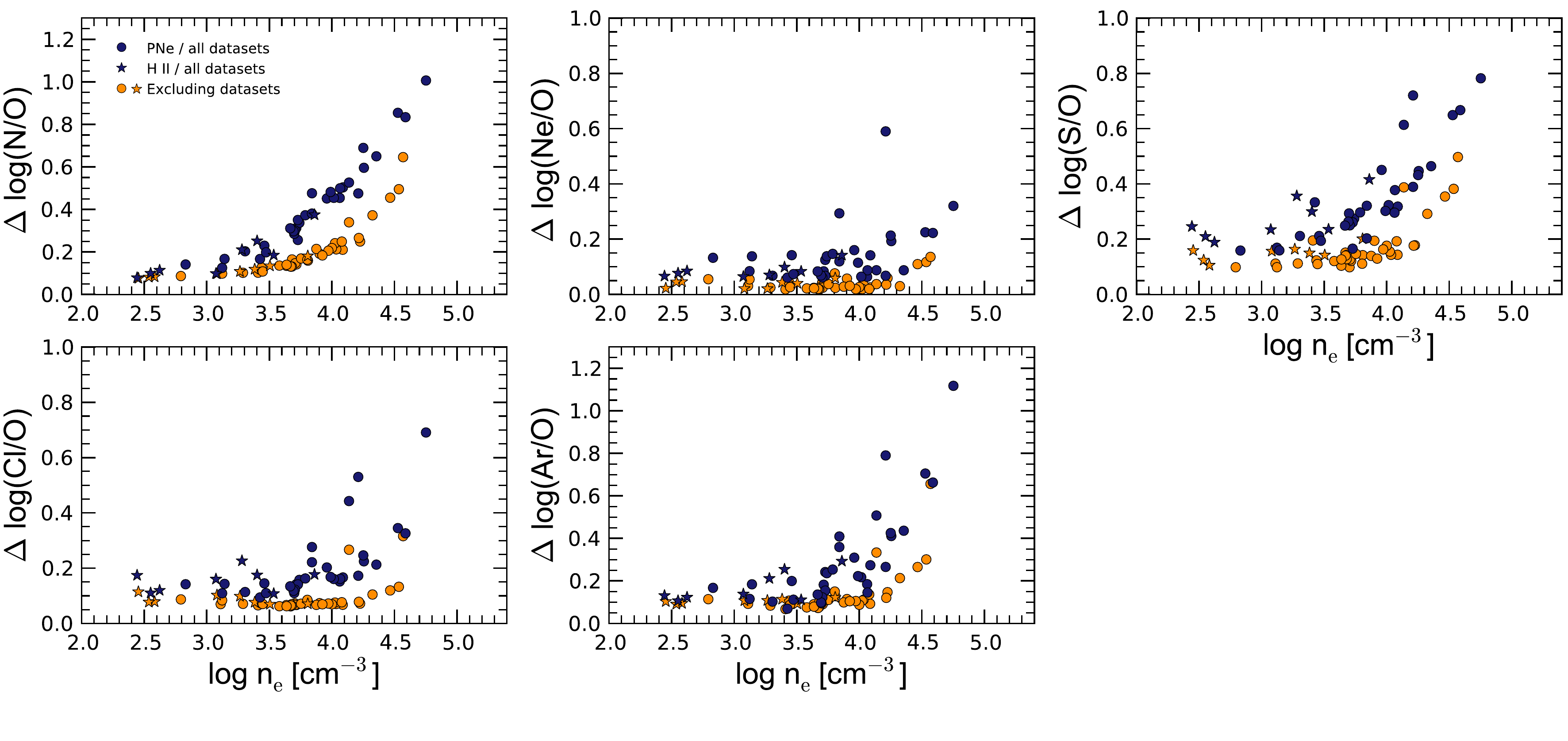}
	\vspace{-0.9cm}
    \caption{Widths of the distributions of the total abundances with respect to oxygen as a function of the median density. The dark (blue) symbols show the results implied by all the available 52 datasets; the light (orange) symbols show the results when we exclude the four datasets that lead to discordant values of density or temperature. A color version of this plot is available in the online article.}
    \label{fig:Comparacion_oxigeno2}
\end{figure*}

Total abundances are often presented with respect to the oxygen abundance, and Fig.\ref{fig:Comparacion_oxigeno2} shows the uncertainties introduced by atomic data on these relative abundances. The uncertainties are especially important for N/O and Ar/O: for these abundance ratios they reach a spread of one order of magnitude at the highest densities we are considering. In fact, for N/O, Ar/O, and also S/O, the uncertainties are larger than for N/H, Ar/H, and S/H, respectively. For neon we have the opposite behaviour, much larger uncertainties for Ne/H than for Ne/O, whereas for chlorine the values of Cl/H only have slightly larger uncertainties than those for Cl/O. At a given density, the uncertainties in all the abundances relative to oxygen are larger for objects with higher values of O$^+$/O.

\section{Summary and conclusions}

We have explored the uncertainties introduced by different atomic data in nebular abundance determinations. We have used high-quality spectra of 36 PNe and eight \ion{H}{ii} regions taken from the literature to compute in a homogeneous way the physical conditions, $n_{\rm{e}}$[\ion{S}{ii}], $n_{\rm{e}}$[\ion{O}{ii}], $n_{\rm{e}}$[\ion{Cl}{iii}], $n_{\rm{e}}$[\ion{Ar}{iv}], $T_{\rm{e}}$[\ion{N}{ii}], and $T_{\rm{e}}$[\ion{O}{iii}], the ionic abundances of O$^{+}$, O$^{++}$, N$^{+}$, Cl$^{++}$, Ar$^{++}$, Ar$^{3+}$, Ne$^{++}$, S$^{+}$, and S$^{++}$, and the total abundances of the corresponding elements using the compilation of atomic data in {\sc pyneb} \citep{Pyneb}. We have sampled all the possible combinations of atomic data and measured the widths of the distributions of values obtained for each quantity.

We find that the choice of atomic data can have a large impact on the derived physical conditions and chemical abundances. As shown in Fig.~\ref{fig:Comparacion_densidad}, the different atomic data lead to variations of 0.2 to 0.3~dex in $n_{\rm{e}}$[\ion{S}{ii}], $n_{\rm{e}}$[\ion{O}{ii}], and $n_{\rm{e}}$[\ion{Cl}{iii}] at low densities. However, the variations increase with density and can reach one order of magnitude at $n_{\rm{e}}\ge10^4$~cm$^{-3}$. The values of $n_{\rm{e}}$[\ion{Ar}{iv}] vary by $\sim0.8$~dex at all densities. The differences in the values calculated for $T_{\rm{e}}$[\ion{O}{iii}] and $T_{\rm{e}}$[\ion{N}{ii}] increase with average density (see Fig.~\ref{fig:Comparacion_Temperaturas}), going from a few per cent at low densities, to 20 per cent for $T_{\rm{e}}$[\ion{O}{iii}] and nearly a factor of 2 for $T_{\rm{e}}$[\ion{N}{ii}] at densities $\sim10^{4.7}$~cm$^{-3}$.

The effect of the different atomic datasets in the derived ionic and total abundances also increases with density, with variations below or around 0.2~dex at $n_{\rm{e}}\le10^3$~cm$^{-3}$, around 0.2 to 0.4~dex at $10^3\mbox{~cm}^{-3}\le{n_{\rm{e}}}\le10^4$~cm$^{-3}$, and around or above 0.4--0.6~dex at $n_{\rm{e}}\ge10^4$~cm$^{-3}$ for most elements (see Figs.~\ref{fig:Comparacion_ab_ionicas} to \ref{fig:Comparacion_oxigeno2}). The values of N/O are especially sensitive to changes in atomic data, with variations that reach one order of magnitude at high density.

Four of the atomic datasets lead to discordant values of density and temperature: the transition probabilities of \citet{TZ10} for S$^+$, the transition probabilities of \citet{WFD96} for O$^+$, the collision strengths of \citet{M83} for Ar$^{+3}$, and the collision strengths of \citet{Pal12} for O$^{++}$. The datasets for S$^+$, O$^+$, and Ar$^{+3}$ imply densities that do not agree with those implied by the other datasets available for the same diagnostic and that differ significantly from the densities inferred from the other density diagnostics in most objects. Besides, the transition probabilities of \citet{TZ10} cannot reproduce the values of the [\ion{S}{ii}] diagnostic in six of our high-density objects. As for the O$^{++}$ collision strengths of \citet{Pal12}, they lead to temperatures that are always lower than those implied by the other O$^{++}$ datasets. This does not necessarily imply that these collision strengths are faulty, but \citet{SSB14} could not reproduce the results of \citet{Pal12}. Hence, we repeated all the calculations of physical conditions and ionic and total abundances excluding these four datasets. The remaining 48 datasets imply uncertainties that are significantly lower, with abundance ratios that differ in less than 0.2--0.3~dex in most objects, although some high-density nebulae still reach uncertainties of 0.4~dex in S/O, 0.5~dex in Ne/H, and 0.6~dex in O/H and N/O.

Most of the uncertainties introduced by atomic data in the determination of chemical abundances arise from the density differences found at high density. We have illustrated the dependence of the density diagnostics on atomic data using a three-level approximation (eq.~\ref{eq1}). This allows us to identify the atomic data that have the largest impact on the results, namely, the transition probabilities of the [\ion{S}{ii}], [\ion{O}{ii}], [\ion{Cl}{iii}], and [\ion{Ar}{iv}] lines used in the density diagnostics, and the collision strengths for Ar$^{3+}$. Improved determinations of these atomic data will reduce the significant impact that atomic data have on the chemical abundances derived for high-density objects. The data for Cl$^{++}$ and Ar$^{3+}$ are especially critical, since these are the ions whose diagnostics work better at high density. In the meantime, our results can be used to estimate the systematic uncertainties introduced by atomic data for any PN or \ion{H}{ii} region whose density is known.

\section*{Acknowledgements}

We thank the referee, P.\ Storey, for several useful comments that helped to improve this paper. We acknowledge support from Mexican CONACYT grant CB-2014-240562. LJdD acknowledges support from CONACYT grant 298356. 



\bibliographystyle{mnras}
\bibliography{Bib} 

\begin{thebibliography}{}
\makeatletter
\relax
\def\mn@urlcharsother{\let\do\@makeother \do\$\do\&\do\#\do\^\do\_\do\%\do\~}
\def\mn@doi{\begingroup\mn@urlcharsother \@ifnextchar [ {\mn@doi@}
  {\mn@doi@[]}}
\def\mn@doi@[#1]#2{\def\@tempa{#1}\ifx\@tempa\@empty \href
  {http://dx.doi.org/#2} {doi:#2}\else \href {http://dx.doi.org/#2} {#1}\fi
  \endgroup}
\def\mn@eprint#1#2{\mn@eprint@#1:#2::\@nil}
\def\mn@eprint@arXiv#1{\href {http://arxiv.org/abs/#1} {{\tt arXiv:#1}}}
\def\mn@eprint@dblp#1{\href {http://dblp.uni-trier.de/rec/bibtex/#1.xml}
  {dblp:#1}}
\def\mn@eprint@#1:#2:#3:#4\@nil{\def\@tempa {#1}\def\@tempb {#2}\def\@tempc
  {#3}\ifx \@tempc \@empty \let \@tempc \@tempb \let \@tempb \@tempa \fi \ifx
  \@tempb \@empty \def\@tempb {arXiv}\fi \@ifundefined
  {mn@eprint@\@tempb}{\@tempb:\@tempc}{\expandafter \expandafter \csname
  mn@eprint@\@tempb\endcsname \expandafter{\@tempc}}}

\bibitem[\protect\citeauthoryear{{Aggarwal} \& {Keenan}}{{Aggarwal} \&
  {Keenan}}{1999}]{AK99}
{Aggarwal} K.~M.,  {Keenan} F.~P.,  1999, \mn@doi [\apjs] {10.1086/313232},
  \href {http://adsabs.harvard.edu/abs/1999ApJS..123..311A} {123, 311}

\bibitem[\protect\citeauthoryear{{Aggarwal} \& {Keenan}}{{Aggarwal} \&
  {Keenan}}{2013}]{Aggarwal13}
{Aggarwal} K.,  {Keenan} F.,  2013, Fusion Science and Technology, \href
  {http://adsabs.harvard.edu/abs/2013arXiv1301.3002A} {63, 3}

\bibitem[\protect\citeauthoryear{{Blum} \& {Pradhan}}{{Blum} \&
  {Pradhan}}{1992}]{BP92}
{Blum} R.~D.,  {Pradhan} A.~K.,  1992, \mn@doi [\apjs] {10.1086/191670}, \href
  {http://adsabs.harvard.edu/abs/1992ApJS...80..425B} {80, 425}

\bibitem[\protect\citeauthoryear{{Butler} \& {Zeippen}}{{Butler} \&
  {Zeippen}}{1989}]{BZ89}
{Butler} K.,  {Zeippen} C.~J.,  1989, \aap, \href
  {http://adsabs.harvard.edu/abs/1989A%26A...208..337B} {208, 337}

\bibitem[\protect\citeauthoryear{{Butler} \& {Zeippen}}{{Butler} \&
  {Zeippen}}{1994}]{BZ94}
{Butler} K.,  {Zeippen} C.~J.,  1994, \aaps, \href
  {http://adsabs.harvard.edu/abs/1994A%26AS..108....1B} {108}

\bibitem[\protect\citeauthoryear{{Copetti} \& {Writzl}}{{Copetti} \&
  {Writzl}}{2002}]{Copetti2002}
{Copetti} M.~V.~F.,  {Writzl} B.~C.,  2002, \mn@doi [\aap]
  {10.1051/0004-6361:20011621}, \href
  {http://adsabs.harvard.edu/abs/2002A%26A...382..282C} {382, 282}

\bibitem[\protect\citeauthoryear{{Delgado-Inglada}, {Morisset}  \&
  {Stasi{\'n}ska}}{{Delgado-Inglada} et~al.}{2014}]{DelIng14}
{Delgado-Inglada} G.,  {Morisset} C.,   {Stasi{\'n}ska} G.,  2014, \mn@doi
  [\mnras] {10.1093/mnras/stu341}, \href
  {http://adsabs.harvard.edu/abs/2014MNRAS.440..536D} {440, 536}

\bibitem[\protect\citeauthoryear{{Delgado-Inglada}, {Rodr{\'{\i}}guez},
  {Peimbert}, {Stasi{\'n}ska}  \& {Morisset}}{{Delgado-Inglada}
  et~al.}{2015}]{DelIng15}
{Delgado-Inglada} G.,  {Rodr{\'{\i}}guez} M.,  {Peimbert} M.,  {Stasi{\'n}ska}
  G.,   {Morisset} C.,  2015, \mn@doi [\mnras] {10.1093/mnras/stv388}, \href
  {http://adsabs.harvard.edu/abs/2015MNRAS.449.1797D} {449, 1797}

\bibitem[\protect\citeauthoryear{{Esteban}, {Peimbert}, {Garc{\'{\i}}a-Rojas},
  {Ruiz}, {Peimbert}  \& {Rodr{\'{\i}}guez}}{{Esteban}
  et~al.}{2004}]{Esteban04}
{Esteban} C.,  {Peimbert} M.,  {Garc{\'{\i}}a-Rojas} J.,  {Ruiz} M.~T.,
  {Peimbert} A.,   {Rodr{\'{\i}}guez} M.,  2004, \mn@doi [\mnras]
  {10.1111/j.1365-2966.2004.08313.x}, \href
  {http://adsabs.harvard.edu/abs/2004MNRAS.355..229E} {355, 229}

\bibitem[\protect\citeauthoryear{{Fang} \& {Liu}}{{Fang} \&
  {Liu}}{2011}]{Fang11}
{Fang} X.,  {Liu} X.-W.,  2011, \mn@doi [\mnras]
  {10.1111/j.1365-2966.2011.18681.x}, \href
  {http://adsabs.harvard.edu/abs/2011MNRAS.415..181F} {415, 181}

\bibitem[\protect\citeauthoryear{Fischer \& Tachiev}{Fischer \&
  Tachiev}{2004}]{FFT04}
Fischer C.~F.,  Tachiev G.,  2004, \mn@doi [At. Data and Nucl. Data Tables]
  {http://dx.doi.org/10.1016/j.adt.2004.02.001}, 87, 1

\bibitem[\protect\citeauthoryear{{Fritzsche}, {Fricke}, {Geschke}, {Heitmann}
  \& {Sienkiewicz}}{{Fritzsche} et~al.}{1999}]{Fal99}
{Fritzsche} S.,  {Fricke} B.,  {Geschke} D.,  {Heitmann} A.,   {Sienkiewicz}
  J.~E.,  1999, \mn@doi [\apj] {10.1086/307328}, \href
  {http://adsabs.harvard.edu/abs/1999ApJ...518..994F} {518, 994}

\bibitem[\protect\citeauthoryear{{Galavis}, {Mendoza}  \& {Zeippen}}{{Galavis}
  et~al.}{1995}]{GMZ95}
{Galavis} M.~E.,  {Mendoza} C.,   {Zeippen} C.~J.,  1995, \aaps, \href
  {http://adsabs.harvard.edu/abs/1995A%26AS..111..347G} {111, 347}

\bibitem[\protect\citeauthoryear{{Galavis}, {Mendoza}  \& {Zeippen}}{{Galavis}
  et~al.}{1997}]{GMZ97}
{Galavis} M.~E.,  {Mendoza} C.,   {Zeippen} C.~J.,  1997, \mn@doi [\aaps]
  {10.1051/aas:1997344}, \href
  {http://adsabs.harvard.edu/abs/1997A%26AS..123..159G} {123}

\bibitem[\protect\citeauthoryear{{Galavis}, {Mendoza}  \& {Zeippen}}{{Galavis}
  et~al.}{1998}]{GMZ98}
{Galavis} M.~E.,  {Mendoza} C.,   {Zeippen} C.~J.,  1998, \mn@doi [\aaps]
  {10.1051/aas:1998435}, \href
  {http://adsabs.harvard.edu/abs/1998A%26AS..131..499G} {131, 499}

\bibitem[\protect\citeauthoryear{{Garc{\'{\i}}a-Rojas}, {Esteban}, {Peimbert},
  {Rodr{\'{\i}}guez}, {Ruiz}  \& {Peimbert}}{{Garc{\'{\i}}a-Rojas}
  et~al.}{2004}]{GarciaR04}
{Garc{\'{\i}}a-Rojas} J.,  {Esteban} C.,  {Peimbert} M.,  {Rodr{\'{\i}}guez}
  M.,  {Ruiz} M.~T.,   {Peimbert} A.,  2004, \mn@doi [\apjs] {10.1086/421909},
  \href {http://adsabs.harvard.edu/abs/2004ApJS..153..501G} {153, 501}

\bibitem[\protect\citeauthoryear{{Garc{\'{\i}}a-Rojas}, {Esteban}, {Peimbert},
  {Costado}, {Rodr{\'{\i}}guez}, {Peimbert}  \& {Ruiz}}{{Garc{\'{\i}}a-Rojas}
  et~al.}{2006}]{GarciaR06}
{Garc{\'{\i}}a-Rojas} J.,  {Esteban} C.,  {Peimbert} M.,  {Costado} M.~T.,
  {Rodr{\'{\i}}guez} M.,  {Peimbert} A.,   {Ruiz} M.~T.,  2006, \mn@doi
  [\mnras] {10.1111/j.1365-2966.2006.10105.x}, \href
  {http://adsabs.harvard.edu/abs/2006MNRAS.368..253G} {368, 253}

\bibitem[\protect\citeauthoryear{{Garc{\'{\i}}a-Rojas}, {Esteban}, {Peimbert},
  {Rodr{\'{\i}}guez}, {Peimbert}  \& {Ruiz}}{{Garc{\'{\i}}a-Rojas}
  et~al.}{2007}]{GarciaR07}
{Garc{\'{\i}}a-Rojas} J.,  {Esteban} C.,  {Peimbert} A.,  {Rodr{\'{\i}}guez}
  M.,  {Peimbert} M.,   {Ruiz} M.~T.,  2007, \rmxaa, \href
  {http://adsabs.harvard.edu/abs/2007RMxAA..43....3G} {43, 3}

\bibitem[\protect\citeauthoryear{{Garc{\'{\i}}a-Rojas}, {Pe{\~n}a}  \&
  {Peimbert}}{{Garc{\'{\i}}a-Rojas} et~al.}{2009}]{GarciaR09}
{Garc{\'{\i}}a-Rojas} J.,  {Pe{\~n}a} M.,   {Peimbert} A.,  2009, \mn@doi
  [\aap] {10.1051/0004-6361:200811185}, \href
  {http://adsabs.harvard.edu/abs/2009A%26A...496..139G} {496, 139}

\bibitem[\protect\citeauthoryear{{Garc{\'{\i}}a-Rojas}, {Pe{\~n}a}, {Morisset},
  {Mesa-Delgado}  \& {Ruiz}}{{Garc{\'{\i}}a-Rojas} et~al.}{2012}]{GarciaR12}
{Garc{\'{\i}}a-Rojas} J.,  {Pe{\~n}a} M.,  {Morisset} C.,  {Mesa-Delgado} A.,
  {Ruiz} M.~T.,  2012, \mn@doi [\aap] {10.1051/0004-6361/201118217}, \href
  {http://adsabs.harvard.edu/abs/2012A%26A...538A..54G} {538, A54}

\bibitem[\protect\citeauthoryear{{Grieve}, {Ramsbottom}, {Hudson}  \&
  {Keenan}}{{Grieve} et~al.}{2014}]{GRHK14}
{Grieve} M.~F.~R.,  {Ramsbottom} C.~A.,  {Hudson} C.~E.,   {Keenan} F.~P.,
  2014, \mn@doi [\apj] {10.1088/0004-637X/780/1/110}, \href
  {http://adsabs.harvard.edu/abs/2014ApJ...780..110G} {780, 110}

\bibitem[\protect\citeauthoryear{{Heise}, {Smith}  \& {Calamai}}{{Heise}
  et~al.}{1995}]{HSC95}
{Heise} C.,  {Smith} P.~L.,   {Calamai} A.~G.,  1995, \mn@doi [\apjl]
  {10.1086/309676}, \href {http://adsabs.harvard.edu/abs/1995ApJ...451L..41H}
  {451, L41}

\bibitem[\protect\citeauthoryear{{Hudson} \& {Bell}}{{Hudson} \&
  {Bell}}{2004}]{HB04}
{Hudson} C.~E.,  {Bell} K.~L.,  2004, \mn@doi [\mnras]
  {10.1111/j.1365-2966.2004.07461.x}, \href
  {http://adsabs.harvard.edu/abs/2004MNRAS.348.1275H} {348, 1275}

\bibitem[\protect\citeauthoryear{{Hudson}, {Ramsbottom}  \& {Scott}}{{Hudson}
  et~al.}{2012}]{HRS12}
{Hudson} C.~E.,  {Ramsbottom} C.~A.,   {Scott} M.~P.,  2012, \mn@doi [\apj]
  {10.1088/0004-637X/750/1/65}, \href
  {http://adsabs.harvard.edu/abs/2012ApJ...750...65H} {750, 65}

\bibitem[\protect\citeauthoryear{{Hyung}}{{Hyung}}{1994}]{Hyung94}
{Hyung} S.,  1994, \mn@doi [\apjs] {10.1086/191860}, \href
  {http://adsabs.harvard.edu/abs/1994ApJS...90..119H} {90, 119}

\bibitem[\protect\citeauthoryear{{Hyung}, {Aller}  \& {Lee}}{{Hyung}
  et~al.}{2001a}]{Hyung01}
{Hyung} S.,  {Aller} L.~H.,   {Lee} W.-b.,  2001a, \mn@doi [\pasp]
  {10.1086/324415}, \href {http://adsabs.harvard.edu/abs/2001PASP..113.1559H}
  {113, 1559}

\bibitem[\protect\citeauthoryear{{Hyung}, {Aller}, {Feibelman}  \&
  {Lee}}{{Hyung} et~al.}{2001b}]{Hyung01b}
{Hyung} S.,  {Aller} L.~H.,  {Feibelman} W.~A.,   {Lee} W.-B.,  2001b, \mn@doi
  [\aj] {10.1086/321171}, \href
  {http://adsabs.harvard.edu/abs/2001AJ....122..954H} {122, 954}

\bibitem[\protect\citeauthoryear{{Kaufman} \& {Sugar}}{{Kaufman} \&
  {Sugar}}{1986}]{KS86}
{Kaufman} V.,  {Sugar} J.,  1986, \mn@doi [Journal of Physical and Chemical
  Reference Data] {10.1063/1.555775}, \href
  {http://adsabs.harvard.edu/abs/1986JPCRD..15..321K} {15, 321}

\bibitem[\protect\citeauthoryear{{Keenan}, {Hibbert}, {Ojha}  \&
  {Conlon}}{{Keenan} et~al.}{1993}]{KHOC93}
{Keenan} F.~P.,  {Hibbert} A.,  {Ojha} P.~C.,   {Conlon} E.~S.,  1993, \mn@doi
  [\physscr] {10.1088/0031-8949/48/2/001}, \href
  {http://adsabs.harvard.edu/abs/1993PhyS...48..129K} {48, 129}

\bibitem[\protect\citeauthoryear{{Kisielius}, {Storey}, {Ferland}  \&
  {Keenan}}{{Kisielius} et~al.}{2009}]{Kal09}
{Kisielius} R.,  {Storey} P.~J.,  {Ferland} G.~J.,   {Keenan} F.~P.,  2009,
  \mn@doi [\mnras] {10.1111/j.1365-2966.2009.14989.x}, \href
  {http://adsabs.harvard.edu/abs/2009MNRAS.397..903K} {397, 903}

\bibitem[\protect\citeauthoryear{{LaJohn} \& {Luke}}{{LaJohn} \&
  {Luke}}{1993}]{LL93}
{LaJohn} L.,  {Luke} T.~M.,  1993, \mn@doi [\physscr]
  {10.1088/0031-8949/47/4/011}, \href
  {http://adsabs.harvard.edu/abs/1993PhyS...47..542L} {47, 542}

\bibitem[\protect\citeauthoryear{{Lennon} \& {Burke}}{{Lennon} \&
  {Burke}}{1994}]{LB94}
{Lennon} D.~J.,  {Burke} V.~M.,  1994, \aaps, \href
  {http://adsabs.harvard.edu/abs/1994A%26AS..103..273L} {103}

\bibitem[\protect\citeauthoryear{{Liu}, {Storey}, {Barlow}, {Danziger}, {Cohen}
   \& {Bryce}}{{Liu} et~al.}{2000}]{Liu00}
{Liu} X.-W.,  {Storey} P.~J.,  {Barlow} M.~J.,  {Danziger} I.~J.,  {Cohen} M.,
   {Bryce} M.,  2000, \mn@doi [\mnras] {10.1046/j.1365-8711.2000.03167.x},
  \href {http://adsabs.harvard.edu/abs/2000MNRAS.312..585L} {312, 585}

\bibitem[\protect\citeauthoryear{{Liu}, {Luo}, {Barlow}, {Danziger}  \&
  {Storey}}{{Liu} et~al.}{2001}]{Liu01}
{Liu} X.-W.,  {Luo} S.-G.,  {Barlow} M.~J.,  {Danziger} I.~J.,   {Storey}
  P.~J.,  2001, \mn@doi [\mnras] {10.1046/j.1365-8711.2001.04676.x}, \href
  {http://adsabs.harvard.edu/abs/2001MNRAS.327..141L} {327, 141}

\bibitem[\protect\citeauthoryear{{Liu}, {Liu}, {Luo}  \& {Barlow}}{{Liu}
  et~al.}{2004}]{Liu04}
{Liu} Y.,  {Liu} X.-W.,  {Luo} S.-G.,   {Barlow} M.~J.,  2004, \mn@doi [\mnras]
  {10.1111/j.1365-2966.2004.08155.x}, \href
  {http://adsabs.harvard.edu/abs/2004MNRAS.353.1231L} {353, 1231}

\bibitem[\protect\citeauthoryear{{Luridiana} \&
  {Garc{\'{\i}}a-Rojas}}{{Luridiana} \&
  {Garc{\'{\i}}a-Rojas}}{2012}]{Luridiana12}
{Luridiana} V.,  {Garc{\'{\i}}a-Rojas} J.,  2012, in {Manchado} A.,
  {Stanghellini} L.,   {Sch\"{o}nberner} D.,  eds,  Proc. IAU Symposium Vol.
  283, Planeraty nebulae: An Eye to the Future. p.~139

\bibitem[\protect\citeauthoryear{{Luridiana}, {Morisset}  \&
  {Shaw}}{{Luridiana} et~al.}{2015}]{Pyneb}
{Luridiana} V.,  {Morisset} C.,   {Shaw} R.~A.,  2015, \mn@doi [\aap]
  {10.1051/0004-6361/201323152}, \href
  {http://adsabs.harvard.edu/abs/2015A%26A...573A..42L} {573, A42}

\bibitem[\protect\citeauthoryear{{McLaughlin} \& {Bell}}{{McLaughlin} \&
  {Bell}}{1993}]{McLB93}
{McLaughlin} B.~M.,  {Bell} K.~L.,  1993, \mn@doi [\apj] {10.1086/172635},
  \href {http://adsabs.harvard.edu/abs/1993ApJ...408..753M} {408, 753}

\bibitem[\protect\citeauthoryear{{McLaughlin} \& {Bell}}{{McLaughlin} \&
  {Bell}}{2000}]{McLB00}
{McLaughlin} B.~M.,  {Bell} K.~L.,  2000, \mn@doi [JPhB]
  {10.1088/0953-4075/33/4/301}, \href
  {http://cdsads.u-strasbg.fr/abs/2000JPhB...33..597M} {33, 597}

\bibitem[\protect\citeauthoryear{{Mendoza}}{{Mendoza}}{1983}]{M83}
{Mendoza} C.,  1983, in {Flower} D.~R.,  ed., Proc. IAU Symp. 103 Planetary
  Nebulae, p. 143.

\bibitem[\protect\citeauthoryear{{Mendoza} \& {Bautista}}{{Mendoza} \&
  {Bautista}}{2014}]{Mendoza14}
{Mendoza} C.,  {Bautista} M.~A.,  2014, \mn@doi [\apj]
  {10.1088/0004-637X/785/2/91}, \href
  {http://adsabs.harvard.edu/abs/2014ApJ...785...91M} {785, 91}

\bibitem[\protect\citeauthoryear{{Mendoza} \& {Zeippen}}{{Mendoza} \&
  {Zeippen}}{1982a}]{MZ82a}
{Mendoza} C.,  {Zeippen} C.~J.,  1982a, \mn@doi [\mnras]
  {10.1093/mnras/198.1.127}, \href
  {http://adsabs.harvard.edu/abs/1982MNRAS.198..127M} {198, 127}

\bibitem[\protect\citeauthoryear{{Mendoza} \& {Zeippen}}{{Mendoza} \&
  {Zeippen}}{1982b}]{MZ82b}
{Mendoza} C.,  {Zeippen} C.~J.,  1982b, \mn@doi [\mnras]
  {10.1093/mnras/199.4.1025}, \href
  {http://adsabs.harvard.edu/abs/1982MNRAS.199.1025M} {199, 1025}

\bibitem[\protect\citeauthoryear{{Mendoza} \& {Zeippen}}{{Mendoza} \&
  {Zeippen}}{1983}]{MZ83}
{Mendoza} C.,  {Zeippen} C.~J.,  1983, \mn@doi [\mnras]
  {10.1093/mnras/202.4.981}, \href
  {http://cdsads.u-strasbg.fr/abs/1983MNRAS.202..981M} {202, 981}

\bibitem[\protect\citeauthoryear{{Munoz Burgos}, {Loch}, {Ballance}  \&
  {Boivin}}{{Munoz Burgos} et~al.}{2009}]{MB09}
{Munoz Burgos} J.~M.,  {Loch} S.~D.,  {Ballance} C.~P.,   {Boivin} R.~F.,
  2009, \mn@doi [\aap] {10.1051/0004-6361/200911743}, \href
  {http://cdsads.u-strasbg.fr/abs/2009A\%26A...500.1253M} {500, 1253}

\bibitem[\protect\citeauthoryear{{Nussbaumer} \& {Rusca}}{{Nussbaumer} \&
  {Rusca}}{1979}]{NR79}
{Nussbaumer} H.,  {Rusca} C.,  1979, \aap, \href
  {http://adsabs.harvard.edu/abs/1979A%26A....72..129N} {72, 129}

\bibitem[\protect\citeauthoryear{Osterbrock \& Ferland}{Osterbrock \&
  Ferland}{2006}]{Osterbrock}
Osterbrock D.~E.,  Ferland G.~J.,  2006, Astrophysics of Gaseous Nebulae and
  Active Galactic Nuclei.
University Science Books, Mill Valley, CA

\bibitem[\protect\citeauthoryear{{Palay}, {Nahar}, {Pradhan}  \&
  {Eissner}}{{Palay} et~al.}{2012}]{Pal12}
{Palay} E.,  {Nahar} S.~N.,  {Pradhan} A.~K.,   {Eissner} W.,  2012, \mn@doi
  [\mnras] {10.1111/j.1745-3933.2012.01252.x}, \href
  {http://adsabs.harvard.edu/abs/2012MNRAS.423L..35P} {423, L35}

\bibitem[\protect\citeauthoryear{{Peimbert}}{{Peimbert}}{2003}]{PeimbertA03}
{Peimbert} A.,  2003, \mn@doi [\apj] {10.1086/345793}, \href
  {http://adsabs.harvard.edu/abs/2003ApJ...584..735P} {584, 735}

\bibitem[\protect\citeauthoryear{Podobedova, Kelleher  \& Wiese}{Podobedova
  et~al.}{2009}]{PKW09}
Podobedova L.~I.,  Kelleher D.~E.,   Wiese W.~L.,  2009, \mn@doi [Journal of
  Physical and Chemical Reference Data] {10.1063/1.3032939}, 38, 171

\bibitem[\protect\citeauthoryear{{Porter}, {Ferland}, {Storey}  \&
  {Detisch}}{{Porter} et~al.}{2012}]{Porter12}
{Porter} R.~L.,  {Ferland} G.~J.,  {Storey} P.~J.,   {Detisch} M.~J.,  2012,
  \mn@doi [\mnras] {10.1111/j.1745-3933.2012.01300.x}, \href
  {http://adsabs.harvard.edu/abs/2012MNRAS.425L..28P} {425, L28}

\bibitem[\protect\citeauthoryear{{Porter}, {Ferland}, {Storey}  \&
  {Detisch}}{{Porter} et~al.}{2013}]{Porter13}
{Porter} R.~L.,  {Ferland} G.~J.,  {Storey} P.~J.,   {Detisch} M.~J.,  2013,
  \mn@doi [\mnras] {10.1093/mnrasl/slt049}, \href
  {http://adsabs.harvard.edu/abs/2013MNRAS.433L..89P} {433, L89}

\bibitem[\protect\citeauthoryear{{Pradhan}}{{Pradhan}}{1976}]{P76}
{Pradhan} A.~K.,  1976, \mn@doi [\mnras] {10.1093/mnras/177.1.31}, \href
  {http://adsabs.harvard.edu/abs/1976MNRAS.177...31P} {177, 31}

\bibitem[\protect\citeauthoryear{{Pradhan}, {Montenegro}, {Nahar}  \&
  {Eissner}}{{Pradhan} et~al.}{2006}]{P06}
{Pradhan} A.~K.,  {Montenegro} M.,  {Nahar} S.~N.,   {Eissner} W.,  2006,
  \mn@doi [\mnras] {10.1111/j.1745-3933.2005.00119.x}, \href
  {http://adsabs.harvard.edu/abs/2006MNRAS.366L...6P} {366, L6}

\bibitem[\protect\citeauthoryear{{Ramsbottom} \& {Bell}}{{Ramsbottom} \&
  {Bell}}{1997}]{RB97}
{Ramsbottom} C.~A.,  {Bell} K.~L.,  1997, \mn@doi [At. Data and Nucl. Data
  Tables] {10.1006/adnd.1997.0741}, \href
  {http://adsabs.harvard.edu/abs/1997ADNDT..66...65R} {66, 65}

\bibitem[\protect\citeauthoryear{{Ramsbottom}, {Bell}  \&
  {Stafford}}{{Ramsbottom} et~al.}{1996}]{RBS96}
{Ramsbottom} C.~A.,  {Bell} K.~L.,   {Stafford} R.~P.,  1996, \mn@doi [At. Data
  and Nucl. Data Tables] {10.1006/adnd.1996.0009}, \href
  {http://adsabs.harvard.edu/abs/1996ADNDT..63...57R} {63, 57}

\bibitem[\protect\citeauthoryear{{Rodr{\'{\i}}guez}}{{Rodr{\'{\i}}guez}}{1999}%
]{rod99}
{Rodr{\'{\i}}guez} M.,  1999, \aap, \href
  {http://adsabs.harvard.edu/abs/1999A%26A...351.1075R} {351, 1075}

\bibitem[\protect\citeauthoryear{{Rubin}}{{Rubin}}{1986}]{Rubin86}
{Rubin} R.~H.,  1986, \mn@doi [\apj] {10.1086/164606}, \href
  {http://adsabs.harvard.edu/abs/1986ApJ...309..334R} {309, 334}

\bibitem[\protect\citeauthoryear{{Sharpee}, {Williams}, {Baldwin}  \& {van
  Hoof}}{{Sharpee} et~al.}{2003}]{Sharpee03}
{Sharpee} B.,  {Williams} R.,  {Baldwin} J.~A.,   {van Hoof} P.~A.~M.,  2003,
  \mn@doi [\apjs] {10.1086/378321}, \href
  {http://adsabs.harvard.edu/abs/2003ApJS..149..157S} {149, 157}

\bibitem[\protect\citeauthoryear{{Stasi{\'n}ska}, {Pe{\~n}a}, {Bresolin}  \&
  {Tsamis}}{{Stasi{\'n}ska} et~al.}{2013}]{Stasinska13}
{Stasi{\'n}ska} G.,  {Pe{\~n}a} M.,  {Bresolin} F.,   {Tsamis} Y.~G.,  2013,
  \mn@doi [\aap] {10.1051/0004-6361/201220345}, \href
  {http://adsabs.harvard.edu/abs/2013A%26A...552A..12S} {552, A12}

\bibitem[\protect\citeauthoryear{{Storey} \& {Hummer}}{{Storey} \&
  {Hummer}}{1995}]{SH95}
{Storey} P.~J.,  {Hummer} D.~G.,  1995, \mn@doi [\mnras]
  {10.1093/mnras/272.1.41}, \href
  {http://adsabs.harvard.edu/abs/1995MNRAS.272...41S} {272, 41}

\bibitem[\protect\citeauthoryear{{Storey} \& {Zeippen}}{{Storey} \&
  {Zeippen}}{2000}]{SZ00}
{Storey} P.~J.,  {Zeippen} C.~J.,  2000, \mn@doi [\mnras]
  {10.1046/j.1365-8711.2000.03184.x}, \href
  {http://adsabs.harvard.edu/abs/2000MNRAS.312..813S} {312, 813}

\bibitem[\protect\citeauthoryear{{Storey}, {Sochi}  \& {Badnell}}{{Storey}
  et~al.}{2014}]{SSB14}
{Storey} P.~J.,  {Sochi} T.,   {Badnell} N.~R.,  2014, \mn@doi [\mnras]
  {10.1093/mnras/stu777}, \href
  {http://adsabs.harvard.edu/abs/2014MNRAS.441.3028S} {441, 3028}

\bibitem[\protect\citeauthoryear{{Tayal}}{{Tayal}}{2007}]{T07}
{Tayal} S.~S.,  2007, \mn@doi [\apjs] {10.1086/513107}, \href
  {http://adsabs.harvard.edu/abs/2007ApJS..171..331T} {171, 331}

\bibitem[\protect\citeauthoryear{{Tayal}}{{Tayal}}{2011}]{T11}
{Tayal} S.~S.,  2011, \mn@doi [\apjs] {10.1088/0067-0049/195/2/12}, \href
  {http://adsabs.harvard.edu/abs/2011ApJS..195...12T} {195, 12}

\bibitem[\protect\citeauthoryear{{Tayal} \& {Gupta}}{{Tayal} \&
  {Gupta}}{1999}]{TG99}
{Tayal} S.~S.,  {Gupta} G.~P.,  1999, \mn@doi [\apj] {10.1086/307971}, \href
  {http://adsabs.harvard.edu/abs/1999ApJ...526..544T} {526, 544}

\bibitem[\protect\citeauthoryear{{Tayal} \& {Zatsarinny}}{{Tayal} \&
  {Zatsarinny}}{2010}]{TZ10}
{Tayal} S.~S.,  {Zatsarinny} O.,  2010, \mn@doi [\apjs]
  {10.1088/0067-0049/188/1/32}, \href
  {http://adsabs.harvard.edu/abs/2010ApJS..188...32T} {188, 32}

\bibitem[\protect\citeauthoryear{{Verner}, {Verner}  \& {Ferland}}{{Verner}
  et~al.}{1996}]{VVF96}
{Verner} D.~A.,  {Verner} E.~M.,   {Ferland} G.~J.,  1996, \mn@doi [At. Data
  and Nucl. Data Tables] {10.1006/adnd.1996.0018}, \href
  {http://adsabs.harvard.edu/abs/1996ADNDT..64....1V} {64, 1}

\bibitem[\protect\citeauthoryear{{Wang} \& {Liu}}{{Wang} \&
  {Liu}}{2007}]{Wang07}
{Wang} W.,  {Liu} X.-W.,  2007, \mn@doi [\mnras]
  {10.1111/j.1365-2966.2007.12198.x}, \href
  {http://adsabs.harvard.edu/abs/2007MNRAS.381..669W} {381, 669}

\bibitem[\protect\citeauthoryear{{Wang}, {Liu}, {Zhang}  \& {Barlow}}{{Wang}
  et~al.}{2004}]{Wang04}
{Wang} W.,  {Liu} X.-W.,  {Zhang} Y.,   {Barlow} M.~J.,  2004, \mn@doi [\aap]
  {10.1051/0004-6361:20041470}, \href
  {http://adsabs.harvard.edu/abs/2004A%26A...427..873W} {427, 873}

\bibitem[\protect\citeauthoryear{{Wesson}, {Liu}  \& {Barlow}}{{Wesson}
  et~al.}{2005}]{Wesson05}
{Wesson} R.,  {Liu} X.-W.,   {Barlow} M.~J.,  2005, \mn@doi [\mnras]
  {10.1111/j.1365-2966.2005.09325.x}, \href
  {http://adsabs.harvard.edu/abs/2005MNRAS.362..424W} {362, 424}

\bibitem[\protect\citeauthoryear{Wiese, Fuhr  \& Deters}{Wiese
  et~al.}{1996}]{WFD96}
Wiese W.~L.,  Fuhr J.~R.,   Deters T.~M.,  1996, Journal of Physical and
  Chemical Reference Data, Monograph 7, 403

\bibitem[\protect\citeauthoryear{{Zeippen}}{{Zeippen}}{1982}]{Z82}
{Zeippen} C.~J.,  1982, \mn@doi [\mnras] {10.1093/mnras/198.1.111}, \href
  {http://adsabs.harvard.edu/abs/1982MNRAS.198..111Z} {198, 111}

\bibitem[\protect\citeauthoryear{{Zeippen}, {Butler}  \& {Le
  Bourlot}}{{Zeippen} et~al.}{1987}]{ZBL87}
{Zeippen} C.~J.,  {Butler} K.,   {Le Bourlot} J.,  1987, \aap, \href
  {http://adsabs.harvard.edu/abs/1987A%26A...188..251Z} {188, 251}

\makeatother
\end{thebibliography}





\bsp	
\label{lastpage}
\end{document}